\documentclass[11pt]{article}
\pdfoutput=1
\usepackage{amsmath,amssymb,graphicx,multirow,xspace}
\usepackage[dvipsnames]{xcolor}
\usepackage[most]{tcolorbox}
\usepackage[colorlinks=true,linktocpage=true]{hyperref}
\hypersetup{                   %
    colorlinks = true,         
    linkcolor  = NavyBlue, 
    urlcolor   = NavyBlue,     
    citecolor  = NavyBlue,  
    anchorcolor = NavyBlue,
    pagecolor = NavyBlue
}
\usepackage[compress,numbers,sort]{natbib}
\usepackage{braket}
\usepackage{slashed}
\usepackage{physics}
\usepackage{bm}
\usepackage{indentfirst}
\usepackage{appendix}
\usepackage{verbatim}
\usepackage{cancel}

\usepackage[T1]{fontenc}
\long\def\symbolfootnote[#1]#2{\begingroup%
\def\thefootnote{\fnsymbol{footnote}}\footnote[#1]{#2}\endgroup}

\newcommand{\gev}{\mathrm{GeV}}
\newcommand{\tev}{\mathrm{TeV}}
\newcommand{\ol}{\overline}

\input prepictex
\input pictex
\input postpictex
\newdimen\tdim
\tdim=\unitlength

\begin{document}

\begin{titlepage}
 \hfill{FERMILAB-PUB-26-0493-T}
    \begin{center}
        {\LARGE\bf
            Accidentally Stable Dark Matter in\\ a Parity Solution to the Strong CP Problem \par}
    \end{center}

    \vspace{0.2cm}
    \begin{center}
        {\large
            Matthew J.~Baldwin,$^{1}$\symbolfootnote[1]{mjbaldwin@uchicago.edu}
            Keisuke Harigaya,$^{1,2}$\symbolfootnote[2]{kharigaya@uchicago.edu}
            Isaac R. Wang,$^{3,4}$\symbolfootnote[3]{isaacw@fnal.gov}
        }\\
        \vspace{0.5cm}
        \textit{
            $\,^1$ Department of Physics, Enrico Fermi Institute, Leinweber Institute for Theoretical Physics, and Kavli Institute for Cosmological Physics, University of Chicago, Chicago, IL 60637, USA \\
            \vspace{0.5cm}
            $\,^2$ Kavli Institute for the Physics and Mathematics of the Universe (WPI),\\
            The University of Tokyo Institutes for Advanced Study,\\
            The University of Tokyo, Kashiwa, Chiba 277-8583, Japan \\
            \vspace{0.5cm}
            $\,^3$ New High Energy Theory Center, \\
            Department of Physics and Astronomy,\\
            Rutgers University, NJ 08854, USA\\
            \vspace{0.5cm}
            $\,^4$ Theory Division,\\
            Fermi National Accelerator Laboratory,\\
            IL 60510, USA}\\
    \end{center}

    \vspace{0.4cm}

    \begin{abstract}
        {\large
            Parity symmetry, with an extended gauge group $SU(3)_c \times SU(2)_L \times SU(2)_R \times U(1)_X$, can solve the strong CP problem. In particular, the model where $SU(2)_R\times U(1)_X$ is broken by the Parity partner of the Standard Model Higgs solves the strong CP problem without the necessity of introducing extra symmetry. We discuss the possibility of accidentally stable dark matter in this framework and show that  $SU(2)_L \times SU(2)_R$ bi-triplet fermions can be stable over cosmological timescales. We compute the relic abundance of the bi-triplet dark matter and derive constraints on the parameter space from collider, direct-detection, and indirect-detection experiments. The $SU(2)_R\times U(1)_X$ symmetry breaking scale is required to be below 150 TeV, and most of the parameter space can be probed by near-future indirect-detection experiments. 
        }
    \end{abstract}

\end{titlepage}

\vspace{0.2cm}
\noindent

\noindent\makebox[\linewidth]{\rule{\textwidth}{1pt}}
\tableofcontents
\noindent\makebox[\linewidth]{\rule{\textwidth}{1pt}}
\newpage

\section{Introduction}

The absence of CP violation in the strong interaction of the Standard Model (SM)-- the strong CP problem-- has remained a longstanding puzzle. 
Although the CP-violating phase in the quark mass matrix is expected to generate observable CP-violating effects~\cite{Bell:1969ts,Adler:1969gk,tHooft:1976rip,tHooft:1976snw}, the upper limit on the neutron electric dipole moment, $d_n \leq 10^{-26}~e\cdot\rm cm$~\cite{nEDM:2020crw}, requires the strong CP phase to be nearly vanishing. Such a puzzle strongly motivates theories with Parity symmetry. 
In the minimal setup, the SM gauge group is extended to $SU(3)_c\times SU(2)_L\times SU(2)_R\times U(1)_X$, and Parity symmetry exchanges $SU(2)_L$-charged particles with their $SU(2)_R$-charged partners~\cite{Beg:1978mt,Mohapatra:1978fy,Babu:1988mw,Babu:1989rb,Kuchimanchi:1995rp,Mohapatra:1995xd,Hall:2018let}. Parity-symmetric models with different gauge groups can be found in Refs.~\cite{Barr:1991qx,Dunsky:2019api,Babu:2023dzz}.

Among these Parity-symmetric scenarios, we focus on \emph{the minimal Higgs model} where $SU(2)_R\times U(1)_X$ is broken to $U(1)_Y$ by the condensation of the Parity partner of the SM Higgs~\cite{Babu:1988mw,Babu:1989rb,Hall:2018let}. The model can indeed solve the strong CP problem without introducing an additional symmetry~\cite{Hall:2018let,Hisano:2023izx}, such as supersymmetry~\cite{Kuchimanchi:1995rp,Mohapatra:1995xd} or CP symmetry~\cite{Kuchimanchi:2023imj}. Furthermore, the hierarchy problem associated with the $SU(2)_R\times U(1)_X$ symmetry breaking scale can be avoided~\cite{Hall:2018let,Blinov:2016kte,Baldwin:2025oqt}.
The minimal Higgs model has been shown to be consistent with solutions of other problems in the frontier of particle physics; neutrino masses can be generated in various ways~\cite{Babu:1988yq,Harigaya:2021txz,Babu:2022ikf,Dror:2020jzy,Hall:2023vjb,Harigaya:2025zru}, and baryon number asymmetry can be generated from an $SU(2)_R$ phase transition~\cite{Harigaya:2022wzt} or leptogenesis~\cite{Gu:2010yf,Dunsky:2020dhn,Carrasco-Martinez:2023nit,Babu:2024glr,Babu:2025vjm}. 

A drawback of the minimal Higgs model, however, is the absence of an unbroken subgroup of $U(1)_X$ that can explain the stability of dark matter (DM). (See Refs.~\cite{Heeck:2015qra,Garcia-Cely:2015quu} for models in which $SU(2)_R\times U(1)_X$ is broken by a Higgs in $({\bf 3},1)$, leaving an unbroken subgroup of $U(1)_X$ that ensures the absolute stability of DM.) 
Therefore, previous studies on DM in the minimal Higgs model introduce an ad hoc $\mathbb{Z}_2$ symmetry~\cite{Baldwin:2024bob,Baldwin:2025oqt} or small parameters~\cite{Dror:2020jzy} to stabilize DM. Studies of DM in theories with different Parity-symmetric gauge groups can be found in Refs.~\cite{Kawamura:2018kut,Dunsky:2019api}.

In this paper, we investigate an alternative way to stabilize DM without imposing additional symmetries or requiring the smallness of certain parameters. We find that a fermionic WIMP candidate can be accidentally stabilized by the gauge symmetry $SU(3)_c\times SU(2)_L\times SU(2)_R\times U(1)_X$ when embedded in either $({\bf 1},{\bf 3},{\bf 3},0)$ or $({\bf 1},{\bf 3},{\bf 3},1)$. Such an accidental stability also arises in the so-called \emph{minimal dark matter} scenario~\cite{Cirelli:2005uq}, where DM is stabilized by the electroweak gauge symmetry. 

We compute the relic freeze-out abundance of DM as a function of the two free parameters of the model -- the DM mass and the $SU(2)_R\times U(1)_X$ breaking scale $v_R$. For sufficiently light DM, annihilation during freeze-out proceeds efficiently through SM electroweak interactions.
Alternatively, new gauge bosons resulting from $SU(2)_R\times U(1)_X$ breaking allow for efficient resonant annihilation when the DM mass is around half of the new gauge boson mass. As we will see, this case requires $v_R \lesssim 150$~TeV for the bi-triplet embeddings. The viable parameter region can be probed by collider, direct-detection, and indirect-detection experiments. In particular, the search for line and continuum gamma-ray signals by the Cherenkov Telescope Array (CTA) observatory will probe most of the parameter space. 

This paper is organized as follows.
In Sec.~\ref{sec:parity}, we briefly review how Parity can solve the strong CP problem.
In Sec.~\ref{sec:stability}, we discuss how bi-triplet fermionic DM is accidentally stabilized by gauge symmetry.
Based on this general discussion, we define our bi-triplet DM model in Sec.~\ref{sec:model}, and compute the relic abundance in Sec.~\ref{sec:annihilation}.
The phenomenology is discussed in Sec.~\ref{sec:pheno}.
Finally, we conclude this work in Sec.~\ref{sec:summary}.

\section{Parity solution to the strong CP problem}
\label{sec:parity}

\begin{table}[tbp]
    \begin{center}
        \begin{tabular}{|c|c|c|c|c|c|c|c|c|c|c|c|c|} \hline
                      & $H_L$          & $H_R$         & $q_i$         & $\ol{q}_i$     & $\ell_i$        & $\ol{\ell}_i$ & $\mathcal{U}_i$         & $\ol{\mathcal{U}}_i$     & $\mathcal{D}_i$          & $\ol{\mathcal{D}}_i$     & $\mathcal{E}_i$   & $\ol{\mathcal{E}}_i$ \\ \hline
            $SU(3)_c$ & {\bf 1}        & {\bf 1}       & {\bf 3}       & ${\bf \ol{3}}$ & {\bf 1}         & {\bf 1}        & {\bf 3}       & ${\bf \ol{3}}$ & {\bf 3}        & ${\bf \ol{3}}$ & {\bf 1} & {\bf 1}     \\
            $SU(2)_L$ & {\bf 2}        & {\bf 1}       & {\bf 2}       & {\bf 1}         & {\bf 2}         & {\bf 1}        & {\bf 1}       & {\bf 1}         & {\bf 1}        & {\bf 1}         & {\bf 1} & {\bf 1}     \\
            $SU(2)_R$ & {\bf 1}        & {\bf 2}       & {\bf 1}       & {\bf 2}         & {\bf 1}         & {\bf 2}        & {\bf 1}       & {\bf 1}         & {\bf 1}        & {\bf 1}         & {\bf 1} & {\bf 1}     \\
            $U(1)_X$  & $-\frac{1}{2}$ & $\frac{1}{2}$ & $\frac{1}{6}$ & $-\frac{1}{6}$  & $- \frac{1}{2}$ & $\frac{1}{2}$  & $\frac{2}{3}$ & $-\frac{2}{3}$  & $-\frac{1}{3}$ & $\frac{1}{3}$   & $-1$    & $1$         \\ \hline
        \end{tabular}
    \end{center}
    \caption{The gauge charges of Higgses and fermions.}
    \label{tab:charges}
\end{table}%

In this section, we review the Parity solution to the strong CP problem, with particular focus on the minimal Higgs model.
The minimal gauge group consistent with Parity symmetry is $SU(3)_c \times SU(2)_L \times SU(2)_R \times U(1)_X$, which is spontaneously broken down to the SM gauge group by the non-zero vacuum expectation value (VEV) $v_R$ of $H_R (\bm{1},\bm{1},\bm{2},-1/2)$, the Parity partner of the SM Higgs field $H_L (\bm{1},\bm{2},\bm{1},1/2)$. This symmetry breaking gives rise to new heavy gauge bosons $W_R$ and $Z_R$, analogous to the SM electroweak gauge bosons $W_L$ and $Z_L$. Searches for $W_R$ at the LHC impose the strongest constraint on the Parity breaking scale of $v_R\gtrsim 14$~TeV, requiring the hierarchy $v_R>v_L$. To obtain this hierarchy, we consider a scalar potential with soft Parity breaking,
\begin{align}
    V = \lambda (|H_R|^4 + |H_L|^4) + \lambda_{LR} |H_R|^2 |H_L|^2 - m^2 (|H_R|^2 + |H_L|^2) - \Delta m^2 (|H_R|^2 - |H_L|^2),
\end{align}
where $\Delta m^2 > 0$ and the last term softly breaks Parity. This term may originate from spontaneous Parity breaking at a scale above $v_R$ through an order parameter $O$ that couples to $H_L$ and $H_R$ as $O(|H_L^2|- |H_R|^2)$. The simplest examples are pure Yang-Mills theories with $\theta=\pi$~\cite{Witten:1980sp,tHooft:1981bkw,Witten:1998uka,Gaiotto:2017yup,Kitano:2020mfk}.

In the minimal Higgs model, Parity partners of the $SU(2)_L$-charged SM fermions and vector-like fermions are introduced.
The minimal fermion and Higgs contents are summarized in Table~\ref{tab:charges}.
Charged fermion masses are generated by the following Yukawa couplings,
\begin{align}
    \label{eq:fermion masses}
    {\cal L} = & x^u_{ij} q_i \ol{\mathcal{U}}_j H_L^\dag + \ol{x}^u_{ij} \ol{q}_i \mathcal{U}_j H_R^\dag + M_{ij}^u \mathcal{U}_i \ol{\mathcal{U}}_j  \nonumber \\
    +          & x^d_{ij} q_i \ol{\mathcal{D}}_j H_L + \ol{x}^d_{ij} \ol{q}_i \mathcal{D}_j H_R + M_{ij}^d \mathcal{D}_i \ol{\mathcal{D}}_j \nonumber            \\
    +          & x^e_{ij} \ell_i \ol{\cal E}_j H_L + \ol{x}^e_{ij} \ol{\ell}_i \mathcal{E}_j H_R + M_{ij}^e \mathcal{E}_i \ol{\mathcal{E}}_j + {\rm h.c.}
\end{align}
To be concrete, here we generate Yukawa couplings via vector-like fermions $(\mathcal{U},\ol{\mathcal{U}})$, $(\mathcal{D},\ol{\mathcal{D}})$, and $(\mathcal{E},\ol{\mathcal{E}})$. Vector-like fermions with different gauge charges can also generate Yukawa couplings; see Ref.~\cite{Hall:2018let}. The DM phenomenology is independent of the choice of the vector-like fermions.
The mass matrix for the up and down-type quarks is given by
\begin{align}
    \label{eq:mass matrix}
    \mathcal{M}_{u,d} = \begin{pmatrix}
                            0                   & x_{ij}^{u,d} v_L \\
                            \ol{x}^{u,d}_{ij}v_R & M^{u,d}_{ij}
                        \end{pmatrix}.
\end{align}
When $M \gg x v_R$, we may integrate out $(\mathcal{U},\ol{\mathcal{U}})$ and $(\mathcal{D},\ol{\mathcal{D}})$ to obtain the mass of SM quarks $\sim x^2 v_R v_L / M$. The SM right-handed quarks mainly come from $\ol{q}$. On the other hand, when $M \ll x v_R$, $\mathcal{U}$ and $\mathcal{D}$ obtain a large Dirac mass $\sim x v_R$ paired with $\ol{q}$.
The SM right-handed quarks mainly come from $\ol{\mathcal{U}}$ and $\ol{\mathcal{D}}$, and the quark masses are $\sim x v_L$. The charged lepton mass has the same structure.
Neutrino mass models can be found in~\cite{Babu:1988yq,Harigaya:2021txz,Babu:2022ikf,Dror:2020jzy,Hall:2023vjb,Harigaya:2025zru}.

The strong CP problem is solved in the following way.
Under Parity symmetry, the gauge fields transform as
\begin{align}
    G_\mu^a(t,\mathbf{x})  \rightarrow & G_\mu^a(t,-\mathbf{x})\times s(\mu),~
    B_\mu(t,\mathbf{x})  \rightarrow B_\mu(t,-\mathbf{x})\times s(\mu),
    \nonumber \\
    W_{L,\mu}^a(t,\mathbf{x})  \rightarrow & W_{R,\mu}^a(t,-\mathbf{x})\times s(\mu),~
    W_{R,\mu}^a(t,\mathbf{x})  \rightarrow W_{L,\mu}^a(t,-\mathbf{x})\times s(\mu),
    \nonumber \\
    &s(\mu)=\begin{cases}
               1  & \mu = 0      \\
               -1 & \mu = 1,2,3.
           \end{cases}
\end{align}
Such a symmetry forbids the CP-violating term $\theta_s G \tilde{G}$. On the other hand, the Higgs and fermion fields transform as
\begin{align}
     & H_L(t,\mathbf{x}) \to H_R^\dagger (t,-\mathbf{x}), \nonumber                                                                                                               \\
     & q(t,\mathbf{x}) \rightarrow i \sigma_2 \ol{q}^*(t,-\mathbf{x}), ~\mathcal{U}(t,\mathbf{x}) \rightarrow i \sigma_2 \ol{\mathcal{U}}^*(t,-\mathbf{x}),  ~\mathcal{D}(t,\mathbf{x}) \rightarrow i \sigma_2 \ol{\mathcal{D}}^*(t,-\mathbf{x}),\nonumber \\
     & \ell(t,\mathbf{x}) \to i \sigma_2 \ol{\ell}^*(t,-\mathbf{x}),~ \mathcal{E}(t,\mathbf{x}) \to i \sigma_2 \ol{\mathcal{E}}^*(t,-\mathbf{x}).
\end{align}
This symmetry imposes $\ol{x}_{ij} = x^*_{ij}$ and $M_{ij} = M^*_{ji}$.%
\footnote{Soft Parity breaking in $M_{ij}$ can be suppressed while that in the Higgs mass explains $v_R\gg v_L$. This is because $|H_{L,R}|^2$ and the fermion bi-linears have different mass dimensions.}
As a result, the determinant of the mass matrix in Eq.~\eqref{eq:mass matrix} is real, and therefore $\theta_q \equiv \arg (\mathcal{M}_u \mathcal{M}_d) = 0$ at tree level.%
\footnote{Here we take $v_L$ and $v_R$ to be real by $U(1)_Y$ and $U(1)_X$ rotation. If they are not taken to be real, the determinants of the mass matrices are complex, but that of up and down have exactly opposite phases, so that the correction to the strong CP phase is absent. The cancellation is guaranteed since the phases of $H_L$ and $H_R$ are gauge degrees of freedom.}
Then it follows that the strong CP phase $\ol{\theta} \equiv \theta_q + \theta_s$ vanishes at tree level.
Quantum corrections to $\ol{\theta}$ are found to be sufficiently small~\cite{Hall:2018let,Hisano:2023izx}, and the strong CP problem is solved.

The minimal Higgs model has two advantages over other types of models where $SU(2)_R\times U(1)_X$ is broken by an $SU(2)_R$ triplet and $SU(2)_L\times U(1)_Y$ is broken by $SU(2)_L\times SU(2)_R$ bi-doublets. First, a possible extra hierarchy problem associated with the intermediate scale $v_R$ is absent in the minimal Higgs model; the fine-tuning to obtain $v_R \ll \Lambda$, where $\Lambda$ is the cutoff scale, is $v_R^2/\Lambda^2$ and that to obtain $v_L \ll v_R$ is $v_L^2/v_R^2$, so the total fine-tuning is $v_L^2/\Lambda^2$, which is as large as that of the SM and may be explained
by environmental selection~\cite{Agrawal:1997gf,Hall:2014dfa,DAmico:2019hih}.
Second, the strong CP problem is straightforwardly solved in the minimal Higgs model as we have shown above. In contrast, when $SU(2)_R\times U(1)_X$ is broken by an $SU(2)_R$ triplet and $SU(2)_L\times U(1)_Y$ is broken by bi-doublets,
the fine-tuning to obtain $v_R$ and $v_L$ is $v_R^2/\Lambda^2$ and $v_L^2/\Lambda^2$, respectively, and the total fine-tuning is much worse than that of the SM unless $\Lambda\sim v_R$. 
Also, the VEVs of bi-doublets generically have a physical CP phase and result in a non-zero strong CP phase, unless extra symmetry is introduced~\cite{Kuchimanchi:1995rp,Mohapatra:1995xd,Kuchimanchi:2023imj}.

\section{Accidental stability of bi-triplet dark matter}
\label{sec:stability}

In this section, we motivate bi-triplet DM by its accidental stability, without the need for additional symmetry or requiring the smallness of any interactions. 

The minimum $U(1)_X$ charge of color-neutral states in the model is $1/2$. If $U(1)_X$ is spontaneously broken by the VEV of a scalar field with an integer $U(1)_X$ charge, there would be a residual $\mathbb{Z}_2$ symmetry that may be used to understand the stability of DM~\cite{Heeck:2015qra,Garcia-Cely:2015quu}. However, in the minimal Higgs model, $H_R$ and $H_L$ have $U(1)_X$ charges of $\pm1/2$ and we cannot understand the stability of DM in this way.

We instead consider the possibility that DM is accidentally stable due to the $SU(2)_L\times SU(2)_R\times U(1)_X$ gauge symmetry. In order to avoid the hierarchy problem associated with scalar masses, we assume that DM is fermionic. In Table~\ref{tab:decay operator}, we list all $SU(2)_L\times SU(2)_R\times U(1)_X$ multiplets ${\mathcal{X}}$ with $SU(2)_{L,R}$ dimensions equal to or smaller than ${\bf 3}$ that contain an electromagnetic (EM) neutral DM candidate, along with the lowest-dimensional operators that lead to their decay. If there exist operators with mass dimension smaller than 6, even when the operator cutoff scale is taken to be as large as the Planck scale, DM decays too rapidly. Therefore, the bi-triplet of $SU(2)_L\times SU(2)_R$ is the lowest dimensional representation in which DM is accidentally stable. Further, for the bi-triplet embedding $(\mathbf{3},\mathbf{3},2)$ the electroweak multiplet containing DM is hypercharged and hence DM has too large a scattering cross section with nucleons via  $Z_L$-boson exchange. We therefore find that the only two viable bi-triplet embeddings of DM are $(\mathbf{3},\mathbf{3},0)$ and $(\mathbf{3},\mathbf{3},1)$, without the need to impose any additional stabilizing symmetry. Indeed, from the dimension-6 operators in Table~\ref{tab:decay operator}, the bi-triplet DM decays into SM particles with a lifetime
\begin{equation}
    \tau_{\mathcal{X}} = \left(\frac{1}{8\pi} \frac{v_R^4 m_{\mathcal{X}}}{\Lambda^4}\right)^{-1} \simeq 10^{31}~ {\rm sec} \left(\frac{10~\rm TeV}{m_{\mathcal{X}}}\right) \left(\frac{40~\rm TeV}{v_R}\right)^4 \left( \frac{\Lambda}{1.2\times 10^{19}~{\rm GeV}} \right)^4,
\end{equation}
where $\Lambda$ is the cutoff scale of the operator and $m_{\mathcal{X}}$ is the DM mass.
As we will see, $m_{\mathcal{X}} = \mathcal{O}(1-10)$~TeV in the viable parameter space, for which the constraint from gamma-ray observations is $\tau_{\mathcal{X}} \gtrsim 10^{28}$ sec~\cite{Cohen:2016uyg,Blanco:2018esa}.
For $v_R< \mathcal{O}(100)$~TeV, which is satisfied in most of the viable parameter space, the constraint is satisfied for $\Lambda$ around the Planck scale.
For $v_R> \mathcal{O}(100)$ TeV, we require $\Lambda$ above the Planck scale or a suppression of the dimension-6 decay operator's coefficient.

We note that the accidental stability of DM relies on the assumption that there are no new particles whose exchange generates the dimension-6 operator with $\Lambda \ll M_{\rm Pl} $. This assumption may not be satisfied in some realizations of the SM Yukawa interactions. Indeed, if the charged-lepton Yukawa is generated by $\Delta({\bf 1},{\bf 2},{\bf 2},0)$, we may introduce the following Lagrangian terms,
\begin{equation}
    \label{eq:L_bidoublet}
    {\cal L} \supset \frac{1}{\Lambda} \mathcal{X} \Delta H_L H_R + x \Delta \ell H_R + \frac{1}{2}m_\Delta \Delta^2 + {\rm h.c.}
\end{equation}
After integrating out $\Delta$, we obtain
\begin{equation}
    {\cal L}_{\rm eff} = \frac{x}{ \Lambda m_\Delta} {\mathcal{X}} \ell H_L H_R H_R.
\end{equation}
With $m_\Delta \sim v_R$, this operator leads to too rapid decay of DM.%
\footnote{One could try to make DM more stable by utilizing the flavor symmetry of SM fermions. For example, the smallness of $m_{\Delta}$ may be explained by a chiral symmetry. Treating $m_\Delta$ as a spurion field, the first term in Eq.~\eqref{eq:L_bidoublet} will be suppressed by $(m_\Delta/\Lambda)^{1/2}$. The resulting decay rate, however, is still too large for $\Lambda\sim M_{\rm Pl}$.}
To avoid this problem, we must assume that the charged-lepton Yukawa is not generated by $\Delta({\bf 1},{\bf 2},{\bf 2},0)$.

\begin{table}[t!]
    \centering
    \begin{tabular}{|c|c|c|}
    \hline
        DM $\chi$ Embedding& Decay Operator& Dimension \\
        \hline
        $({\bf 1},{\bf 2},-1/2)$ & $\mathcal{X} \ol{\ell}$             & 3         \\\hline
        $({\bf 1},{\bf 3},0)$    & $\mathcal{X} H_R^* \ol{\ell}$       & 4         \\
        $({\bf 1},{\bf 3},1)$    & $\mathcal{X} H_R \ol{\ell}$         & 4         \\
        \hline
        $({\bf 2},{\bf 2},0)$    & $\mathcal{X} H_R \ell$               & 4         \\
        $({\bf 2},{\bf 2},1)$    & $\mathcal{X} H_R^* \ell$             & 4         \\
        \hline
        $({\bf 2},{\bf 3},1/2)$  & $\mathcal{X} H_R H_R^* \ell $        & 5         \\
        \hline
        $({\bf 3},{\bf 3},0)$    & $\mathcal{X} H_R H_R^* \ell H_L^*$   & 6         \\
        $({\bf 3},{\bf 3},1)$    & $\mathcal{X} H_R^* H_R^* \ell H_L^*$ & 6         \\
        $({\bf 3},{\bf 3},2)$    & $\mathcal{X} H_R^* H_R^* \ell H_L$   & 6         \\
        \hline
    \end{tabular}
    \caption{The lowest dimension operators that lead to DM decays for $SU(2)_{L,R}$ representations of dimension $\mathbf{3}$ or smaller. Bi-triplet DM decays only via dimension-6 or higher dimensional operators, accidentally stabilizing DM.}
    \label{tab:decay operator}
\end{table}

\section{Bi-triplet dark matter}
\label{sec:model}
In this section, we describe bi-triplet DM in the minimal Parity-symmetric left-right model for both $(\mathbf{1},\mathbf{3},\mathbf{3},0)$ and $(\mathbf{1},\mathbf{3},\mathbf{3},1)$ embeddings. Hereafter, we suppress the DM $SU(3)_c$ singlet charge, showing only $SU(2)_L\times SU(2)_R\times U(1)_X$ charges for brevity. We denote the EM neutral weak gauge boson from $SU(2)_L\times U(1)_Y$ breaking as $Z$ and the analogous gauge boson from $SU(2)_R\times U(1)_X$ breaking as $Z'$. As discussed in App.~\ref{app:gauge_coup}, these neutral bosons mix with each other, forming two mass eigenstates, the lighter of which being $Z_L$ and the heavier $Z_R$, although the mixing is important only for DM direct detection. The charged weak gauge bosons $W_L$ and $W_R$ do not mix with each other at tree level.

\subsection{Bi-triplet in $(\mathbf{3},\mathbf{3},0)$}\label{sec:330_Lagrangian}
DM can be embedded into a Weyl fermion in $\mathcal{X}(\mathbf{3},\mathbf{3},0)$, which is its own Parity partner. The gauge, kinetic, and mass terms for $\mathcal{X}$ are
\begin{align}
    \label{eq:Lagrangian}
    \mathcal{L} = \Tr[\mathcal{X}^\dag\ol{\sigma}^\mu\left(i\partial_\mu \mathcal{X}+gW_{L\mu}\mathcal{X}-g \mathcal{X} W_{R\mu}\right)] - \frac{1}{2} m_{\mathcal{X}} (\mathcal{X} \mathcal{X} + \rm{h.c.}),
\end{align}
where $m_{\mathcal{X}}$ is the DM mass. Note that the $SU(2)_{L,R}$ gauge couplings are equal due to Parity. 

After $SU(2)_R\times U(1)_X\rightarrow U(1)_Y$ breaking,
\begin{equation}
\mathcal{X}(\mathbf{3},\mathbf{3},0)\rightarrow \mathcal{X}^+(\mathbf{3},1)\oplus \mathcal{X}^-(\mathbf{3},-1)\oplus \mathcal{X}^0(\mathbf{3},0),
\end{equation}
where gauge charges under $SU(2)_L \times U(1)_Y$ are shown in parentheses. The fields $\mathcal{X}^{\pm,0}$ denote the components of $\mathcal{X}$ with hypercharges $\pm 1,0$, respectively. $\mathcal{X}^\pm$ are the Dirac partners of each other.

After $SU(2)_L\times U(1)_Y\rightarrow U(1)_{\text{EM}}$ breaking,
\begin{equation}
\begin{split}
    \mathcal{X}^+(\mathbf{3},1)&\rightarrow \mathcal{X}^{++}(2)\oplus \mathcal{X}^{+0}(1)\oplus \mathcal{X}^{+-}(0),\\
    \mathcal{X}^-(\mathbf{3},-1)&\rightarrow  \mathcal{X}^{-+}(0)\oplus \mathcal{X}^{-0}(-1)\oplus \mathcal{X}^{--}(-2),\\
    \mathcal{X}^0(\mathbf{3},0)&\rightarrow \mathcal{X}^{0+}(1)\oplus \mathcal{X}^{00}(0)\oplus \mathcal{X}^{0-}(-1).
\end{split}
\end{equation}
The EM charge is shown in parentheses.
These particles can be further combined into:
\begin{itemize}
    \item Neutral Majorana fermion $\chi^0 \equiv (\mathcal{X}^{00},i\sigma^2 \mathcal{X}^{00*})$.
    \item Dirac fermions with EM charge $-1$: $\chi^- \equiv (\mathcal{X}^{0-}, i \sigma^2\mathcal{X}^{0+*})$, $E = (\mathcal{X}^{-0}, i \sigma^2\mathcal{X}^{+0*})$.
    \item Dirac fermion with EM charge $-2$: $D\equiv(\mathcal{X}^{--}, i \sigma^2\mathcal{X}^{++*})$.
    \item Neutral Dirac fermion: $N \equiv (\mathcal{X}^{-+}, i \sigma^2 \mathcal{X}^{+-*})$.
\end{itemize}
As we will discuss in Sec.~\ref{sec:mass_splitting}, $\chi^0$ is the lightest state and the DM candidate with mass $m_{\chi^0}\equiv m_{\mathcal{X}}$. 
The interaction Lagrangian for these gauge eigenstates after electroweak symmetry breaking is
\begin{align}
    \label{eq:lagrangian final}
    \mathcal{L} \supset & g s_L A_\mu\left[-\ol{\chi^-}\gamma^\mu\chi^--2 \ol{D}\gamma^\mu D-\ol{E}\gamma^\mu E\right]\\
    &+g c_L Z_{\mu}\left[-\ol{\chi^-}\gamma^\mu\chi^--\frac{1-2s_L^2}{c_L^2}\ol{D}\gamma^\mu D+t_L^2\ol{E}\gamma^\mu E+\frac{1}{c_L^2}\ol{N}\gamma^\mu N\right] \nonumber\\
    &+g c_R Z'_{\mu}\left[-\ol{D}\gamma^\mu D-\ol{E}\gamma^\mu E-\ol{N}\gamma^\mu N\right]\nonumber\\
    & + g W_{L\mu}^- \left[\ol{D}\gamma^\mu E + \ol{\chi}^-\gamma^\mu\chi^0 -\ol{E}\gamma^\mu N\right] + \rm{h.c.}\nonumber\\
    & + g W_{R\mu}^- \left[\ol{E} \gamma^\mu\chi^0 + \ol{D}\gamma^\mu \chi^- + \ol{N} \gamma^\mu(i \gamma^2 (\chi^-)^*)\right] + \rm{h.c.}.  \nonumber 
\end{align}
We define $\alpha_2 \equiv g^2/(4\pi)$, where $g$ is the common $SU(2)_{L, R}$ gauge coupling.
We also define $s_L \equiv \sin \theta_L$, $c_L \equiv \cos \theta_L$, and $t_L \equiv \tan \theta_L$, where $\theta_L$ is the Weinberg angle.
The corresponding $SU(2)_R$ quantities are denoted by $s_R$, $c_R$, and $t_R$.

\subsection{Bi-triplet in $(\mathbf{3},\mathbf{3},1)$}
DM can also be embedded into a Dirac fermion ${\mathcal{X}}(\mathbf{3},\mathbf{3},1)$ composed of two Weyl fields ${\mathcal{X}}^\pm(\mathbf{3},\mathbf{3},\pm1)$, which are the Parity partners of one another. Their Parity transformations are
\begin{equation}
    {\mathcal{X}}^+(t,\mathbf{x})\rightarrow i\sigma^2 {\mathcal{X}}^{-*}(t,-\mathbf{x}),~~{\mathcal{X}}^-(t,\mathbf{x})\rightarrow i\sigma^2 {\mathcal{X}}^{+*}(t,-\mathbf{x}).
\end{equation}
The gauge, kinetic, and mass terms for ${\mathcal{X}}^\pm$ are
\begin{equation}\label{eq:L_bitripletXcharged}
    \begin{split}
        \mathcal{L}&=\Tr[{\mathcal{X}}^{+\dag}\ol{\sigma}^\mu\left(i\partial_\mu {\mathcal{X}}^++gW_{L\mu}{\mathcal{X}}^+-g{\mathcal{X}}^+ W_{R\mu}-g_X B_{X\mu} {\mathcal{X}}^+\right)]\\
        &+\Tr[{\mathcal{X}}^{-\dag}\ol{\sigma}^\mu\left(i\partial_\mu {\mathcal{X}}^-+gW_{L\mu}{\mathcal{X}}^--g{\mathcal{X}}^- W_{R\mu}+g_X B_{X\mu} {\mathcal{X}}^-\right)]\\
        &-m_{\mathcal{X}}\left({\mathcal{X}}^{+T} {\mathcal{X}}^-+{\rm h.c.}\right)\,,
    \end{split}
\end{equation}
where $m_{\mathcal{X}}$ is the mass of ${\mathcal{X}}$.

After $SU(2)_R\times U(1)_X\rightarrow U(1)_Y$ breaking,
\begin{equation}
{\mathcal{X}}(\mathbf{3},\mathbf{3},1)\rightarrow {\mathcal{X}}_2(\mathbf{3},2)\oplus {\mathcal{X}}_1(\mathbf{3},1)\oplus {\mathcal{X}}_0(\mathbf{3},0),
\end{equation}
where ${\mathcal{X}}_{Y_i}$ denotes the multiplet of ${\mathcal{X}}$ with hypercharge $Y_i$.

After $SU(2)_L\times U(1)_Y\rightarrow U(1)_{\text{EM}}$ breaking,
\begin{equation}
\begin{split}
    \mathcal{X}_2(\mathbf{3},2)&\rightarrow R_3(3)\oplus R_2(2)\oplus R_1(1),\\
    \mathcal{X}_1(\mathbf{3},1)&\rightarrow S_2(2)\oplus S_1(1)\oplus S_0(0),\\
    \mathcal{X}_0(\mathbf{3},0)&\rightarrow \psi^{+}(1)\oplus \psi^{-}(-1)\oplus \psi^{0}(0),
\end{split}
\end{equation}
where $R_{Q_i}$ and $S_{Q_i}$ denote the components of $\mathcal{X}_i$ with EM charge $Q_i$, and $\psi^{\pm,0}$ denote the components of the multiplet ${\mathcal{X}}_0$ with EM charge $\pm1,0$.
Again, as shown in the next subsection, the lightest state $\psi^0$ is the DM candidate with mass $M_{\psi^0} \equiv m_{\mathcal{X}}$.

The interaction Lagrangian of these gauge eigenstates after electroweak symmetry breaking is
\begin{equation}
\begin{split}
\mathcal{L}\supset g s_L A_\mu&\left[\ol{\psi^+}\gamma^\mu\psi^+-\ol{\psi^-}\gamma^\mu\psi^-+\ol{S_1}\gamma^\mu S_1+2\ol{S_2}\gamma^\mu S_2+\ol{R_1}\gamma^\mu R_1+2\ol{R_2}\gamma^\mu R_2+3\ol{R_3}\gamma^\mu R_3\right]\\
+g c_L Z_{\mu}&\left[\ol{\psi^+}\gamma^\mu\psi^+-\ol{\psi^-}\gamma^\mu\psi^--(1+t_L^2)\ol{S_0}\gamma^\mu S_0-t_L^2\ol{S_1}\gamma^\mu S_1+(1-t_L^2)\ol{S_2}\gamma^\mu S_2\right.\\&\left.-(1+2t_L^2)\ol{R_1}\gamma^\mu R_1-2t_L^2\ol{R_2}\gamma^\mu R_2+(1-2t_L^2)\ol{R_3}\gamma^\mu R_3\right]\\
+\frac{g}{c_R}Z'_{\mu}&\left[-\ol{\psi^0}\gamma^\mu\psi^0-\ol{\psi^+}\gamma^\mu\psi^+-\ol{\psi^-}\gamma^\mu\psi^--t_L^2\ol{S_0}\gamma^\mu S_0-t_L^2\ol{S_1}\gamma^\mu S_1-t_L^2\ol{S_2}\gamma^\mu S_2\right.\\&\left.+(1-2t_L^2)\ol{R_1}\gamma^\mu R_1+(1-2t_L^2)\ol{R_2}\gamma^\mu R_2+(1-2t_L^2)\ol{R_3}\gamma^\mu R_3\right]\\
+g W_{L\mu}^-&\left[-\ol{\psi^-}\gamma^\mu\psi^0-\ol{\psi^0}\gamma^\mu\psi^++\ol{S_0}\gamma^\mu S_1+\ol{S_1}\gamma^\mu S_2-\ol{R_1}\gamma^\mu R_2-\ol{R_2}\gamma^\mu R_3\right]+\text{h.c.}\\
+g W_{R\mu}^-&\left[\ol{\psi^0}\gamma^\mu S_1-\ol{\psi^-}\gamma^\mu S_0-\ol{\psi^+}\gamma^\mu S_2-\ol{S_0}\gamma^\mu R_1-\ol{S_2}\gamma^\mu R_3+\ol{S_1}\gamma^\mu R_2\right]+\text{h.c.}.
\end{split}
\end{equation}
\subsection{Mass splitting}\label{sec:mass_splitting}
In both bi-triplet DM embeddings, mass splittings are generated via quantum corrections involving the gauge bosons from $SU(2)_R\times U(1)_X$ and $SU(2)_L\times U(1)_Y$ breaking. The mass splitting $\delta M^R_{ij}=M_{Y_i}-M_{Y_j}$ between multiplets with hypercharges $Y_i$ and $Y_j$, with masses $M_{Y_i}$ and $M_{Y_j}$, respectively, from the same multiplet with $U(1)_X$ charge $X$, is given by
\begin{equation}
   \delta M^R_{ij}=\frac{\alpha_2 M_{Y_j}}{4\pi}\left[(Y_i^2-Y_j^2)s_R^2  {\cal F}\left(\frac{M_{Z_R}}{M_{Y_j}}\right)+(Y_i-Y_j)(Y_i+Y_j-2X)\left( {\cal F}\left(\frac{M_{W_R}}{M_{Y_j}}\right)- {\cal F}\left(\frac{M_{Z_R}}{M_{Y_j}}\right)\right)\right],
\end{equation}
where
\begin{equation}
    {\cal F}(z)=\frac{1}{2} z \left( 2 z^3 \ln z - 2 z + 
   \sqrt{z^2 - 4} (z^2 + 2) \ln(\frac{1}{2} (z^2 - 2 - z \sqrt{z^2 - 4}))\right).
\end{equation}
The mass splitting $\delta M^L_{ij}=M_{Q_i}-M_{Q_j}$ between particles with charge $Q_i$ and $Q_j$, with masses $M_{Q_i}$ and $M_{Q_j}$, respectively, from the same multiplet with hypercharge $Y$, is given by
\begin{equation}
   \delta M^L_{ij}=\frac{\alpha_2 M_{Q_j}}{4\pi}\left[(Q_i^2-Q_j^2)s_L^2  {\cal F}\left(\frac{M_{Z_L}}{M_{Q_j}}\right)+(Q_i-Q_j)(Q_i+Q_j-2Y)\left( {\cal F}\left(\frac{M_{W_L}}{M_{Q_j}}\right)- {\cal F}\left(\frac{M_{Z_L}}{M_{Q_j}}\right)\right)\right].
\end{equation}

\begin{figure}[t!]
    \centering
    \includegraphics[width=0.49\linewidth]{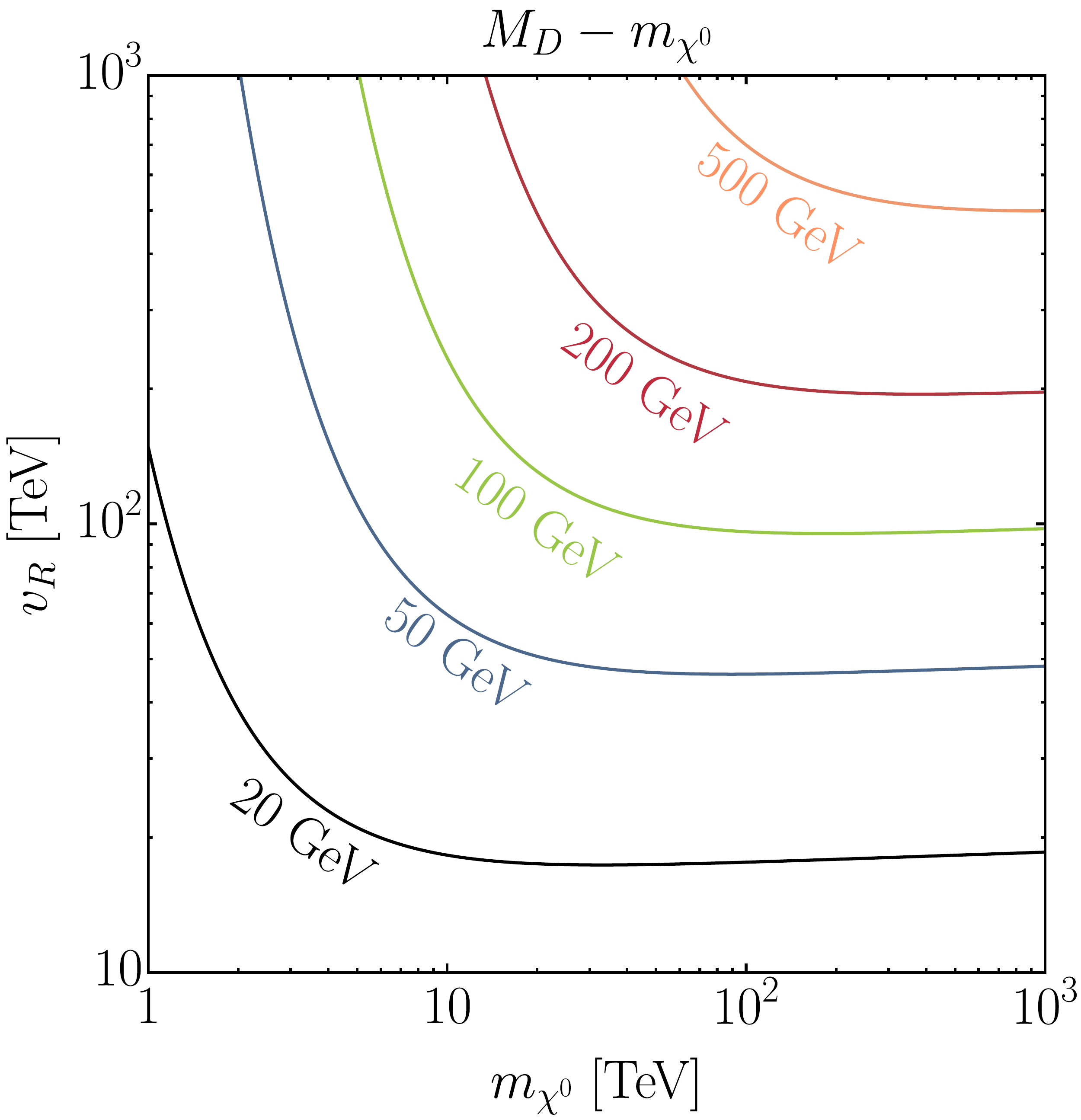}
    \includegraphics[width=0.49\linewidth]{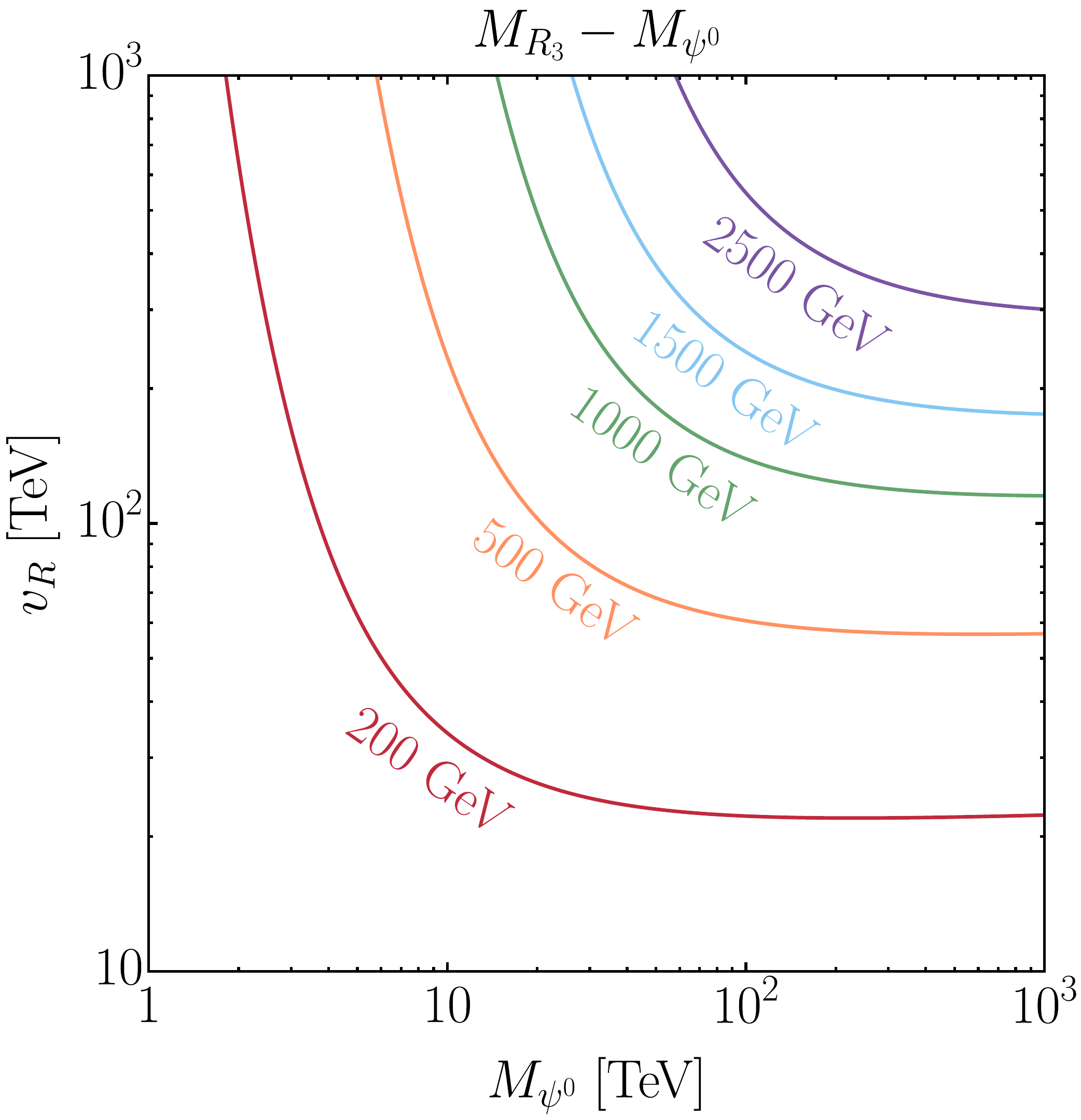}
    \caption{Mass splitting between the heaviest and lightest states of the $(\mathbf{3},\mathbf{3},0)$ (left) and $(\mathbf{3},\mathbf{3},1)$ (right) multiplets.}
    \label{fig:running}
\end{figure}

The mass splitting between the heaviest and lightest states of the $(\mathbf{3},\mathbf{3},0)$ and $(\mathbf{3},\mathbf{3},1)$ models are shown in Fig.~\ref{fig:running}. For $(\mathbf{3},\mathbf{3},0)$, $\chi^0$ is the lightest state and is the DM candidate.%
\footnote{
When $SU(2)_R\times U(1)_X$ is broken by a Higgs in $({\bf 1},{\bf 3},1)$, $M_{Z_R}/M_{W_R}$ becomes larger. Then, unlike in our setup, for a sufficiently large DM mass, the hypercharge and EM neutral state is no longer the lightest one~\cite{Heeck:2015qra,Garcia-Cely:2015quu}.
}
For $(\mathbf{3},\mathbf{3},1)$, $\psi^0$ is the lightest state and is the DM candidate.

\section{Relic abundance of bi-triplet dark matter}
\label{sec:annihilation}
In this section, we compute the relic abundance of DM in the $(\mathbf{3},\mathbf{3},0)$ and $(\mathbf{3},\mathbf{3},1)$ multiplets.

\subsection{Dark matter annihilation}\label{sec:DM_ann}

The relic abundance of DM, with mass $m_{\cal X}$ and co-moving number density $n_{\cal X}$, is determined by the freeze-out process.
To compute the DM relic abundance, we solve for the DM yield $Y \equiv n_{\cal X}/s$, where $s = g_{*s} (2 \pi^2/45) (m_{\cal X}/x)^3$ is the entropy density, and $g_{*s}$ is the effective degrees of freedom in entropy. The DM yield is obtained by solving the Boltzmann equation for coannihilation~\cite{Griest:1990kh}
\begin{align}
    \frac{dY}{dx} = - \frac{\langle \sigma_{\rm eff}v \rangle}{H x} \left( 1 - \frac{x}{3 g_{* s}} \frac{d g_{* s}}{dx} \right) s \left(Y^2 - Y_{\rm eq}^2\right),
\end{align}
where $x \equiv m_{\cal X}/T$. The Hubble expansion rate and the equilibrium yield are given by 
\begin{align}
    H (x)= \sqrt{\frac{g_*}{90}} \frac{\pi}{M_{\rm Pl}} \left( \frac{m_{\cal X}}{x} \right)^2,~~
    Y_{\rm eq}(x)  = g_{\rm eff}(x) \frac{m_{\cal X}^3}{s} \frac{1}{(2 \pi x)^{3/2}}e^{-x},
\end{align}
respectively, where $g_*$ is the effective number of relativistic degrees of freedom.
The total thermally averaged effective cross section is given by 
\begin{align}
\label{eq:sigmaeff}
    \langle \sigma_{\rm eff} v \rangle = \sum_{ij} \langle \sigma_{ij} v \rangle \frac{g_i(x) g_j(x)}{g_{\rm eff}(x)^2}\,,
\end{align}
where the sum runs over initial particle species $i$ and $j$ whose interconversion rates exceed the Hubble expansion rate during freeze-out.
Here,
\begin{align}
    \label{eq:dof}
    g_i(x)\equiv \mathfrak{g}_i (1+\Delta_i)^{3/2} \exp(-x \Delta_i),~~g_{\rm eff}(x) = \sum_i g_i(x)\,,
\end{align}
where  $\Delta_i \equiv m_i - m_{\cal X}$ is the mass splitting between particle $i$ with mass $m_i$ and DM.
The intrinsic degrees of freedom $\mathfrak{g}_i=2$ for Majorana fermions, and $\mathfrak{g}_i=4$ for Dirac fermions. 
$\langle \sigma_{ij} v \rangle$ is the thermally averaged annihilation cross section and is computed from
\begin{align}\label{eq:thermal_avg}
    \langle \sigma_{ij} v \rangle = \left( \frac{x}{4 \pi} \right)^{3/2} \int dv (\sigma_{ij} v) 4 \pi v^2 \exp(- \frac{x v^2}{4})\,,
\end{align}
where $\sigma_{ij}v$ are the total annihilation cross sections.

When the initial-state particles are much heavier than the electroweak gauge bosons mediating their interactions, long-range non-perturbative effects can significantly modify the initial-state wavefunction, which is known as the Sommerfeld effect~\cite{Sommerfeld:1931qaf}.
This effect has profound implications on the DM freeze-out abundance and indirect-detection signals~\cite{Hisano:2004ds,Hisano:2006nn,Cassel:2009wt,Arkani-Hamed:2008hhe,Slatyer:2009vg}. We provide a brief review of the Sommerfeld effect in App.~\ref{App:Sommerfeld}. The annihilation cross sections of all processes \emph{before} including the Sommerfeld effect are given in App.~\ref{sec:cross section} for $(\mathbf{3},\mathbf{3},0)$ DM and App.~\ref{sec:cross section 331} for $(\mathbf{3},\mathbf{3},1)$ DM. The long-range force potentials that modify these cross sections through the Sommerfeld effect are shown in App.~\ref{App:long_range_potentials_330} for $(\mathbf{3},\mathbf{3},0)$ DM and App.~\ref{App:long_range_potentials_331} for $(\mathbf{3},\mathbf{3},1)$ DM.

\subsection{Relic abundance of $(\mathbf{3},\mathbf{3},0)$ DM}

Interactions among particles within the same electroweak multiplet are mediated by $W_L$. 
These interactions remain efficient throughout DM freeze-out, allowing particles within the same electroweak multiplet to coannihilate.
Interactions between particles in different electroweak multiplets are mediated by $W_R$.
The scattering rate via $W_R$ exchange at temperature $T$ is about $T^5/(8\pi v_R^4)$. This rate exceeds the Hubble expansion rate when
\begin{equation}\label{eq:coan_condition}
    v_R < 300~{\rm TeV} \left(\frac{T/m_{\chi^0}}{0.05}\right)^{3/4} \left(\frac{m_{\chi^0}}{1~{\rm TeV}}\right)^{3/4}.
\end{equation}
This condition is always satisfied during freeze-out in the parameter region where resonant annihilation via $W_R$ and $Z_R$ determines the freeze-out abundance.
By contrast, when annihilation via $SU(2)_L\times U(1)_Y$ interactions determines the freeze-out abundance, $v_R$ can be large enough that $W_R$ exchange decouples before freeze-out is complete.
We therefore consider the following two cases:
\begin{itemize}
    \item Coannihilation: $W_R$-mediated interactions are effective during freeze-out, and particles within different electroweak multiplets can coannihilate with one another. The DM relic abundance is given by the relic abundance of $\chi^0$, that is $\Omega_{\rm DM}h^2= \Omega_{\chi^0}h^2$.
    \item Non-coannihilation: $W_R$-mediated interactions are ineffective at the time of freeze-out. Particles can coannihilate only with other particles within the same electroweak multiplet, populating the lightest state within each multiplet. $\chi^0$ and $\chi^\pm$ annihilate with one another to set $\Omega_{\chi^0}h^2$, while $D$, $E$ and $N$ annihilate with one another to set $\Omega_N h^2$. $N$ later decays into $\chi^0$. The DM relic abundance in this case is given by $\Omega_{\rm DM}h^2 = \Omega_{\chi^0}h^2 + m_{\chi^0} \Omega_N h^2/M_N $.
\end{itemize}

\begin{figure}[t!]
    \centering
    \hspace*{-0.5cm}
    \includegraphics[width=0.49\linewidth]{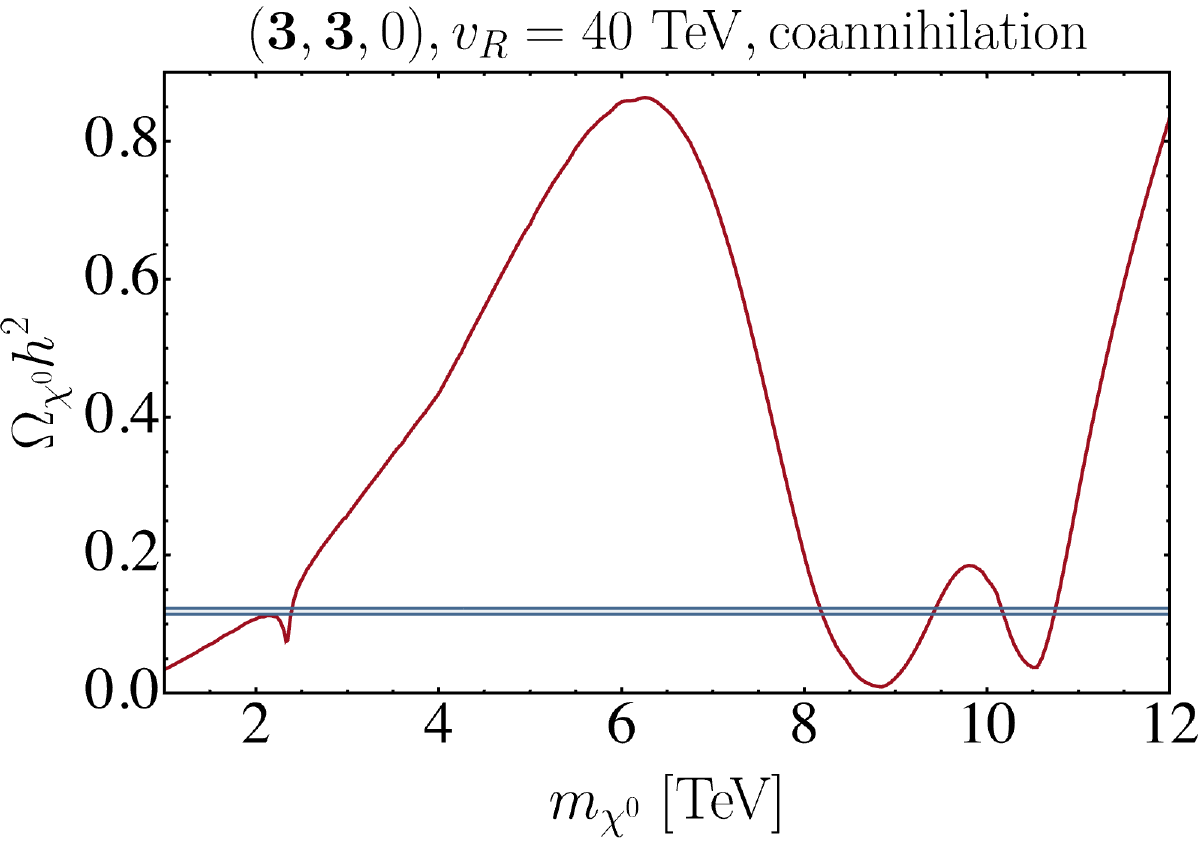}
    \includegraphics[width=0.49\linewidth]{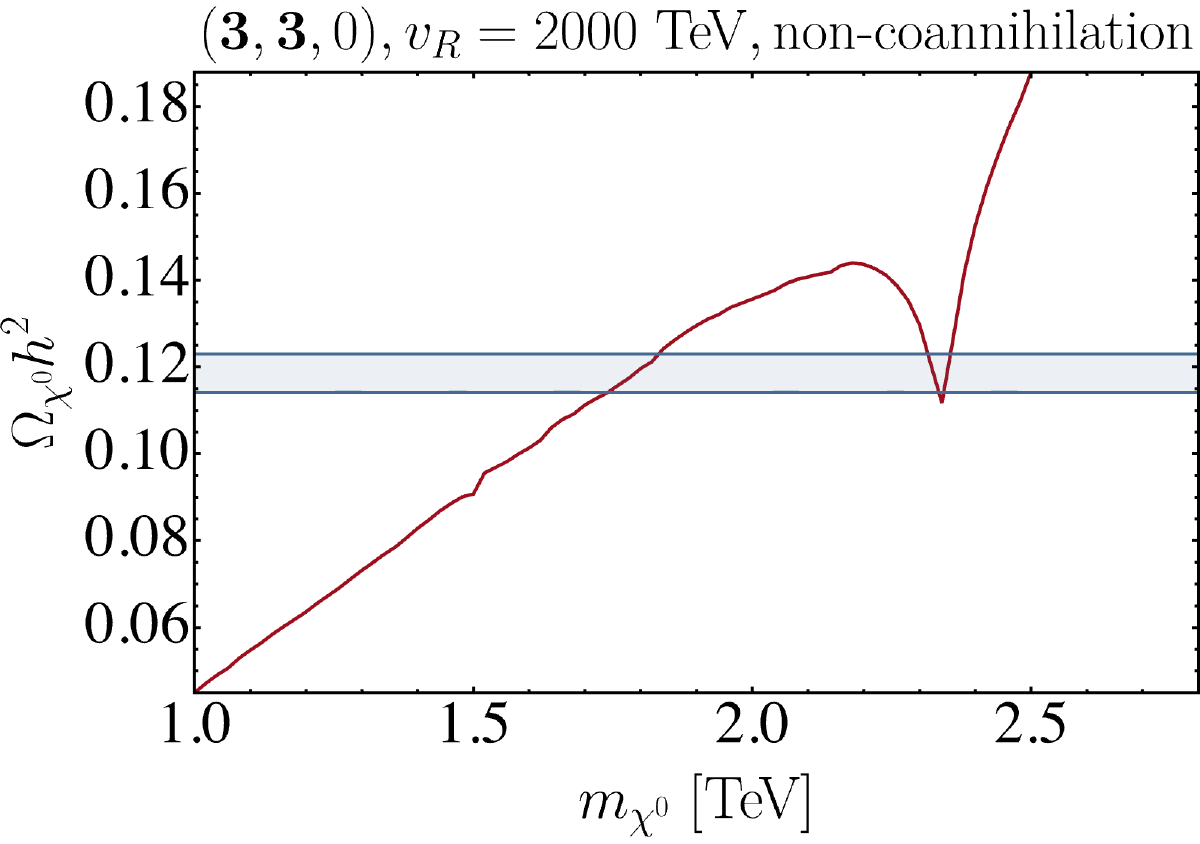}
    \caption{The DM relic abundance as a function of $m_{\chi^0}$ with fixed $v_R$. Left: coannihilation case with $v_R=40$ TeV. Right: non-coannihilation case with $v_R=2000$ TeV. The horizontal blue band is the observed DM relic abundance.}
    \label{fig:fix vR}
\end{figure}

Fig.~\ref{fig:fix vR} shows the DM relic abundance as a function of $m_{\chi^0}$ with $v_R$ fixed, for coannihilation (left) and non-coannihilation (right). The blue horizontal band corresponds to the observed DM relic abundance.
In the \emph{coannihilation case}, annihilation through electroweak interactions produces the observed DM relic abundance for $m_{\chi^0} \simeq 2.4~\tev$, slightly below the wino DM mass $m_{\chi^0} \simeq 2.9~\tev$~\cite{Hisano:2006nn}.
This difference arises from the larger number of degrees of freedom in the bi-triplet compared to the wino.
These extra degrees of freedom contribute significantly to the total effective degrees of freedom (Eq.~\eqref{eq:dof}), thereby diluting the effective annihilation cross section (Eq.~\eqref{eq:sigmaeff}).
A smaller DM mass is therefore required to enhance the effective annihilation cross section.
As we will see in Sec.~\ref{sec:indirect}, this lower DM mass, closer to the Sommerfeld resonance than wino DM, is disfavored by indirect-detection constraints. Instead, the higher masses that produce the correct DM relic abundance via $W_R$ and $Z_R$ resonances are of primary interest.
As $v_R$ increases, the $W_R$ and $Z_R$ masses increase, and the DM masses satisfying the resonance conditions also increase. At the same time, the annihilation cross section becomes suppressed for larger $v_R$, leading to an upper bound on $v_R$ when neither the $W_R$ nor the $Z_R$ resonance can produce the observed DM relic abundance. In the \emph{non-coannihilation case}, the correct DM relic abundance is obtained for DM masses close to the Sommerfeld resonance peaks of $\chi^0 \chi^0$ annihilation.
Three solutions around $2$ TeV are predicted. However, as discussed in Sec.~\ref{sec:indirect}, this case is disfavored by indirect-detection constraints.

\begin{figure}
    \centering
    \includegraphics[width=0.55\linewidth]{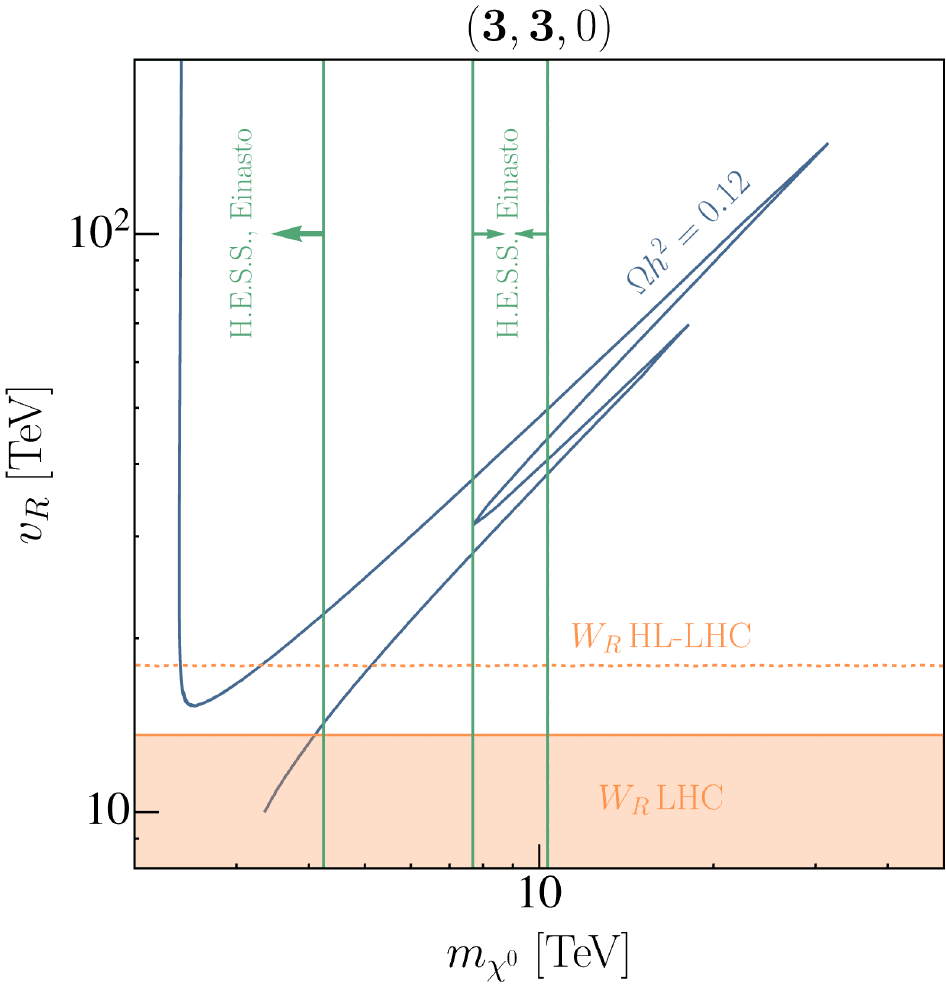}
    \includegraphics[width=0.49\linewidth]{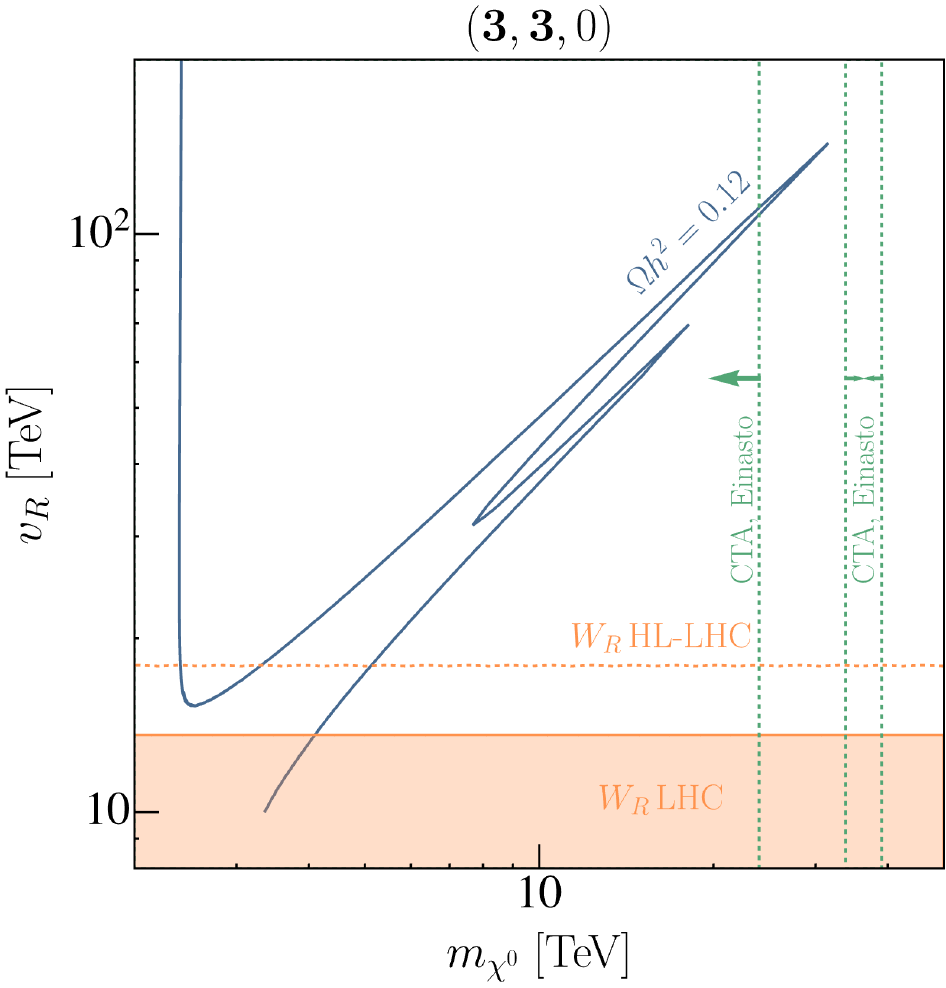}
    \includegraphics[width=0.49\linewidth]{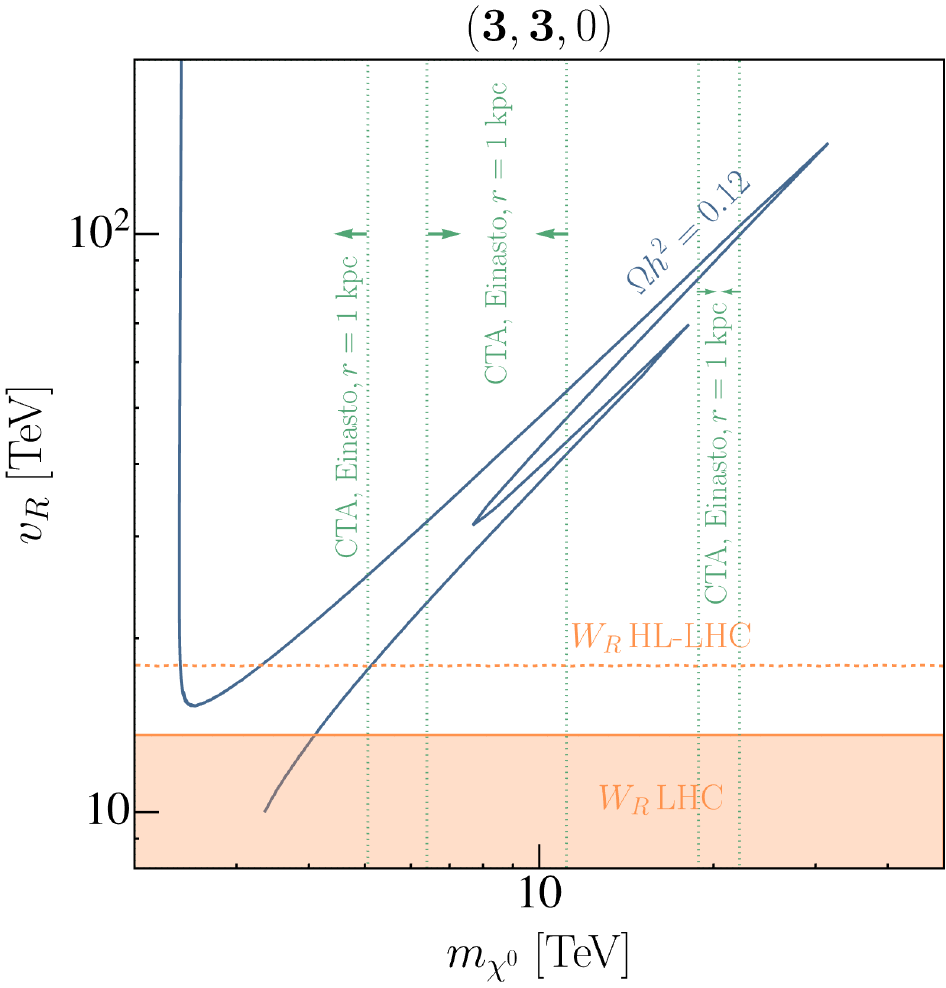}
    \caption{The $\left(m_{\chi^0},v_R\right)$ parameter space for $(\mathbf{3},\mathbf{3},0)$ DM. The blue contour corresponds to the observed DM relic abundance. $W_R$ and $Z_R$ resonance branches are observed, and an upper bound on the Parity breaking scale is obtained from the $W_R$ branch of around $v_R\lesssim 150$~TeV. Constraints from LHC $W_R$ searches and projected sensitivity of the HL-LHC are shown in solid and dashed orange lines, respectively. H.E.S.S.~indirect-detection constraints (assuming the Einasto density profile) are shown in solid green lines in the top panel. The projected sensitivity of the CTA is shown in the bottom two panels, assuming the Einasto profile with no core (left) and $r=1$~kpc core (right). Arrows point inwards towards the regions disfavoured by indirect detection.}
    \label{fig:main}
\end{figure}

Fig.~\ref{fig:main} shows the parameter space that produces the correct DM relic abundance in the $(m_{\chi^0},v_R)$ plane. The coannihilation case applies for the range of $v_R$ shown in the figure. At low DM masses, the correct abundance is obtained for $m_{\chi^0} \simeq 2.4~\tev$.
For higher $v_R$, coannihilation becomes inefficient and the prediction on the DM mass gradually becomes smaller down to around $2$ TeV.

For higher DM masses, the $W_R$ and $Z_R$ resonance branches are observed.%
\footnote{We observe small deformations of the resonance branches in our numerical results that are not visible in Fig.~\ref{fig:main}-- for example at~$m_{\chi^0}\approx20~\text{TeV}$ on the $Z_R$ branch. These correspond to regions of parameter space modified by the Sommerfeld effect.}
When $v_R$ is sufficiently small, the $W_R$ and $Z_R$ resonance branches merge.
The $Z_R$ resonance branch can produce the correct DM relic abundance for $v_R \lesssim 100~\tev$, while the $W_R$ gauge boson resonance branch can extend to $v_R$ as large as $150~\tev$.
The difference arises because the $W_R$-mediated annihilation channels contain fermions in both the $(\mathbf{3},0)$ and $(\mathbf{3},1)$ electroweak multiplets, while fermions in the $Z_R$-mediated annihilation channels only come from $(\mathbf{3},1)$. The latter channels receive two Boltzmann suppression factors (see Eq.~\eqref{eq:dof}) and therefore have a smaller annihilation rate.

Constraints on the parameter space by LHC searches for $W_R$ are shown in the orange shaded region, with projected sensitivity of the HL-LHC shown by the dotted orange line. Indirect-detection constraints by the High Energy Stereoscopic System (H.E.S.S.), assuming an Einasto profile, are shown by green lines, with arrows pointing inward towards the disfavoured region.
The projected sensitivity of the CTA for the Einasto profile and a cored Einasto profile are shown in dashed and dotted green lines, respectively. These constraints are discussed in detail in Sec.~\ref{sec:pheno}.

\begin{figure}[t!]
    \centering
    \includegraphics[width=0.49\linewidth]{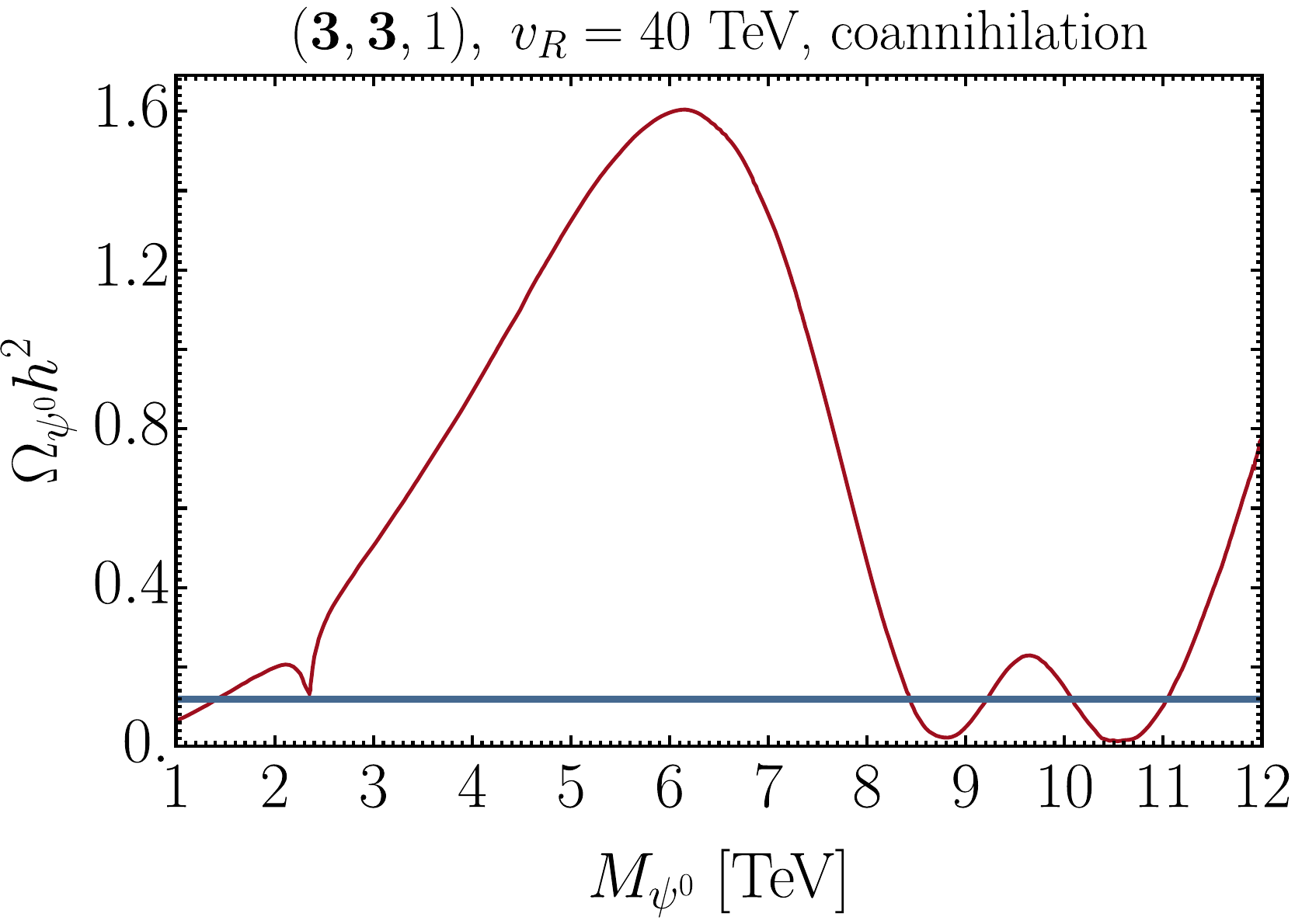}
    \includegraphics[width=0.49\linewidth]{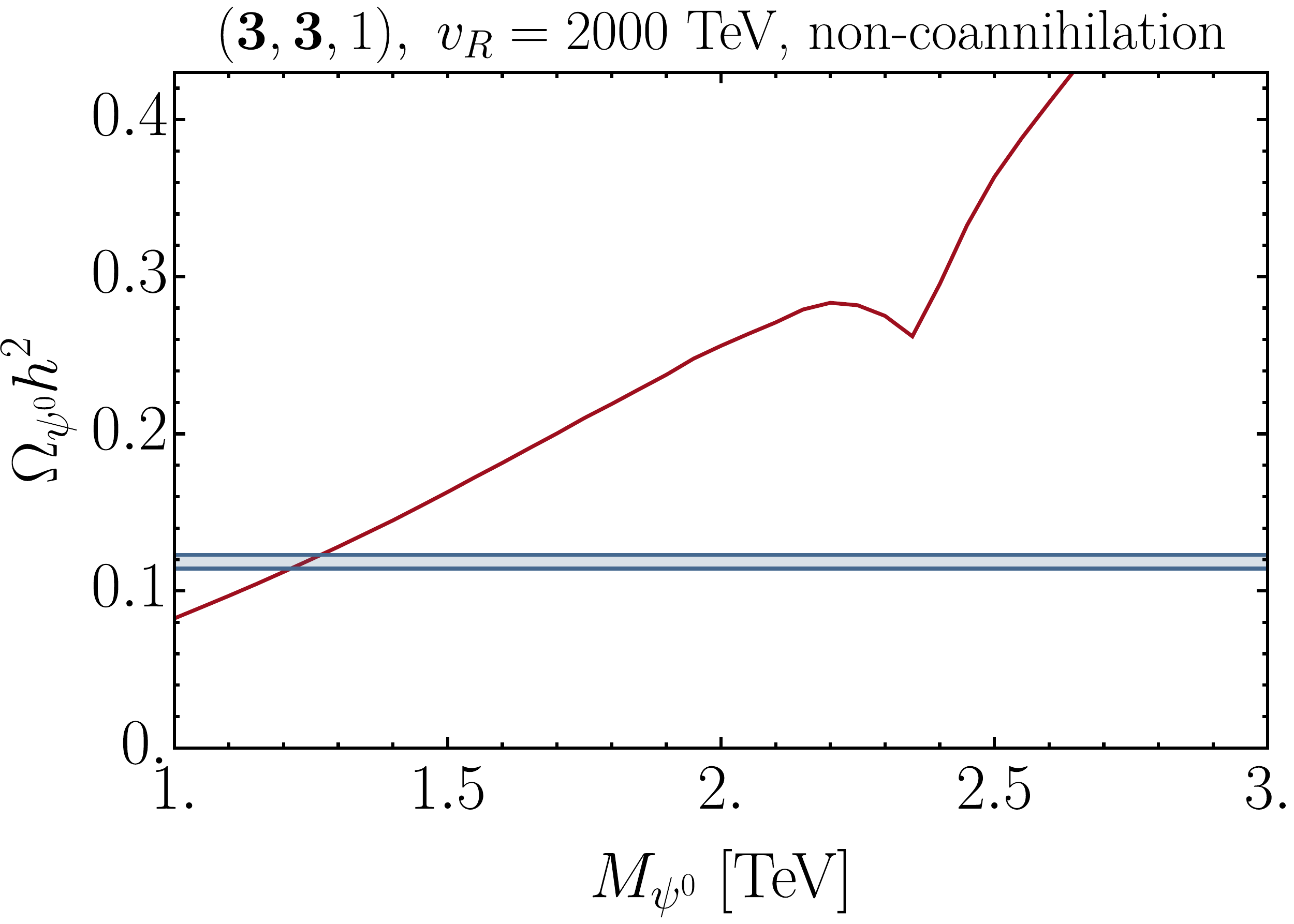}
    \caption{The DM relic abundance as a function of $M_{\psi^0}$ with fixed $v_R$. Left: coannihilation case with $v_R=40$ TeV. Right: non-coannihilation case with $v_R=2000$ TeV. The horizontal blue band is the observed DM relic abundance.}
    \label{fig:fix vR 331}
\end{figure}

\subsection{Relic abundance of $(\mathbf{3},\mathbf{3},1)$ DM}
\begin{figure}
    \centering
    \includegraphics[width=0.55\linewidth]{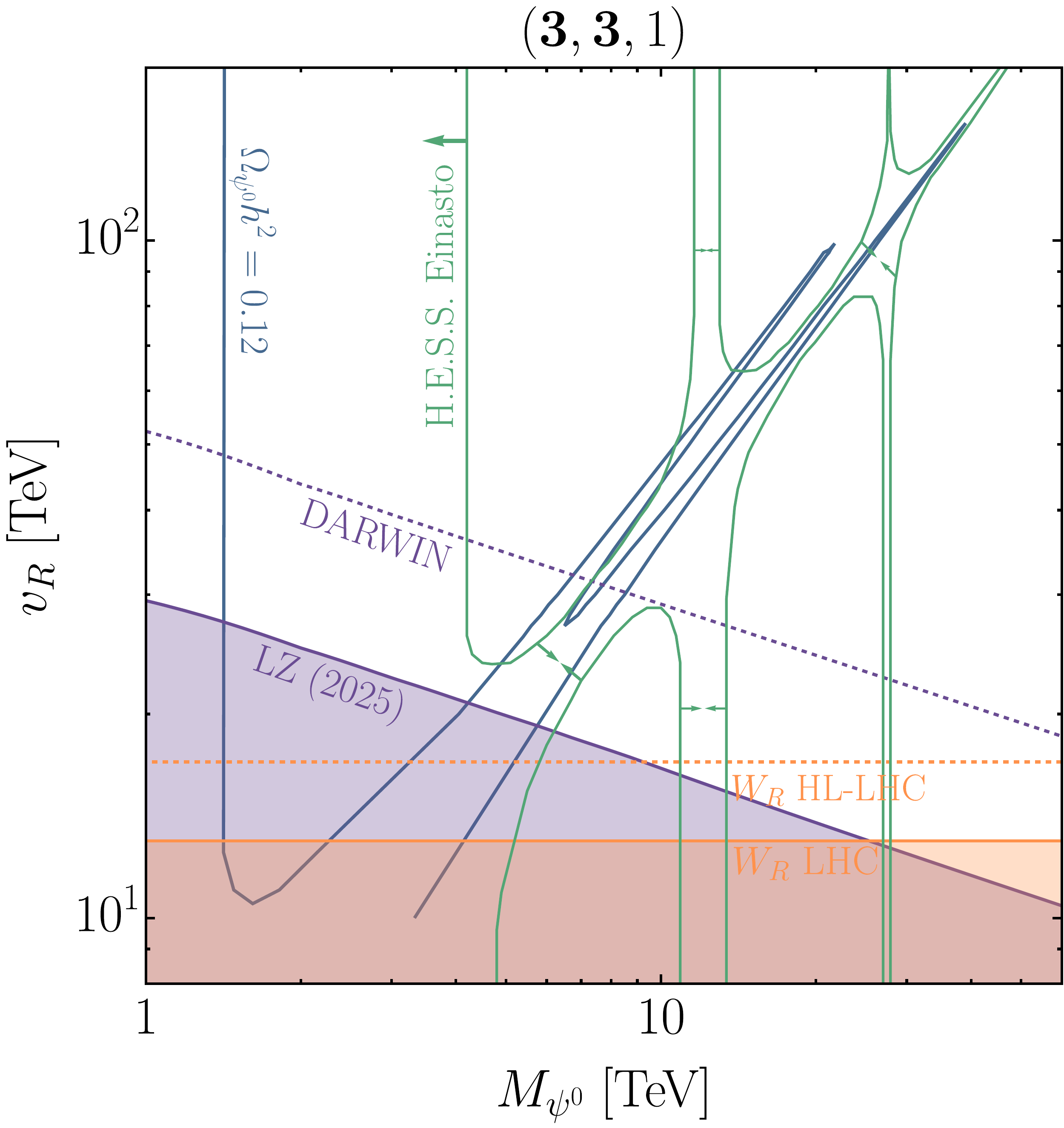}
    \includegraphics[width=0.49\linewidth]{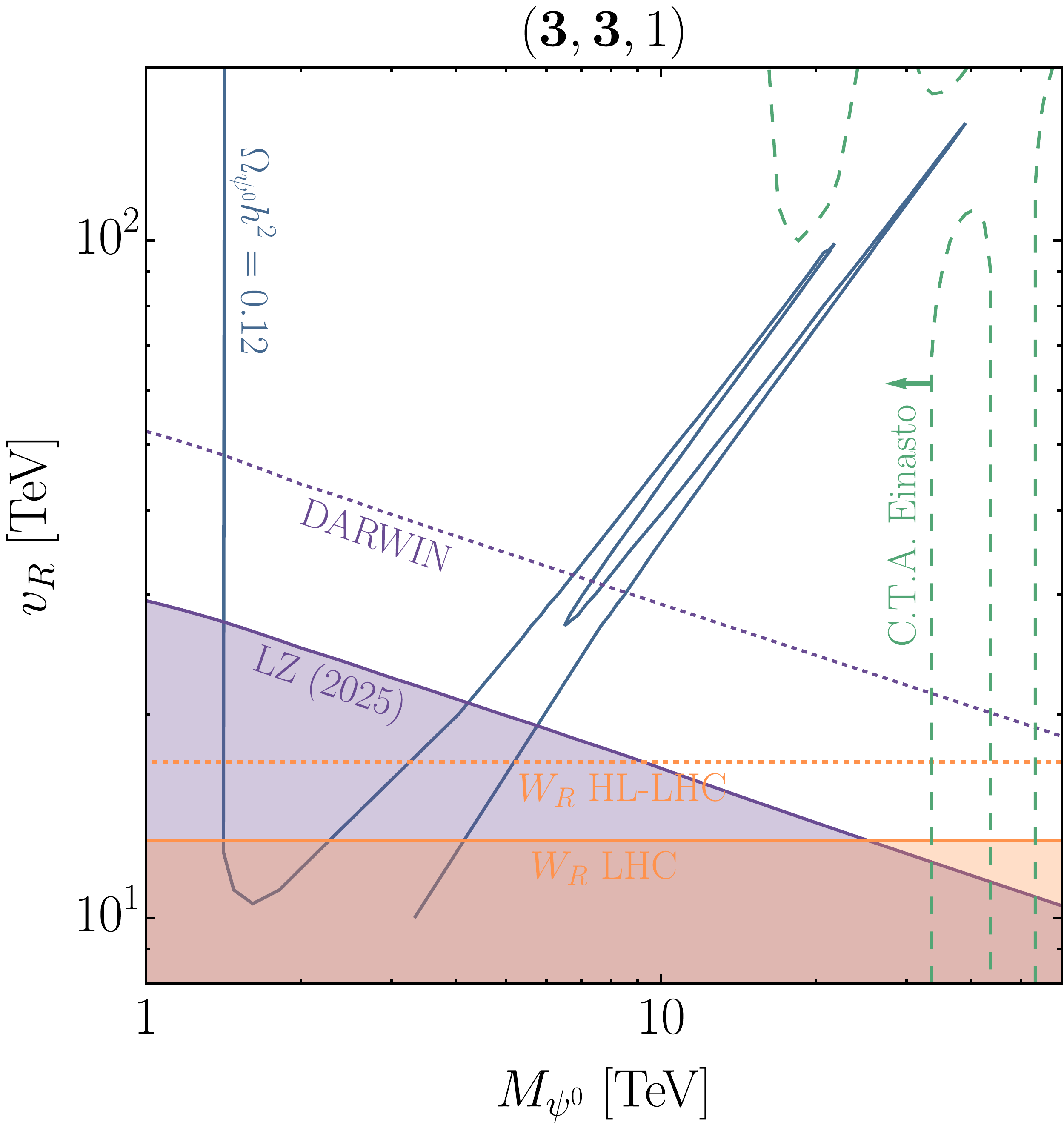}
    \includegraphics[width=0.49\linewidth]{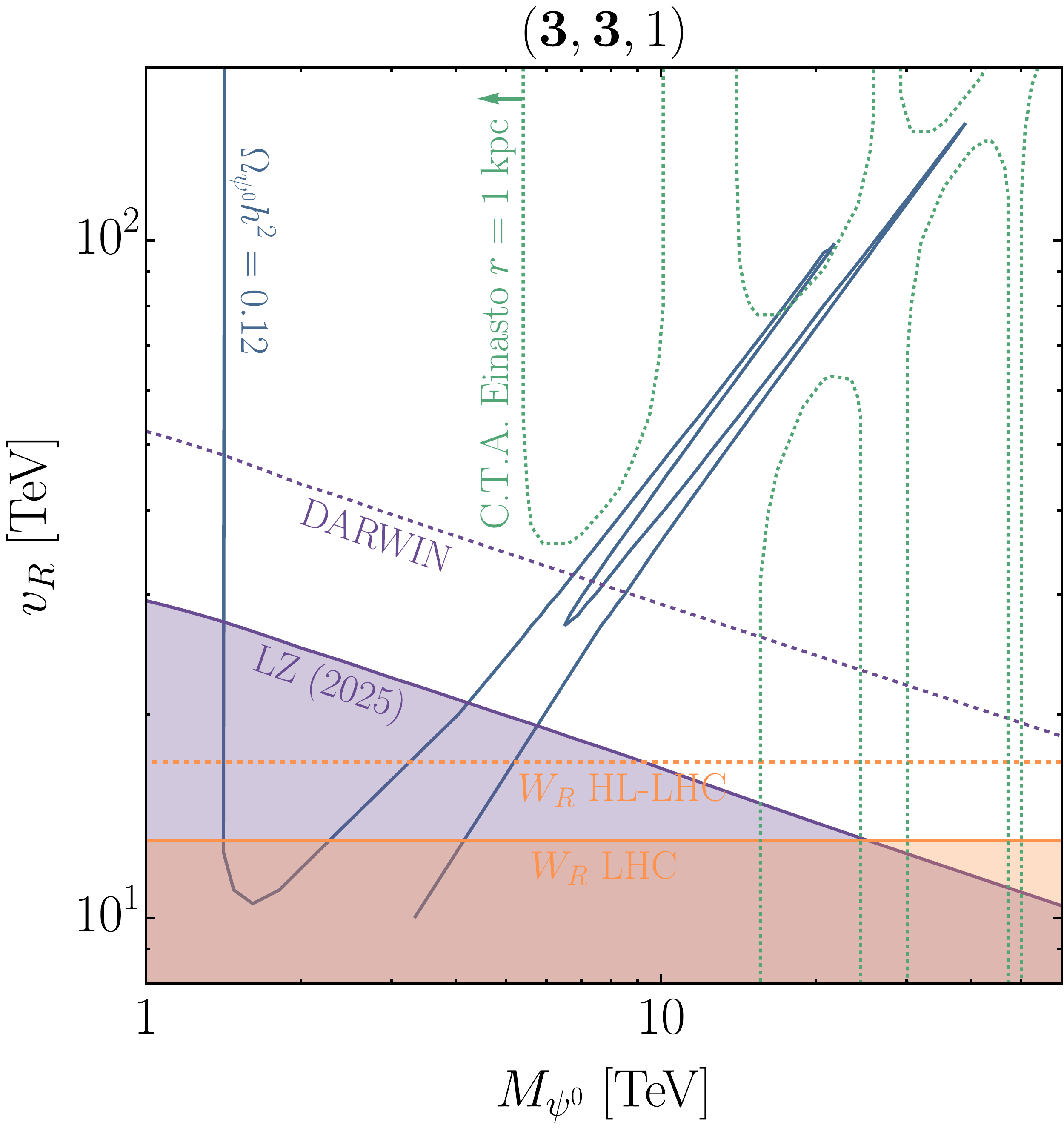}
    \caption{The $\left(M_{\psi^0},v_R\right)$ parameter space for $(\mathbf{3},\mathbf{3},1)$ DM. The blue contour corresponds to the observed DM relic abundance, with $W_R$ and $Z_R$ resonance branches. An upper bound on the Parity breaking scale is obtained from the $Z_R$ branch of around $v_R\lesssim150$~TeV. Constraints from LHC $W_R$ searches and projected sensitivity of the HL-LHC are shown in solid and dashed orange lines, respectively. H.E.S.S.~indirect-detection constraints (assuming the Einasto density profile) are shown in solid green lines in the top panel. The projected sensitivity of the CTA is shown in the bottom two panels, assuming the Einasto profile with no core (left) and $r=1$~kpc core (right). Arrows point inwards towards the regions disfavoured by indirect detection.}
    \label{fig:main_331}
\end{figure}

For $(\mathbf{3},\mathbf{3},1)$ DM, the coannihilation condition is also given by Eq.~\eqref{eq:coan_condition}.
The DM relic abundance in the coannihilation case is $\Omega_{\rm DM}h^2=\Omega_{\psi^0}h^2$. In the non-coannihilation case, $\psi^{0,\pm}$ annihilate with one another to set $\Omega_{\psi^0}h^2$, while $S_{0,1,2}$ annihilate with one another to set $\Omega_{S_0} h^2$, and $R_{1,2,3}$ annihilate with one another to set $\Omega_{R_1} h^2$. Then $S_0$ and $R_1$ later decay to $\psi^0$. The DM relic abundance in this case is given by $\Omega_{\rm DM}h^2 = \Omega_{\psi^0}h^2 + M_{\psi^0} \Omega_{S_0} h^2/M_{S_0}+ M_{\psi^0} \Omega_{R_1} h^2/M_{R_1}$.

Fig.~\ref{fig:fix vR 331} shows the DM relic abundance as a function of $M_{\psi^0}$ with $v_R$ fixed, for coannihilation (left) and non-coannihilation (right). The features of these plots are understood analogously to the $(\mathbf{3},\mathbf{3},0)$ results.

In Fig.~\ref{fig:main_331}, we show the parameter space that produces the correct DM relic abundance in the $(M_{\psi^0},v_R)$ plane. The features are also analogous to the $(\mathbf{3},\mathbf{3},0)$ coannihilation result, except for the relative effectiveness of the $W_R$ and $Z_R$ resonances. In this case, the $W_R$ resonance branch can produce the correct abundance for $v_R\lesssim 100$~TeV, while the $Z_R$ resonance branch can further extend to $v_R$ as large at $150$~TeV. This can be understood as follows. In contrast to $(\mathbf{3},\mathbf{3},0)$, the electroweak multiplet containing DM couples to $Z_R$ and there exist annihilation channels via $Z_R$ without Boltzmann suppression, whereas annihilation via $W_R$ is at least singly Boltzmann suppressed.
In addition to the collider and indirect-detection constraints shown in orange and green, respectively, direct-detection constraints from LZ~\cite{LZCollaboration:2024lux} are shown in the solid purple line with the projected sensitivity of DARWIN~\cite{DARWIN:2016hyl} shown in the purple dashed line. The details of these constraints are discussed in the following section.

\section{Phenomenology}
\label{sec:pheno}

\subsection{Collider searches for new gauge bosons}
The breaking of the extended $SU(2)_R \times U(1)_X$ gauge symmetry predicts new $W_R$ and $Z_R$ gauge bosons, similar to those in the SM electroweak sector. Neglecting the running of the gauge couplings, their masses can be approximated as
\begin{align}
    M_{W_R} & = \frac{v_R}{v_L}M_{W_L} = 13.8 \left(\frac{v_R}{30~\tev}\right)~\tev, \nonumber \\
    M_{Z_R}  & = \frac{v_R}{v_L}M_{Z_L} = 15.4 \left(\frac{v_R}{30~\tev}\right)~\tev, 
\end{align}
with $v_L \simeq 174~\gev$. In numerical computations, we include running effects and evaluate the gauge couplings at $v_R$.

Currently, $W_R$ searches provide a more stringent bound on $v_R$ than $Z_R$ searches. $W_R$ gauge boson masses below $6.0~\tev$ are excluded by $W_R\rightarrow l\nu$ searches~\cite{ATLAS:2019lsy}, corresponding to $v_R \gtrsim 13~\tev$. The High-Luminosity LHC can probe $W_R$ with masses up to $7.9~\tev$~\cite{ATL-PHYS-PUB-2018-044}, or $v_R \simeq 17~\tev$. The current constraint and future prospect are shown by the orange shaded regions and orange dotted lines, respectively, in Figs.~\ref{fig:main} and \ref{fig:main_331}.

\subsection{Collider searches for charged fermions}

The DM in most of the predicted parameter space is too heavy to be produced at the LHC.
However, the non-resonance branch of the $(\bm{3}, \bm{3}, 1)$ DM, where $M_{\psi^0}=1.4$ TeV for coannihilation and $M_{\psi^0}=1.2$~TeV for non-coannihilation, can be probed by long-lived charged particle searches if the corresponding lifetime of charged particles is longer than 10~ns~\cite{ATLAS:2025fdm}.

The relevant decay channel is $R_1$ decaying into $S_0$, whose rate is
\begin{equation}
    \Gamma_{R_1\rightarrow S_0}=N_f\frac{2 G_{F_R}^2}{15\pi^3}\left(M_{R_1}-M_{S_0}\right)^5\approx(100~{\rm ns})^{-1} \frac{N_f}{9} \left(\frac{M_{R_1}-M_{S_0}}{200~ {\rm GeV}}\right)^5\left(\frac{4\times 10^3~{\rm TeV}}{v_R}\right)^4,
\end{equation}
where $G_{F_R}=1/(2\sqrt{2}v_R^2)$ is the $SU(2)_R$ Fermi constant and $N_f$ is the number of final states.
With the three generations of leptons and the first two generations of quarks, $N_f=9$.
The lifetime of $R_1$ greater than $10$ ns corresponds to the non-coannihilation case.
Comparing the lifetime as a function of $v_R$ for $M_{\psi^0}=1.2$~TeV with the constraints in~\cite{ATLAS:2025fdm}, we find that the non-coannihilation case with $v_R \gtrsim 1100$ TeV is already excluded.

\subsection{Direct detection}
Bi-triplet DM in $(\mathbf{3},\mathbf{3},1)$ couples to $Z_L$ and $Z_R$ gauge bosons. The effective Lagrangian of DM and SM quarks is
\begin{equation}
    \mathcal{L}_\text{eff}=\frac{\sqrt{2}}{3}G_{F_R}\ol{\psi^0}\gamma^\mu\psi^0\left[\left(3-2 t_L^2-8s_L^2t_L^2\right)\ol{u}\gamma_\mu u+\left(-3-2 t_L^2+4s_L^2t_L^2\right)\ol{d}\gamma_\mu d\right].
\end{equation}
These interactions lead to the following scattering cross section between DM and nucleons
\begin{equation}
    \sigma_\text{DD}=\frac{2G_{F_R}^2 m_n^2}{\pi}\frac{1}{A^2}\left[(A-Z)\left(1+2 t_L^2\right)-Z\left(1-2t_L^2-4s_L^2t_L^2\right)\right]^2,
\end{equation}
where $m_n$ is the nucleon mass, and $A$ and $Z$ are the mass and atomic numbers, respectively. This direct-detection cross section is constrained by Xenon scattering measurements by the LZ experiment~\cite{LZCollaboration:2024lux}, with a total exposure of 4.2 tonne-years, placing a lower bound on $v_R$. The DARWIN experiment~\cite{DARWIN:2016hyl} will improve these constraints with a total exposure of $1000$ tonne-years.
The current constraint and future prospect are shown by the purple shaded region and purple dotted line, respectively, in Fig.~\ref{fig:main_331}.
As we see, direct detection can surpass the collider constraints.

\subsection{Indirect detection}
\label{sec:indirect}
The annihilation of bi-triplet DM at the galactic center predicts observable gamma-ray signals. When DM directly annihilates to photons, a monochromatic line signal is produced, whereas annihilation to other SM particles results in a shower that produces a continuum signal.

At tree level, the annihilation cross section of DM into monochromatic photons is zero. However, through the Sommerfeld effect, initial DM states mix with charged states that have a non-zero annihilation cross section to photons. The Sommerfeld enhanced DM annihilation cross section to photons for $(\mathbf{3},\mathbf{3},0)$ DM is
\begin{align}\label{eq:xsec_line}
    \langle \sigma v \rangle_{\chi^0 \chi^0\rightarrow\gamma \gamma} = & \frac{1}{N_0}\int_0^{v_{\rm esc}} dv~ 4 \pi v^2 \exp(-\frac{v^2}{4v_0^2}) \left[ (\sigma v)_{\chi^+ \chi^- \to \gamma \gamma} + \frac{1}{2}(\sigma v)_{\chi^+ \chi^- \to Z \gamma} \right] |d_{21}|^2\,,
\end{align}
where $v_0 \simeq 220~\rm km/s$ is the average DM velocity, $v_{\rm esc} \simeq 550~\rm km/s$ is the escape velocity, $d_{ij}$ are the Sommerfeld factors, computed as described in App.~\ref{App:Sommerfeld}, and $N_0=\int_0^{v_{\rm esc}}4\pi v^2\exp(-v^2/(4 v_0^2))$.

Annihilation into other SM states produces a continuum spectrum of gamma rays via the showering of the final-state SM particles. This continuum signal is computed by summing over all non-photon final states. For example, the Sommerfeld enhanced DM annihilation cross section to $W_LW_L$ for $(\mathbf{3},\mathbf{3},0)$ DM is
\begin{align}\label{eq:xsec_continuum}
    \langle \sigma v \rangle_{\chi^0 \chi^0\rightarrow WW} = \frac{1}{N_0}\int_0^{v_{\rm esc}} dv~ 4 \pi v^2 e^{-\frac{v^2}{4v_0^2}} &\left[(\sigma v)_{\chi^0 \chi^0} |d_{22}|^2 +  (\sigma v)_{\chi^+ \chi^-}  |d_{21}|^2\right.\nonumber\\
    &\left.+ 2\Gamma_{\chi^0 \chi^0 \leftrightarrow \chi^+ \chi^-} \Re(d_{22} d_{12}^*)\right]\,,
\end{align}
where $\Gamma_{\chi^0 \chi^0 \leftrightarrow \chi^+ \chi^-}$ is the component of the absorptive matrix encoding the DM and charged initial state mixing, details and definition of which are given in App.~\ref{App:Sommerfeld}. Equivalent expressions of Eqs.~\eqref{eq:xsec_line} and~\eqref{eq:xsec_continuum} are also obtained for $(\mathbf{3},\mathbf{3},1)$ DM. 

\begin{figure}[t!]
    \centering
    \includegraphics[width=0.49\linewidth]{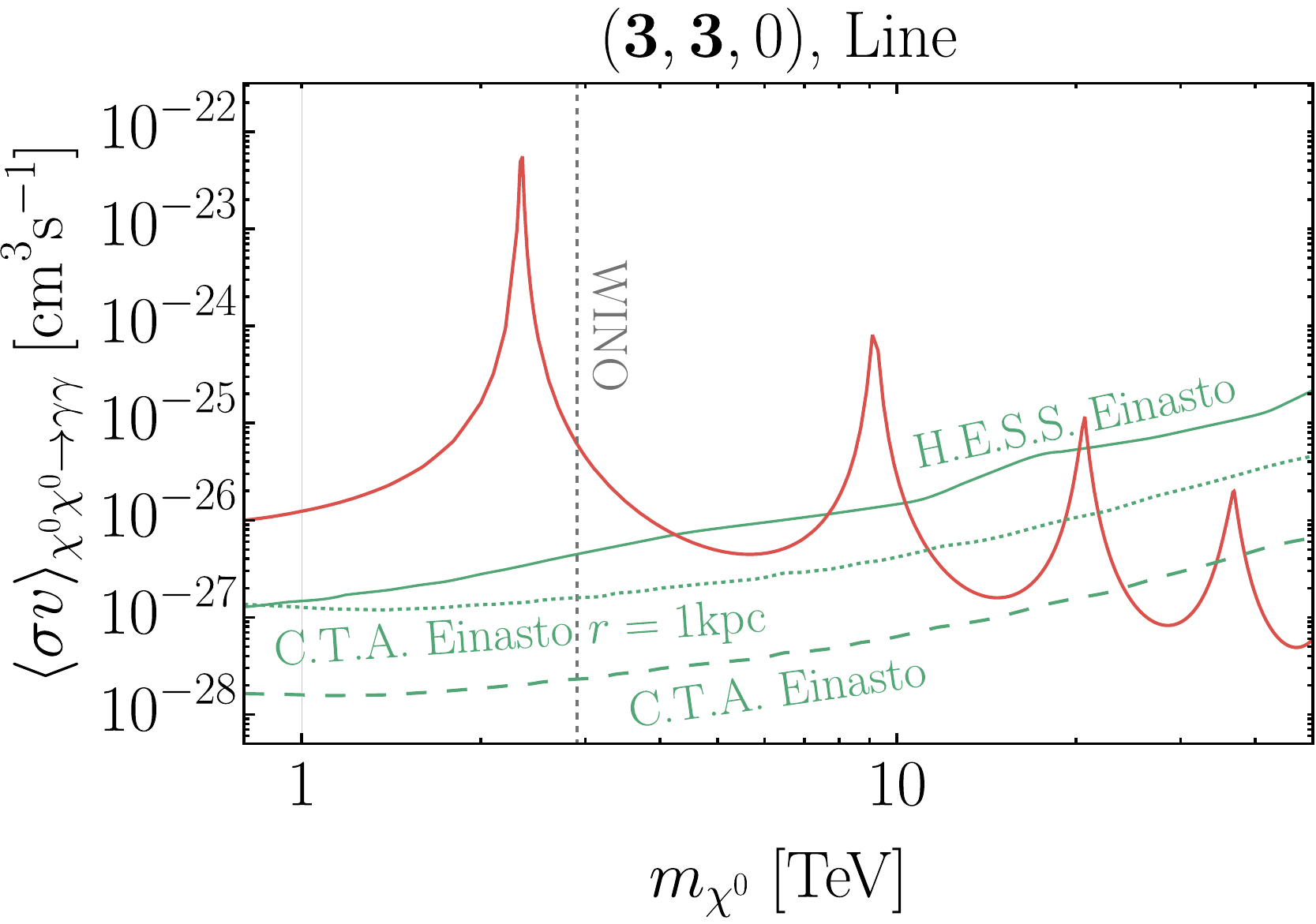}
    \includegraphics[width=0.49\linewidth]{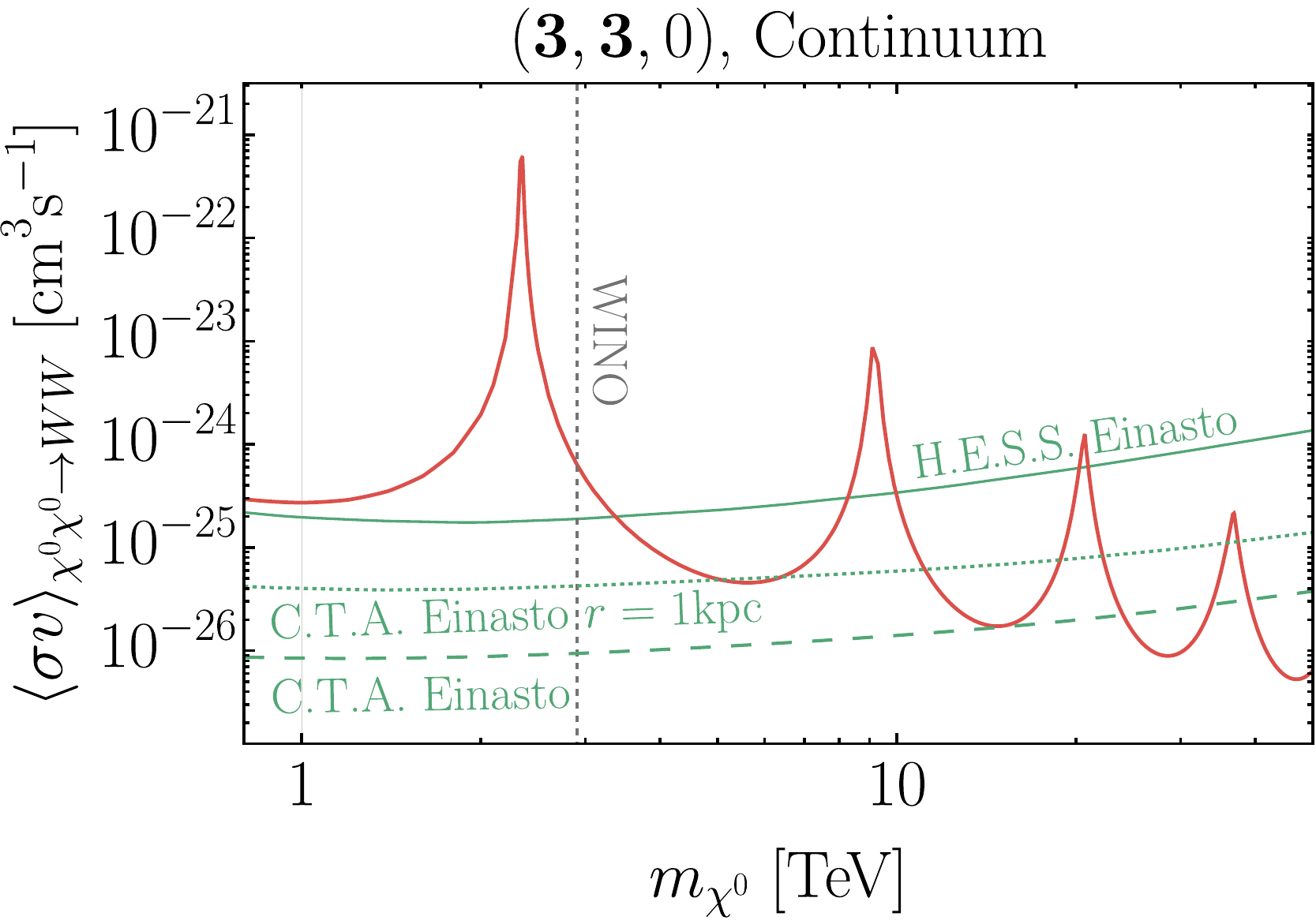}
    \includegraphics[width=0.49\linewidth]{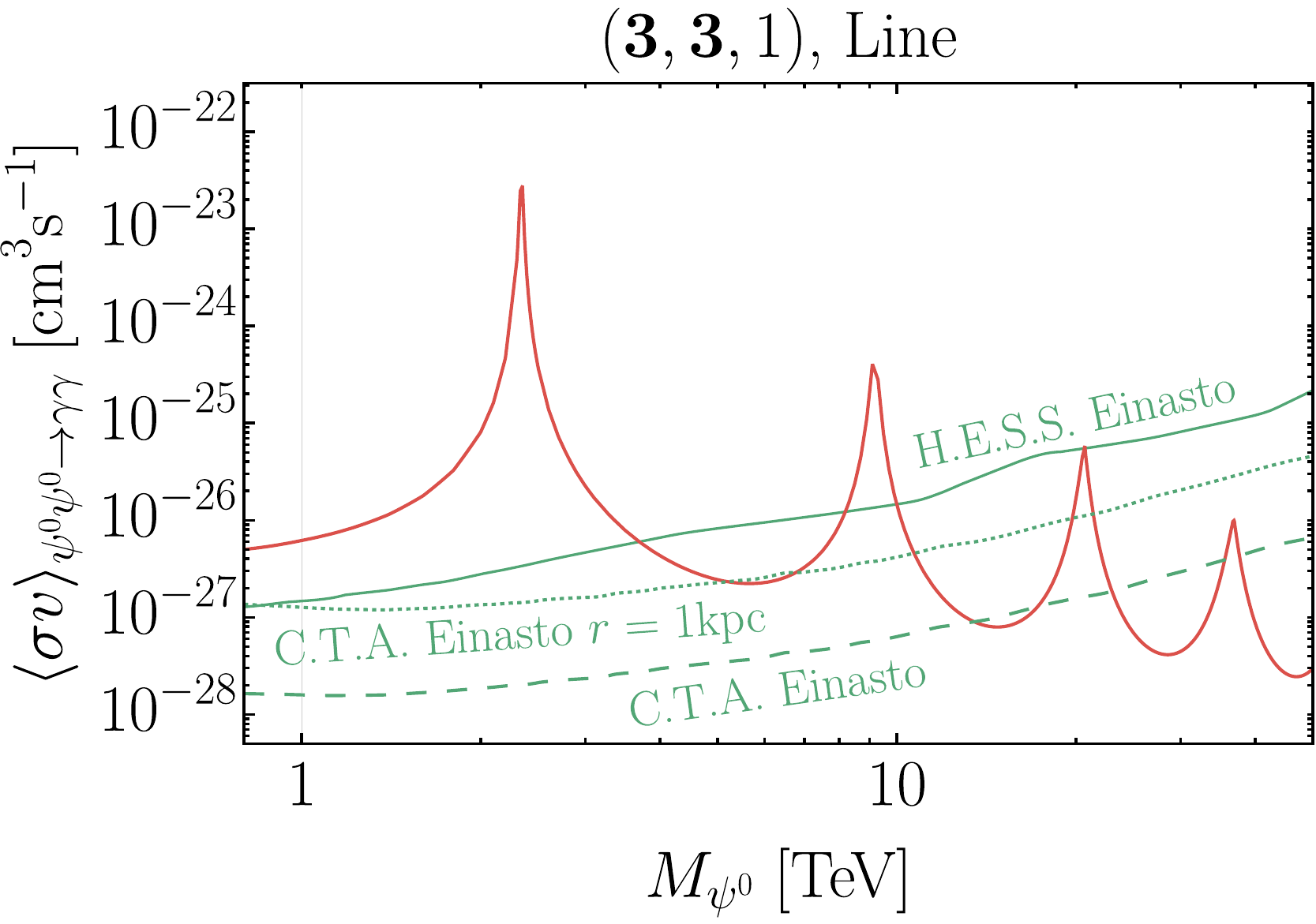}
    \includegraphics[width=0.49\linewidth]{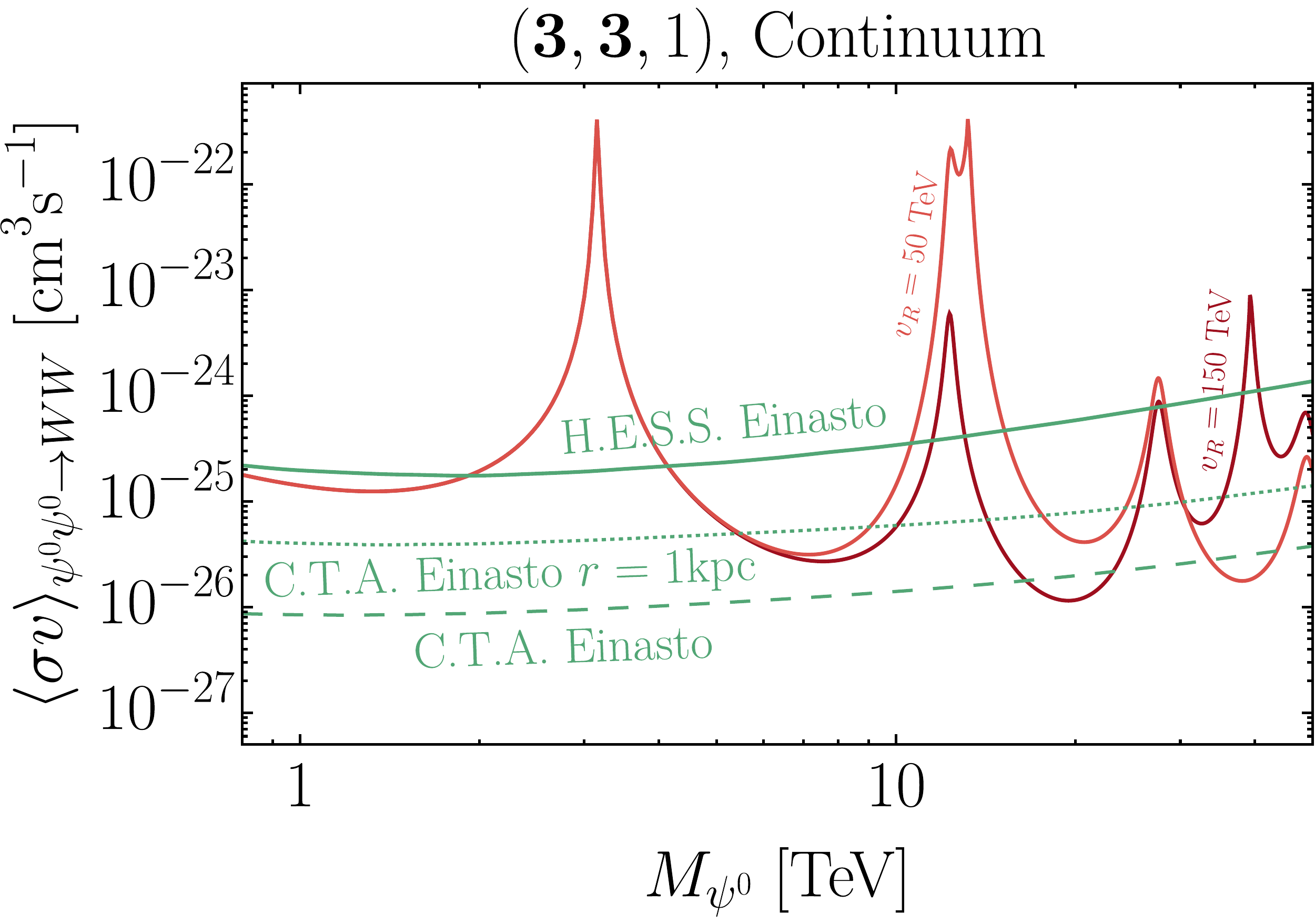}
    \caption{DM annihilation cross sections at the galactic center. The top and bottom panels show cross sections for $(\mathbf{3},\mathbf{3},0)$ and $(\mathbf{3},\mathbf{3},1)$ DM, respectively. In each panel, the left plot shows the photon line signal from DM annihilation directly into photons, and the right panel shows the photon continuum signal from DM annihilation into other SM particles. Two curves are shown for the $(\mathbf{3},\mathbf{3},1)$ DM continuum signal, corresponding to two values of $v_R$; for each, the $Z_R$ resonance is observed in addition to the Sommerfeld resonances. The green solid lines show the H.E.S.S. constraint for the Einasto profile (with refinement~\cite{HESS:2018cbt,HESS:2022ygk,Rodd:2024qsi}). The CTA projected sensitivity for the Einasto profile with no core and $1$~kpc core are shown in green dashed and dotted lines, respectively.}
    \label{fig:indirect}
\end{figure}

Fig.~\ref{fig:indirect} shows the predicted line and continuum gamma-ray signals for $(\mathbf{3},\mathbf{3},0)$ and $(\mathbf{3},\mathbf{3},1)$ DM.
Sommerfeld resonances are observed in all cases.
The line signals for both $(\mathbf{3},\mathbf{3},0)$ and $(\mathbf{3},\mathbf{3},1)$ DM, as well as the continuum signal for $(\mathbf{3},\mathbf{3},0)$ DM, do not involve heavy gauge bosons and are therefore independent of $v_R$.
By contrast, the continuum signal for $(\mathbf{3},\mathbf{3},1)$ involves $Z_R$ exchange, producing a corresponding resonance in addition to the Sommerfeld resonances.
We plot this signal for two values of $v_R$ to illustrate this feature.
For comparison, the $(\mathbf{3},\mathbf{3},0)$ panels include a vertical grey dashed line, corresponding to the wino DM prediction, $m_{\chi^0}=2.9$~TeV.

Currently, the most stringent constraints of line and continuum gamma-ray signals come from searches by H.E.S.S.~\cite{HESS:2022ygk}, whose results were refined in Ref.~\cite{Rodd:2024qsi}, assuming the Einasto DM density profile.
The CTA is projected to improve these constraints by more than an order of magnitude~\cite{CTA:2020qlo,CTAO:2024wvb}.
In Figs.~\ref{fig:main},~\ref{fig:main_331}, and~\ref{fig:indirect}, we show the refined H.E.S.S. constraints for the Einasto profile, together with the projected CTA sensitivities for Einasto profiles without a core and with a $1$~kpc core.%
\footnote{We choose the core size of $1$ kpc since the Einasto profile with this core size has a similar $J$-factor to the Thelma profile~\cite{Safdi:2025sfs} that is suggested by some galaxy simulations~\cite{Cautun:2019eaf}.}
Current viable parameter regions must lie away from both Sommerfeld and heavy gauge boson resonances. The low-mass regions, where the electroweak interaction determines the freeze-out abundance, are disfavored by indirect-detection constraints. With a more cored profile, however, these regions are viable. In particular, we expect that $M_{\psi^0}=1.2$ and $1.4$ TeV for $(\mathbf{3},\mathbf{3},1)$ DM are viable if the core size is as large as $1$ kpc.
Assuming an Einasto profile with a core size of around $1$~kpc, the CTA will probe nearly all of the parameter space for $(\mathbf{3},\mathbf{3},1)$ DM, while the parameter space for $(\mathbf{3},\mathbf{3},0)$ DM can be nearly all probed for less cored DM profiles.

\subsection{BBN}

Within the $(\mathbf{3},\mathbf{3},0)$ DM multiplet, $N$ is the only particle whose main decay channel is mediated by $W_R$ gauge boson, i.e., $N \to \chi^+ W_R^{-*}$, with $\chi^+$ eventually decaying into $\chi^0$ and SM fermions mediated by $W_L$.
The decay rate is
\begin{equation}
    \Gamma_{N} = N_f \frac{2 G^2_{F_R}}{15\pi^2} \left(M_N-M_{\chi^-}\right)^5 
    \approx(1~{\rm sec})^{-1} \frac{N_f}{9} \left(\frac{M_{N}-M_{\chi^-}}{100~ {\rm GeV}}\right)^5\left(\frac{10^5~{\rm TeV}}{v_R}\right)^4.
\end{equation}
If $N$ decays after BBN, the highly energetic decay products destroy the light elements, ruining the predictions of BBN. In the parameter region where the DM relic abundance is determined by the resonance annihilation through $W_R$ or $Z_R$, the decay of $N$ occurs much before BBN begins. When annihilation via the electroweak interactions dominate, $v_R$ can be large and the BBN constraint exists. However, this case is disfavored by indirect-detection experiments.
The same conclusion holds for $(\mathbf{3},\mathbf{3},1)$ DM, although the indirect detection bound is less stringent. In addition, the parameter region excluded by BBN is already excluded by the LHC search for long-lived charged particle searches.

\section{Summary and outlook}
\label{sec:summary}

In this paper, we have studied accidentally stable DM in a Parity solution to the strong CP problem. We have found that the bi-triplet DM embeddings enjoy an accidental stability because their leading decay operators are dimension six. The freeze-out abundance of DM was computed for the two viable bi-triplet embeddings.

When the DM mass is much below the new gauge boson mass, the abundance is determined solely by electroweak interactions. This case, however, is disfavored by indirect-detection experiments unless the DM profile of the galactic center is highly cored. When the DM mass is comparable to the new gauge boson masses, annihilation involving the new gauge bosons changes the predicted DM mass.
In particular, resonant annihilation allows for DM masses well above $\mathcal{O}(1)$ TeV. Interestingly, in this branch, the $SU(2)_R\times U(1)_X$ symmetry breaking scale is bounded from above to be $v_R\lesssim150$~TeV. The free parameters of the model-- the DM mass and the $SU(2)_R\times U(1)_X$ breaking scale-- are probed by a combination of collider, direct-detection, and indirect-detection experiments. 
The HL-LHC can probe the parameter space with $v_R< 17$ TeV.
For $({\bf 3},{\bf 3},1)$ DM, near-future direct-detection experiments can probe $v_R \lesssim 30$~TeV. Complementary to the $\mathcal{O}(10)$~TeV reach of collider and direct-detection experiments, indirect-detection searches from H.E.S.S. can probe larger values of $v_R$ for certain galactic center DM density profiles.
Future indirect-detection experiments can extend even further.
If the DM profile is as cuspy as the Einasto profile, the CTA can probe the entire parameter space of $({\bf 3},{\bf 3},1)$ DM and almost all of the parameter space of $({\bf 3},{\bf 3},0)$ DM. Even assuming a core as large as $1$~kpc, the CTA can probe almost all of the parameter space of $({\bf 3},{\bf 3},1)$ DM and a significant fraction of that of $({\bf 3},{\bf 3},0)$ DM.

The relatively low $v_R$ may lead to signals beyond those from DM and new gauge bosons. For example, since the UV completion of the Yukawa interactions is around $v_R$, rare processes such as $\mu \rightarrow e \gamma$ and the electron electric dipole moment may be observable.
The rates of these processes will depend on the details of the UV Yukawa structure, which we leave for future work.

\section*{Acknowledgments}

We thank Nicholas Rodd for helpful discussions on the indirect-detection signal.
Fermilab is operated by Fermi Forward Discovery Group, LLC under Contract No. 89243024CSC000002 with the U.S.~Department of Energy, Office of Science, Office of High Energy Physics.
This work was supported by the U.S.~Department of Energy under Grant  No.~DE-SC0009924 (MJB, KH), DOE distinguished scientist fellowship grant FNAL 22-33 (IRW), and the World Premier International Research Center Initiative (WPI), MEXT, Japan (Kavli IPMU) (KH). 

\appendix

\section{Gauge interactions}\label{app:gauge_coup}
In this appendix, we show the gauge interactions of SM fermions, Higgses, and DM multiplets.

\subsection{Mass mixing and interaction with the SM Higgs}
The SM Higgs couples not only to SM gauge bosons but also to the new heavy gauge boson $Z_R$. We denote the SM neutral weak gauge boson as $Z$ and the analogous gauge boson from $SU(2)_R\times U(1)_X$ breaking as $Z'$. Then, as we will see, these bosons mix, forming two mass eigenstates, the lighter of which we call $Z_L$ and the heavier $Z_R$. The charged weak gauge bosons do not mix at tree level.

The coupling to $W_L$ is given by
\begin{align}
    \mathcal{L} \supset g M_{W_L} h W_L^+ W_L^- + \frac{g^2}{4} h^2 W_L^+ W_L^-.
\end{align}
The coupling to the SM $Z$ gauge boson is given by
\begin{align}
    \mathcal{L} \supset \frac{1}{8}(g^2 + g_Y^2) h^2 Z^2 + g \frac{M_{Z}^2}{M_{W_L}}h Z^2.
\end{align}

$H_L$ also couples to the $Z'$ gauge boson associated with $SU(2)_R\times U(1)_X$ breaking due to its $U(1)_X$ charge,
\begin{align}
    \mathcal{L} \supset \frac{1}{2}g\frac{s_R^4}{c_R^2} M_{W_L} h Z'^2 + \frac{1}{8} g^2 \frac{s_R^4}{c_R^2} h^2 Z'^2.
\end{align}
This $Z'$ mixes with the SM $Z$ boson via the following terms
\begin{align}
    \mathcal{L} \supset \frac{g^2}{4} \frac{s_L^2}{c_L^2 \sqrt{1-2s_L^2}} h^2 Z Z' + g \frac{s_L^2}{c_L \sqrt{1-2s_L^2}}  M_{Z} h Z Z' + \frac{s_L^2}{\sqrt{1-2s_L^2}}  M_{Z}^2 Z Z'.
\end{align}
The mixing is given by the last term as
\begin{align}
    \sin \theta_{Z Z'} \simeq  \frac{M_{Z}^2}{M_{Z'}^2} \frac{s_L^2}{\sqrt{1-2s_L^2}}.
\end{align}

\subsection{Interactions with SM fermions}

We assume that the SM fermions and right-handed neutrinos dominantly come from the $SU(2)_L$ or $SU(2)_R$ doublets
\begin{align}
    q = \begin{pmatrix}
            u \\ d
        \end{pmatrix},~
    \ol{q} = \begin{pmatrix}
                  \ol{d} \\ - \ol{u}
              \end{pmatrix},~
    \ell = \begin{pmatrix}
               \nu \\ e
           \end{pmatrix},~
    \ol{\ell} = \begin{pmatrix}
                     \ol{e} \\ -\ol{N}
                 \end{pmatrix}.
\end{align}
The gauge interactions of SM fermions and right-handed neutrinos are given by
\begin{align}
    \mathcal{L} & =\frac{g}{\sqrt{2}} W_{L\mu}^- \left(d^\dag \ol{\sigma}^\mu u + e^\dag \ol{\sigma}^\mu \nu  \right)+   \frac{g}{\sqrt{2}} W_{L\mu}^+ \left(u^\dag \ol{\sigma}^\mu d + \nu^\dag \ol{\sigma}^\mu e \right) \nonumber                                              \\
                & -   \frac{g}{\sqrt{2}} W_{R\mu}^- \left(\ol{u}^\dag \ol{\sigma}^\mu \ol{d} + \ol{N}^\dag \ol{\sigma}^\mu \ol{e} \right) -\frac{g}{\sqrt{2}} W_{R\mu}^+ \left(\ol{d}^\dag \ol{\sigma}^\mu \ol{u} + \ol{e}^\dag \ol{\sigma}^\mu \ol{N}  \right) \nonumber \\
                & + \frac{g}{c_R}Z'_{\mu}\sum_{f} \left(I_{3R,f}-s_R^2 Y_f\right) f^\dag \ol{\sigma}^\mu f  +
    \frac{g}{c_L}Z_\mu\sum_{f} \left(I_{3L,f}-s_L^2 Q_f\right) f^\dag \ol{\sigma}^\mu f +
    e A_\mu \sum_{f} Q_f f^\dag\ol{\sigma}^\mu f,
\end{align}
where $f$ are Weyl fermions.
In terms of four-component fields,
\begin{align}
    \mathcal{L} & =\frac{g}{\sqrt{2}}W_{L\mu}^-\left( \ol{d}\gamma^\mu P_L u + \ol{e}\gamma^\mu P_L \nu \right) + \frac{g}{\sqrt{2}}W_{L\mu}^+\left( \ol{u}\gamma^\mu P_L d + \ol{\nu}\gamma^\mu P_L e \right) \nonumber \\
                & +  \frac{g}{\sqrt{2}}W_{R\mu}^-\left( \ol{d}\gamma^\mu P_R u + \ol{e}\gamma^\mu P_R N \right) + \frac{g}{\sqrt{2}}W_{R\mu}^+\left( \ol{u}\gamma^\mu P_R d + \ol{N }\gamma^\mu P_R e \right) \nonumber  \\
                & + \frac{g}{c_R}Z_\mu' \sum_F \ol{F} \gamma^\mu \left(s_R^2\left(I_{3F}- Q_F\right)P_L + \left(I_{3F}-s_R^2Q_F\right) P_R \right) F \nonumber                                                              \\
                & +  \frac{g}{c_L} Z_\mu \sum_F \ol{F} \gamma^\mu \left(\left(I_{3F}- s_L^2 Q_F\right)P_L - s_L^2Q_F P_R \right) F +
    e A_\mu \sum_F Q_F \ol{F}\gamma^\mu F,
\end{align}
where $F$ are Dirac fermions.

\section{The Sommerfeld effect}\label{App:Sommerfeld}
In this appendix, we briefly review the Sommerfeld effect. We set the DM mass to be $m_{\cal X}$, and define $x=m_{\cal X}/T$. By the time of freeze-out at $x\approx 20$, DM is non-relativistic. We therefore work in the non-relativistic limit, where $s$-wave contributions dominate the annihilation cross section. For analytical calculations, we define the center-of-mass (CoM) frame assuming both initial states have the same mass. This approximation can be justified since the mass differences between initial state particles, in the relevant parameter space, are at most $1-10\%$ of the DM mass, as shown in Fig.~\ref{fig:running}. In numerical calculations, $m_{\cal X}$ is set to be the average mass of the two initial state particles.

Consider $N$ incoming initial states $a = 1,2,3,\dots, N$ with identical total charge and spin, each containing two particles that annihilate.
The annihilation cross section for a certain initial state $a$ can be written as
\begin{align}
    \label{eq:sigma v}
    (\sigma v)_a = c_a \sum_{kl} \Gamma_{kl} d_{ak} d_{al}^*\,,
\end{align}
where $c_a = 2$ for identical particles in the initial state, and $c_a = 1$ otherwise.
The absorptive matrix $\Gamma$ has the form
\begin{align}
    \Gamma_{kl}|_{k=l}      = \frac{1}{c_k} (\sigma v)_k \,,~~\Gamma_{kl}|_{k \neq l}  = \frac{1}{\sqrt{c_k c_l}} (\sigma v)_{k \leftrightarrow l}\,.
\end{align}
The off-diagonal elements of the absorptive matrix encode the transition amplitude between 2 different initial states. For two initial states $k$ and $l$, the transition amplitude is computed via the optical theorem as
\begin{align}
    \label{eq:optical}
    (\sigma v)_{k \leftrightarrow l} = \frac{1}{2}\sum_{j} \sqrt{(\sigma v)_{k \rightarrow j} \times (\sigma v)_{l \rightarrow j}},
\end{align}
where the summation is performed over all common annihilation final states $j$.

$d_{ak}$ are the Sommerfeld factors and can be computed by solving the following Schr\"{o}dinger equation
\begin{align}
    - \frac{1}{m_{\cal X}} \frac{d^2}{dr^2} \mathbf{g}(r) + V(r) \cdot \mathbf{g}(r) = \frac{m_{\cal X} v^2}{4} \mathbf{g}(r)\,,
\end{align}
where both the wave function $\mathbf{g}(r)$ and the long-range potentials $V(r)$ are $N \times N$ matrices and $r$ is the radial coordinate. The long-range potentials are provided in App.~\ref{App:long_range_potentials_330} and App.~\ref{App:long_range_potentials_331}.
Two boundary conditions are required to solve this equation.
First, only the outgoing wave exists at spatial infinity,
\begin{align}
    \left.\mathbf{g}_{ab}\right|_{r \rightarrow \infty} \to \sum_k \delta_{ak} d_{kb}(v) \exp(i \frac{m_{\cal X} v r}{2})\,.
\end{align}
Second, the wave function is regular at any $r$.
The Sommerfeld factor $d_{ak}$ is then computed by numerically solving the Schrodinger equation with the above two boundary conditions.
For a cross-check, we also run the variable phase method~\cite{Ershov:2011zz} to compare with the result of directly solving the Schr\"odinger equation.
The difference is found to be below $1\%$.

\section{Annihilation cross sections for $(\mathbf{3},\mathbf{3},0)$}
\label{sec:cross section}

In this appendix, we show the annihilation cross sections for $(\mathbf{3},\mathbf{3},0)$ DM. We only include $s$-wave annihilation, which is a good approximation during freeze-out. For conciseness, we show the result with the leading-order dependence on $v_L$. We further ignore the resonance width and take $s=4m_{\chi^0}^2$ in this section. In the numerical computation, however, the full result is used.
In particular, the dependence of the s-channel $W_R$ and $Z_R$ propagators on the velocity of DM is included.

\subsection{$\chi^0$ and $\chi^\pm$}
\subsubsection{$\chi^0 \chi^-$}
For an initial state with spin-0,
\begin{align}
    \sigma_0 v \vert_{W_L Z_L} =    \frac{2 \pi \alpha_2^2 c_L^2}{m_{\chi^0}^2}\,,~
    \sigma_0 v \vert_{W_L \gamma} =  \frac{2 \pi \alpha_2^2 s_L^2}{m_{\chi^0}^2}\,.
\end{align}

For an initial state with spin-0,
\begin{align}
    \sigma_0 v \vert_{f \ol{f}'} = \frac{\pi \alpha_2^2}{3m_{\chi^0}^2}\,,~ \sigma_0 v \vert_{W_L h_L} = \frac{\pi \alpha_2^2}{12 m_{\chi^0}^2}\,~ \sigma_0 v \vert_{W_L Z_L} = \frac{\pi \alpha_2^2}{12 m_{\chi^0}^2}\,.
\end{align}

\subsubsection{ $\chi^+ \chi^-$}
For an initial state with spin-0,
\begin{align}
    \sigma_0 v \vert_{\gamma \gamma} = \frac{4 \pi \alpha_2^2 s_L^4}{m_{\chi^0}^2}\,,~ \sigma_0 v \vert_{W_L W_L} = \frac{2 \pi \alpha_2^2}{m_{\chi^0}^2}\,,~
    \sigma_0 v \vert_{Z_LZ_L} = \frac{4 \pi \alpha_2^2 c_L^4}{m_{\chi^0}^2}\,,~
    \sigma_0 v \vert_{Z_L \gamma} = \frac{8 \pi \alpha_2^2 c_L^2 s_L^2}{m_{\chi^0}^2}\,.
\end{align}

For an initial state with spin-1,
\begin{align}
    \sigma_0 v \vert_{W_LW_L} =  \frac{\pi \alpha_2^2}{12 m_{\chi^0}^2}\,,~
    \sigma_0 v \vert_{Z_Lh_L} =   \frac{\pi \alpha_2^2}{12 m_{\chi^0}^2}\,,~
    \sigma_0 v \vert_{f \ol{f}} = \frac{2\pi \alpha_2^2 I_{3f}^2}{3 m_{\chi^0}^2}\,.
\end{align}

\subsubsection{  $\chi^- \chi^-$}
The identical particle $\chi^-$ only forms a spin-0 initial state. For an initial state with spin-0,
\begin{align}
    \sigma_0 v \vert_{W_LW_L} = \frac{4 \pi \alpha_2^2}{m_{\chi^0}^2}\,.
\end{align}

\subsubsection{$\chi^0$ $\chi^0$ }
The initial particle pair $\chi^0 \chi^0$ will always form a spin-0 state. For an initial state with spin-0,
\begin{align}
    \sigma_0 v \vert_{W_LW_L} = \frac{8 \pi \alpha_2^2}{m_{\chi^0}^2}\,.
\end{align}

\subsection{$D$, $E$ and $N$}

\subsubsection{  $\ol{N} D$}
For an initial state with spin-0,
\begin{align}
    \sigma_0 v \vert_{W_L W_L} = \frac{4 \pi \alpha_2^2}{m_{\chi^0}^2}\,.
\end{align}

\subsubsection{$\ol{D} D$}
For an initial state with spin-0,
\begin{align}
    \sigma_0 v \vert_{\gamma \gamma} = & \frac{64 \pi \alpha_2^2 s_L^4}{m_{\chi^0}^2}\,,~
    \sigma_0 v \vert_{W_L W_L} =  \frac{2 \pi \alpha_2^2}{m_{\chi^0}^2}\,, ~                                                                             
    \sigma_0 v \vert_{Z_LZ_L} =             \frac{4 \pi \alpha_2^2}{m_{\chi^0}^2} \frac{(c_L^2 - s_L^2)^4}{c_L^4}\,, \\
    \sigma_0 v \vert_{Z_L \gamma} = &\frac{32\pi \alpha_2^2 s_L^2}{m_{\chi^0}^2} \frac{(c_L^2 - s_L^2)^2}{c_L^2}\,,  ~                         
    \sigma_0 v \vert_{Z_R \gamma} =   \frac{32 \pi \alpha_2^2c_R^2 s_L^2}{m_{\chi^0}^2} \left[1 - \frac{M_{Z_R}^2}{4m_{\chi^0}^2}\right]\,,                                                                                            \nonumber \\
    \sigma_0 v \vert_{Z_R Z_L} =          & \frac{8 \pi \alpha_2^2 c_R^2 (1-2c_L^2)^2}{c_L^2 m_{\chi^0}^2} \left[1 - \frac{M_{Z_R}^2}{4 m_{\chi^0}^2}\right]^3 \left[1 - \frac{M_{Z_R}^2}{4 m_{\chi^0}^2} + \frac{M_{Z_R}^4}{32 m_{\chi^0}^4}\right]^{-2}\,. \nonumber
\end{align}

For an initial state with spin-1,
\begin{align}
    \sigma_0 v \vert_{W_L W_L} =   & \frac{\pi \alpha_2^2}{12 m_{\chi^0}^2 c_L^4} \left[1 + \frac{2 \pi  \alpha_2^2 c_R s_L^2 c_L}{\sqrt{1-2s_L^2}} \left(1 - \frac{M_{Z_R}^2}{4m_{\chi^0}^2}\right)^{-1} + \frac{\pi  \alpha_2^2 c_R^2 s_L^4 c_L^2}{1-2s_L^2}\left(1 - \frac{M_{Z_R}^2}{4m_{\chi^0}^2}\right)^{-2}\right]\,, \nonumber\\
    \sigma_0 v \vert_{Z_L h_L} =     & \frac{\pi \alpha_2^2}{12 c_L^4 m_{\chi^0}^2} \left[(1-2c_L^2)^2 + 4 c_R c_L s_L^2\sqrt{2 c_L^2-1} \left(1 - \frac{M_{Z_R}^2}{4m_{\chi^0}^2}\right)^{-1}  + 4 \frac{c_R^2 s_L^4  c_L^2}{2c_L^2 - 1} \left(1 - \frac{M_{Z_R}^2}{4m_{\chi^0}^2}\right)^{-2}   \right]\,,                                                                                                                     \nonumber                                  \\
    \sigma_0 v \vert_{f \ol{f}} = & \frac{2 \pi  \alpha_2^2}{3 m_{\chi^0}^2 c_L^4}\left[\left(I_{3 f}^2\left(1-2 c_L^2\right)^2-2 I_{3 f} Q_f\left(2 c_L^4-3 c_L^2+1\right)+2 s_L^4Q_f^2\right) \right.\nonumber                                                                                             \\
                                   & \left.  2 c_L^2\left(s_R^2 I_{3 f}^2\left(2 c_L^2-1\right) - I_{3 f} Q_f \left(c_L^2 \left(3 s_R^2+1\right) - 2 s_R^2-1\right) -2 s_R^2 s_L^2 Q_f^2 \right) \left(1 - \frac{M_{Z_R}^2}{m_{\chi^0}^2}\right)^{-1}\right.\nonumber                                          \\
                                   & \left.2 c_L^4 \left((1+s_R^4)I_{3 f}^2-2 I_{3 f} Q_f\left(s_R^2+s_R^4\right) + 2 s_R^4 Q_f^2\right) \left(1 - \frac{M_{Z_R}^2}{m_{\chi^0}^2}\right)^{-2}\right]\,.\o
\end{align}

\subsubsection{$\ol{E}E$}
For an initial state with spin-0,
\begin{align}
    \sigma_0 v \vert_{\gamma \gamma} = & \frac{4 \pi \alpha_2^2 s_L^4}{m_{\chi^0}^2}\,,~
    \sigma_0 v \vert_{W_LW_L} =  \frac{8 \pi \alpha_2^2}{m_{\chi^0}^2}\,, ~                                                                                             
    \sigma_0 v \vert_{Z_LZ_L} =           \frac{4 \pi \alpha_2^2 s_L^8}{c_L^4 m_{\chi^0}^2}\,,~
    \sigma_0 v \vert_{Z_L \gamma} =  \frac{8\pi \alpha_2^2 s_L^6}{c_L^2 m_{\chi^0}^2}\,,                                                                                                                                     \\
    \sigma_0 v \vert_{Z_R \gamma} =     & \frac{8 \pi \alpha_2^2c_R^2 s_L^2}{m_{\chi^0}^2} \left[1 - \frac{M_{Z_R}^2}{4m_{\chi^0}^2}\right]\,, ~                                        
    \sigma_0 v \vert_{Z_L Z_R} =           \frac{8 \pi \alpha_2^2 c_R^2 s_L^4}{c_L^2 m_{\chi^0}^2} \left[1 - \frac{M_{Z_R}^2}{4 m_{\chi^0}^2}\right]^3 \left[1 - \frac{M_{Z_R}^2}{4 m_{\chi^0}^2} + \frac{M_{Z_R}^4}{32 m_{\chi^0}^4}\right]^{-2}\,.\nonumber
\end{align}

For an initial state with spin-1,
\begin{align}
    \sigma_0 v \vert_{W_L W_L} =   & \frac{\pi  \alpha _2^2 s_L^4}{12 m_{\chi^0}^2 c_L^4} \left[1 + \frac{2c_R c_L}{\sqrt{2c_L^2 - 1}} \left(1 - \frac{M_{Z_R}^2}{4m_{\chi^0}^2}\right)^{-1} + \frac{c_R^2 c_L^2}{2c_L^2 - 1} \left(1-  \frac{M_{Z_R}^2}{4m_{\chi^0}^2}\right)^{-2}\right],   \\
    \sigma_0 v \vert_{Z_L h_L} =     & \frac{\pi \alpha_2^2 s_L^4}{12 c_L^4 m_{\chi^0}^2} \left[1 - 4 \frac{c_R c_L}{\sqrt{2c_L^2 - 1}} \left(1 - \frac{M_{Z_R}^2}{4m_{\chi^0}^2}\right)^{-1} + 4 \frac{c_R^2 c_L^2}{2c_L^2 - 1}  \left(1 - \frac{M_{Z_R}^2}{4m_{\chi^0}^2}\right)^{-2}\right], \nonumber\\
    \sigma_0 v \vert_{f \ol{f}} = & \frac{2 \pi  \alpha _2^2}{3 m_{\chi^0}^2 c_L^4} \left[s_L^4\left(I_{3f}^2 + 2 I_{3 f} Q_f+2 Q_f^2\right) - 4 c_L^2 s_R^2 s_L^2 \left(I_{3f}^2 - I_{3f}Q_f + 2Q_f^2\right) \left(1 - \frac{M_{Z_R}^2}{4m_{\chi^0}^2}\right)^{-1}\right. \nonumber \\
                                   & \left. + 2 c_L^4 \left(I_{3f}^2(1+s_R^4) - 2I_{3f} Q_f (1+s_R^2) + 2 s_R^4 Q_f\right) \left(1 - \frac{M_{Z_R}^2}{4m_{\chi^0}^2}\right)^{-2}\right]\,.\nonumber
\end{align}

\subsubsection{$\ol{N} N$}
For an initial state with spin-0,
\begin{align}
    \sigma_0 v \vert_{W_LW_L} =   & \frac{2 \pi \alpha_2^2}{m_{\chi^0}^2}\,,~
    \sigma_0 v \vert_{Z_LZ_L} = \frac{4 \pi \alpha_2^2}{c_L^4 m_{\chi^0}^2}\,,    ~                                          
    \sigma_0 v \vert_{Z_L Z_R} =  \frac{8 \pi \alpha_2^2 c_R^2}{c_L^2 m_{\chi^0}^2} \left[1 - \frac{M_{Z_R}^2}{4 m_{\chi^0}^2}\right]^3 \left[1 - \frac{M_{Z_R}^2}{4 m_{\chi^0}^2} + \frac{M_{Z_R}^4}{32 m_{\chi^0}^4}\right]^{-2}\,.
\end{align}

For an initial state with spin-1,
\begin{align}
    \sigma_0 v \vert_{W_L W_L} =   & \frac{\pi \alpha_2^2}{12 c_L^4 m_{\chi^0}^2} \left[(2c_L^2-1)^2 + 2 c_L s_L^2 c_R \sqrt{(2c_L^2-1)}  \left(1 - \frac{M_{Z_R}^2}{4m_{\chi^0}^2}\right)^{-1} + \frac{c_L^2 s_L^4 c_R^2}{(2c_L^2-1) } \left(1 - \frac{M_{Z_R}^2}{4m_{\chi^0}^2}\right)^{-2}
    \right]\,,     \nonumber                                                                                                                                                                                                                                                             \\
    \sigma_0 v \vert_{Z h_L} =     & \frac{\pi \alpha_2^2}{12 c_L^4 m_{\chi^0}^2} \left[1 - \frac{4 c_R c_L s_L^2}{\sqrt{2c_L^2 - 1}}  \left(1 - \frac{M_{Z_R}^2}{4m_{\chi^0}^2}\right)^{-1} + \frac{4 c_R^2 c_L^2 s_L^4}{2c_L^2 - 1} \left(1 - \frac{M_{Z_R}^2}{4m_{\chi^0}^2}\right)^{-2}\right] \,, \\
    \sigma_0 v \vert_{f \ol{f}} = & \frac{2 \pi \alpha_2^2}{3 c_L^4 m_{\chi^0}^2} \left[\left(I_{3f}^2 -2 s_L^2I_{3 f} Q_f + 2  s_L^4Q_f^2\right) \right. \nonumber                                                                                                                   \\
                                   & \left. + 2 (I_{3 f}^2 s_R^2 -I_{3 f} Q_f \left(s_R^2 \left(s_L^2+1\right) + s_L^2\right)+ 2 s_R^2 s_L^2Q_f^2 )  \left(1 - \frac{M_{Z_R}^2}{4m_{\chi^0}^2}\right)^{-1} \right. \nonumber                                                            \\
                                   & \left. 2(I_{3 f}^2 \left(s_R^4+1\right) - 2 I_{3 f} Q_f \left(s_R^2+1\right) s_R^2 + 2 s_R^4 Q_f^2) \left(1 - \frac{M_{Z_R}^2}{4m_{\chi^0}^2}\right)^{-2}
    \right]\,. \nonumber
\end{align}

\subsubsection{$D\ol{E}$ }

For an initial state with spin-0,
\begin{align}
    \sigma_0 v \vert_{W_L Z_L} =  & \frac{2 \pi \alpha_2^2 (c_L^2 - 2 s_L^2)^2}{c_L^2 m_{\chi^0}^2}\,,~
    \sigma_0 v \vert_{W_L \gamma} = \frac{18 \pi \alpha_2^2 s_L^2}{m_{\chi^0}^2}\,,                                        ~
    \sigma_0 v \vert_{W_L Z_R} =  \frac{8 \pi \alpha_2^2 c_R^2}{m_{\chi^0}^2}  \left(1 - \frac{M_{Z_R}^2}{4m_{\chi^0}^2}\right)\,.
\end{align}

For an initial state with spin-1,
\begin{align}
    \sigma_0 v \vert_{f \ol{f}'} = \frac{\pi \alpha_2^2}{3m_{\chi^0}^2}\,,~
    \sigma_0 v \vert_{W_L h_L} =         \frac{\pi \alpha_2^2}{12 m_{\chi^0}^2}\,,~
    \sigma_0 v \vert_{W_L Z_L} =       \frac{\pi \alpha_2^2}{12 m_{\chi^0}^2}\,.
\end{align}

\subsubsection{$E \ol{N}$}

For an initial state with spin-0,
\begin{align}
    \sigma_0 v \vert_{W_L Z_L} =  & \frac{2 \pi \alpha_2^2 (s_L^2 + 1)^2}{c_L^2 m_{\chi^0}^2}\,,~
    \sigma_0 v \vert_{W_L \gamma} = \frac{2 \pi \alpha_2^2 s_L^2}{m_{\chi^0}^2}\,,                                        ~
    \sigma_0 v \vert_{W_L Z_R} =  \frac{8 \pi \alpha_2^2 c_R^2}{m_{\chi^0}^2}  \left(1 - \frac{M_{Z_R}^2}{4m_{\chi^0}^2}\right)\,.
\end{align}

For an initial state with spin-1,
\begin{align}
    \sigma_0 v \vert_{f \ol{f}'} = \frac{\pi \alpha_2^2}{3m_{\chi^0}^2}\,,~
    \sigma_0 v \vert_{W_L h_L} =        \frac{\pi \alpha_2^2}{12 m_{\chi^0}^2}\,,~
    \sigma_0 v \vert_{W_L Z_L} =       \frac{\pi \alpha_2^2}{12 m_{\chi^0}^2}\,.
\end{align}

\subsection{$\chi^0, \chi^\pm$ and $D,E$, $N$ Mixing}

\subsubsection{$\chi^0 D$}

For an initial state with spin-0,
\begin{align}
    \sigma_0 v \vert_{W_L W_R}= \frac{8 \pi \alpha_2^2}{m_{\chi^0}^2} \left(1 - \frac{M_{W_R}^2}{4m_{\chi^0}^2}\right)^3 \left(1 - \frac{M_{W_R}^2}{4m_{\chi^0}^2} + \frac{M_{W_R}^4}{32 m_{\chi^0}^4}\right)^{-2}\,.
\end{align}

\subsubsection{$\chi^- E$}

For an initial state with spin-0,
\begin{align}
    \sigma_0 v \vert_{W_L W_R}=\frac{8 \pi \alpha_2^2}{m_{\chi^0}^2} \left(1 - \frac{M_{W_R}^2}{4m_{\chi^0}^2}\right)^3 \left(1 - \frac{M_{W_R}^2}{4m_{\chi^0}^2} + \frac{M_{W_R}^4}{32 m_{\chi^0}^4}\right)^{-2}\,.
\end{align}

\subsubsection{$\chi^+ D$}

For an initial state with spin-0,
\begin{align}
    \sigma_0 v \vert_{W_R\gamma}=\frac{9\pi\alpha_2^2}{8m_{\chi^0}^2}s_L^2\left[1-\frac{M_{W_R}^2}{4m_{\chi^0}^2}\right]\,,~
    \sigma_0 v \vert_{Z_L W_R}= \frac{2 \pi \alpha_2^2 (1-3c_L^2)^2}{c_L^2 m_{\chi^0}^2} \left[1 - \frac{M_{W_R}^2}{4m_{\chi^0}^2}\right]\,.
\end{align}

For an initial state with spin-1,
\begin{align}
    \sigma_0 v \vert_{Z_L W_R}= \frac{\pi \alpha_2^2 s_L^4}{12 c_L^2 m_{\chi^0}^2} \frac{M_{Z_L}^2}{M_{W_R}^2} \left[1 + \frac{5 M_{W_R}^2}{2m_{\chi^0}^2}  + \frac{M_{W_R}^4}{16 m_{\chi^0}^4}\right] \left[1 - \frac{M_{W_R}^2}{4m_{\chi^0}^2}\right]^{-1}\,,~
    \sigma_0 v \vert_{f\ol{f}'}=\frac{\pi\alpha_2^2}{12 m_{\chi^0}^2}\left[1-\frac{M_{W_R}^2}{4m_{\chi^0}^2}\right]^{-2}\,.
\end{align}

\subsubsection{$\chi^0 E$}
For an initial state with spin-0,
\begin{align}
    \sigma_0 v \vert_{W_R \gamma}= & \frac{2 \pi \alpha_2^2 s_L^2}{m_{\chi^0}^2} \left[1 - \frac{M_{W_R}^2}{4m_{\chi^0}^2}\right]\,,    ~
    \sigma_0 v \vert_{W_R Z_L} =     \frac{2 \pi \alpha_2^2 s_L^4}{c_L^2 m_{\chi^0}^2} \left[1 - \frac{M_{W_R}^2}{4m_{\chi^0}^2}\right]\,.
\end{align}

For an initial state with spin-1,
\begin{align}
    \sigma_0 v \vert_{Z_L W_R}=       & \frac{\pi \alpha_2^2 s_L^4}{12 c_L^2 m_{\chi^0}^2} \frac{M_{Z_L}^2}{M_{W_R}^2} \left[1 + \frac{5 M_{W_R}^2}{2m_{\chi^0}^2}  + \frac{M_{W_R}^4}{16 m_{\chi^0}^4}\right] \left[1 - \frac{M_{W_R}^2}{4m_{\chi^0}^2}\right]^{-1}\,, \\
    \sigma_0 v \vert_{f \ol{f}'} = & \frac{\pi\alpha_2^2}{12 m_{\chi^0}^2}\left[1-\frac{M_{W_R}^2}{4m_{\chi^0}^2}\right]^{-2}\,.\nonumber
\end{align}

\subsubsection{$N \chi^-$}
For an initial state with spin-0,
\begin{align}
    \sigma_0 v \vert_{W_R \gamma}= & \frac{2 \pi \alpha_2^2 s_L^2}{m_{\chi^0}^2} \left[1 - \frac{M_{W_R}^2}{4m_{\chi^0}^2}\right]\,,            ~
    \sigma_0 v \vert_{Z_L W_R}=       \frac{2 \pi \alpha_2^2 (1+c_L^2)^2}{c_L^2 m_{\chi^0}^2} \left[1 - \frac{M_{W_R}^2}{4m_{\chi^0}^2}\right]\,.
\end{align}

For an initial state with spin-1,
\begin{align}
    \sigma_0 v \vert_{Z_L W_R}=       & \frac{\pi \alpha_2^2 s_L^4}{12 c_L^2 m_{\chi^0}^2} \frac{M_{Z_L}^2}{M_{W_R}^2} \left[1 + \frac{5 M_{W_R}^2}{2m_{\chi^0}^2}  + \frac{M_{W_R}^4}{16 m_{\chi^0}^4}\right] \left[1 - \frac{M_{W_R}^2}{4m_{\chi^0}^2}\right]^{-1}\,,~
    \sigma_0 v \vert_{f \ol{f}'} =  \frac{\pi\alpha_2^2}{12 m_{\chi^0}^2}\left[1-\frac{M_{W_R}^2}{4m_{\chi^0}^2}\right]^{-2}\,.
\end{align}

\subsubsection{$\chi^- \ol{E}$}

For an initial state with spin-0,
\begin{align}
    \sigma_0 v \vert_{W_L W_R}& =\frac{8\pi\alpha_2^2}{m_{\chi^0}^2}\left[1-\frac{M_{W_R}^2}{4m_{\chi^0}^2}\right].
\end{align}

\subsubsection{$\chi^0 \ol{N}$}

For an initial state with spin-0,
\begin{align}
    \sigma_0 v \vert_{W_L W_R}& =\frac{8\pi\alpha_2^2}{m_{\chi^0}^2}\left[1-\frac{M_{W_R}^2}{4m_{\chi^0}^2}\right]
\end{align}

\section{Annihilation cross sections for $(\mathbf{3},\mathbf{3},1)$}
\label{sec:cross section 331}

In this appendix, we show the annihilation cross sections for $(\mathbf{3},\mathbf{3},1)$ DM. Only $s$-wave annihilation is included. For conciseness, we show the result with the leading-order dependence on $v_L$. We further ignore the resonance width and take $s=4M_{\psi^0}^2$ in this section. In the numerical computation, however, the full result is used.

\subsection{$R_1$, $R_2$ and $R_3$}
\subsubsection{$R_1\ol{R_1}$}
For an initial state with spin-0,
\begin{align}
    &\sigma_0v|_{W_L W_L}=\frac{2\pi\alpha_2^2}{M_{\psi^0}^2},~~\sigma_0v|_{W_R W_R}=\frac{2\pi\alpha_2^2}{M_{\psi^0}^2}\left(1-\frac{M_{W_R}^2}{M_{\psi^0}^2}\right)^{3/2}\left(1-\frac{M_{W_R}^2}{2 M_{\psi^0}^2}\right)^{-2},\\
    &\sigma_0v|_{Z_L Z_L}=\frac{4\pi\alpha_2^2}{ M_{\psi^0}^2}\frac{(s_L^2+1)^4}{c_L^2},~~\sigma_0v|_{Z_L Z_R}=\frac{8\pi\alpha_2^2}{M_{\psi^0}^2}\frac{(3s_L^2-1)^2(s_L^2+1)^2}{c_L^4(1-2s_L^2)}\left(1-\frac{M_{Z_R}}{4M_{\psi^0}^2}\right), \nonumber \\
    &\sigma_0v|_{Z_L \gamma}=\frac{8\pi\alpha_2^2}{M_{\psi^0}^2}\frac{s_L^2(s_L^2+1)^2}{c_L^2},~~\sigma_0v|_{Z_R Z_R}=\frac{4\pi\alpha_2^2}{M_{\psi^0}^2}\frac{(3 s_L^2-1)^4}{c_L^4(1-2 s_L^2)^2}\left(1-\frac{M_{Z_R}^2}{M_{\psi^0}^2}\right)^{3/2}\left(1-\frac{M_{Z_R}^2}{2M_{\psi^0}^2}\right)^{-2},\nonumber \\
    &\sigma_0v|_{Z_R \gamma}=\frac{8\pi\alpha_2^2}{M_{\psi^0}^2}\frac{s_L^2(3 s_L^2-1)}{c_L^2(1-2 s_L^2)}\left(1-\frac{M_{Z_R}^2}{4M_{\psi^0}^2}\right),~~\sigma_0v|_{\gamma\gamma}=\frac{4\pi\alpha_2^2 s_L^4}{M_{\psi^0}^2}.\nonumber
\end{align}

For an initial state with spin-1,
\begin{align}
    \sigma_0v|_{f\ol{f}}&=\frac{\pi\alpha_2^2}{3 M_{\psi^0}^2}\frac{1}{(1-2s_L^2)^2}\left(1-\frac{M_{Z_R}^2}{4M_{\psi^0}^2}\right)^{-2}\\
    &\times\left[2s_L^4(2I_3(I_3-3)+2Q_f(Q_f-1)+5)+2s_L^2(2I_3-3)+1\right.\nonumber\\
    &-\left.\frac{M_{W_R}^2}{4 M_{\psi^0}^2}(2s_L^4(I_3(8I_3-15)+Q_f(8Q_f-5)+5)+s_L^2(10I_3-2Q_f-5)+1)\right.\nonumber\\
    &+\left.\frac{M_{W_R}^4}{M_{\psi^0}^4}(s_L^4(8I_3(4I_3-3)+8Q_f(4Q_f-1)+5)+2s_L^2(4I_3-4Q_f-1)+1)\right],\nonumber\\
    \sigma_0v|_{W_LW_L}&=\frac{\pi\alpha_2^2}{12 M_{\psi^0}^2}\frac{(1-3 s_L^2)^2}{(1-2 s_L^2)^2}\left(1-\frac{M_{W_R}^2}{4M_{\psi^0}^2}\right)^{2}\left(1-\frac{M_{Z_R}^2}{4M_{\psi^0}^2}\right)^{-2},\nonumber \\
    \sigma_0v|_{W_RW_R}&=\frac{\pi\alpha_2^2}{12 M_{\psi^0}^2}\frac{c_L^4}{(1-2 s_L^2)^2}\left(1-\frac{M_{W_R}^2}{M_{\psi^0}^2}\right)^{3/2}\left(1-\frac{M_{W_R}^2}{2M_{\psi^0}^2}\right)^{-2}\nonumber\\
    &\times\left(1+5\frac{M_{W_R}^2}{M_{\psi^0}^2}+\frac{3}{4}\frac{M_{W_R}^4}{M_{\psi^0}^4}\right)\left(1-\frac{s_L^2}{c_L^2}\frac{M_{W_R}^2}{M_{\psi^0}^2}\right)^{2}\left(1-\frac{M_{Z_R}^2}{4M_{\psi^0}^2}\right)^{-2}.\nonumber
\end{align}

\subsubsection{$R_1\ol{R_2}$}
For an initial state with spin-0,
\begin{align}
    &\sigma_0v|_{W_LZ_L}=\frac{2\pi\alpha_2^2}{M_{\psi^0}^2}\frac{(2 s_L^2+1)^2}{c_L^2},~~\sigma_0v|_{W_LZ_R}=\frac{8\pi\alpha_2^2}{M_{\psi^0}^2}\frac{(1-3 s_L^2)^2}{c_L^2(1-2s_L^2)}\left(
    1-\frac{M_{Z_R}^2}{4 M_{\psi^0}^2}\right),~~\sigma_0v|_{W_L\gamma}=\frac{18\pi\alpha_2^2s_L^2 }{M_{\psi^0}^2}.
\end{align}
For an initial state with spin-1,
\begin{align}
    &\sigma_0v|_{f\ol{f'}}=\frac{\pi \alpha_2^2}{3 M_{\psi^0}^2},~\sigma_0v|_{W_Lh}=\frac{\pi \alpha_2^2}{12 M_{\psi^0}^2},~\sigma_0v|_{W_LZ_L}=\frac{\pi \alpha_2^2}{12 M_{\psi^0}^2}.
\end{align}

\subsubsection{$R_1\ol{R_3}$}
For an initial state with spin-0,
\begin{align}
    \sigma_0v|_{W_LW_L}=\frac{4\pi \alpha_2^2}{ M_{\psi^0}^2}.
\end{align}

\subsubsection{$R_2\ol{R_2}$}
For an initial state with spin-0,
\begin{align}
    \sigma_0v|_{W_LW_L}&=\frac{8\pi\alpha_2^2}{M_{\psi^0}^2},~~
    \sigma_0v|_{W_RW_R}=\frac{2\pi\alpha_2^2}{M_{\psi^0}^2}\left(1-\frac{M_{W_R}^2}{M_{\psi^0}^2}\right)^{3/2}\left(1-\frac{M_{W_R}^2}{2M_{\psi^0}^2}\right)^{-2},~~
    \sigma_0v|_{Z_LZ_L}=\frac{64\pi\alpha_2^2s_L^8}{c_L^4M_{\psi^0}^2},\nonumber \\
    \sigma_0v|_{Z_LZ_R}&=\frac{32\pi\alpha_2^2}{M_{\psi^0}^2}\frac{s_L^4(1-3s_L^2)^2}{c_L^4(1-2s_L^2)}\left(1-\frac{M_{Z_R}^2}{4M_{\psi^0}^2}\right),~~
    \sigma_0v|_{Z_L\gamma}=\frac{128\pi\alpha_2^2s_L^6}{c_L^2M_{\psi^0}^2},\\
    \sigma_0v|_{Z_RZ_R}&=\frac{4\pi\alpha_2^2}{M_{\psi^0}^2}\frac{(1-3 s_L^2)^4}{c_L^4(1-2 s_L^2)^2}\left(1-\frac{M_{Z_R}^2}{M_{\psi^0}^2}\right)^{3/2}\left(1-\frac{M_{Z_R}^2}{2M_{\psi^0}^2}\right)^{-2},\nonumber\\
    \sigma_0v|_{Z_R\gamma}&=\frac{32\pi\alpha_2^2}{M_{\psi^0}^2}\frac{s_L^2(1-3s_L^2)^2}{c_L^2(1-2s_L^2)}\left(1-\frac{M_{Z_R}^2}{4M_{\psi^0}^2}\right),~~
    \sigma_0v|_{\gamma\gamma}=\frac{64\pi\alpha_2^2s_L^4}{M_{\psi^0}^2}. \nonumber
\end{align}

For an initial state with spin-1,
\begin{equation}
    \begin{split}
    \sigma_0v|_{f\ol{f}}&=\frac{\pi\alpha_2^2}{6 M_{\psi^0}^2}\frac{1}{(1-2s_L^2)^2}\left(1-\frac{M_{Z_R}^2}{4M_{\psi^0}^2}\right)^{-2}\\
    &\times \left[8s_L^4(I_3^2-2I_3+Q_f(Q_f-2)+2)+4s_L^2(I_3+Q_f-2)+1\right.\\
    &\left.-\frac{M_{W_R}^2}{M_{\psi^0}^2}s_L^2(2s_L^2(4I_3^2-5I_3+Q_f(4Q_f-5)+2)+2I_3+2Q_f-1)\right.\\
    &\left.+\frac{M_{W_R}^4}{4M_{\psi^0}^4}s_L^4(8I_3^2-4I_3+4Q_f(2Q_f-1)+1)\right],\\
    \sigma_0v|_{W_LW_L}&=\frac{\pi\alpha_2^2}{12M_{\psi^0}^2}\frac{s_L^4}{(1-2s_L^2)^2}\left(1-\frac{M_{W_R}^2}{2M_{\psi^0}^2}\right)^{2}\left(1-\frac{M_{Z_R}^2}{4M_{\psi^0}^2}\right)^{-2},\\
    \sigma_0v|_{W_RW_R}&=\frac{\pi\alpha_2^2}{12 M_{\psi^0}^2}\frac{c_L^4}{(1-2 s_L^2)^2}\left(1-\frac{M_{W_R}^2}{M_{\psi^0}^2}\right)^{3/2}\left(1-\frac{M_{W_R}^2}{2M_{\psi^0}^2}\right)^{-2}\\
    &\times\left(1+5\frac{M_{W_R}^2}{M_{\psi^0}^2}+\frac{3}{4}\frac{M_{W_R}^4}{M_{\psi^0}^4}\right)\left(1-\frac{s_L^2}{c_L^2}\frac{M_{W_R}^2}{M_{\psi^0}^2}\right)^{2}\left(1-\frac{M_{Z_R}^2}{4M_{\psi^0}^2}\right)^{-2},\\
    \sigma_0v|_{Z_Lh}&=\frac{\pi\alpha_2^2}{12M_{\psi^0}^2}\frac{s_L^4}{(1-2s_L^2)^2}\left(1-\frac{M_{W_R}^2}{2M_{\psi^0}^2}\right)^{2}\left(1-\frac{M_{Z_R}^2}{4M_{\psi^0}^2}\right)^{-2}.
    \end{split}
\end{equation}

\subsubsection{$R_2\ol{R_3}$}
For an initial state with spin-0,
\begin{equation}
    \sigma_0v|_{W_LZ_L}=\frac{2\pi\alpha_2^2}{M_{\psi^0}^2}\frac{(5s_L^2-1)^2}{c_L^2},~
    \sigma_0v|_{W_LZ_R}=\frac{8\pi\alpha_2^2}{M_{\psi^0}^2}\frac{(3s_L^2-1)^2}{c_L^2(1-2s_L^2)}\left(1-\frac{M_{Z_R}^2}{4M_{\psi^0}^2}\right),~
    \sigma_0v|_{W_L\gamma}=\frac{50\pi\alpha_2^2s_L^2}{M_{\psi^0}^2}.
\end{equation}
For an initial state with spin-1,
\begin{align}
    \sigma_0v|_{f\ol{f'}}&=\frac{\pi\alpha_2^2}{3M_{\psi^0}^2},~\sigma_0v|_{W_Lh}=\frac{\pi\alpha_2^2}{12M_{\psi^0}^2},~
    \sigma_0v|_{W_LZ_L}=\frac{\pi\alpha_2^2}{12M_{\psi^0}^2}.
\end{align}

\subsubsection{$R_3\ol{R_3}$}
For an initial state with spin-0,
\begin{align}
    \sigma_0v|_{W_LW_L}&=\frac{2\pi\alpha_2^2}{M_{\psi^0}^2},~
    \sigma_0v|_{W_RW_R}=\frac{2\pi\alpha_2^2}{M_{\psi^0}^2}\left(1-\frac{M_{W_R}^2}{M_{\psi^0}^2}\right)^{3/2}\left(1-\frac{M_{W_R}^2}{2M_{\psi^0}^2}\right)^{-2}, \\
    \sigma_0v|_{Z_LZ_L}&=\frac{4\pi\alpha_2^2}{M_{\psi^0}^2}\frac{(3 s_L^2-1)^4}{c_L^4},~
    \sigma_0v|_{Z_LZ_R}=\frac{8\pi\alpha_2^2}{M_{\psi^0}^2}\frac{(1-3s_L^2)^4}{c_L^4(1-2s_L^2)}\left(1-\frac{M_{Z_R}^2}{4M_{\psi^0}^2}\right),\nonumber \\
    \sigma_0v|_{Z_L\gamma}&=\frac{72\pi\alpha_2^2}{M_{\psi^0}^2}\frac{s_L^2(3 s_L^2-1)^2}{c_L^2},~
    \sigma_0v|_{Z_RZ_R}=\frac{4\pi\alpha_2^2}{M_{\psi^0}^2}\frac{(1-3 s_L^2)^4}{c_L^4(1-2 s_L^2)^2}\left(1-\frac{M_{Z_R}^2}{M_{\psi^0}^2}\right)^{3/2}\left(1-\frac{M_{Z_R}^2}{2M_{\psi^0}^2}\right)^{-2},\nonumber\\
    \sigma_0v|_{Z_R\gamma}&=\frac{72\pi\alpha_2^2}{M_{\psi^0}^2}\frac{s_L^2(1-3s_L^2)^2}{c_L^2(1-2s_L^2)}\left(1-\frac{M_{Z_R}^2}{4M_{\psi^0}^2}\right),~
    \sigma_0v|_{\gamma\gamma}=\frac{324\pi\alpha_2^2s_L^4}{M_{\psi^0}^2}.\nonumber
\end{align}

For an initial state with spin-1,
\begin{align}
    \sigma_0v|_{f\ol{f}}&=\frac{\pi\alpha_2^2}{3 M_{\psi^0}^2}\frac{1}{(1-2s_L^2)^2}\left(1-\frac{M_{Z_R}^2}{4M_{\psi^0}^2}\right)^{-2}\\
    &\times\left[2s_L^4(2I_3(I_3-1)+2Q_f(Q_f-3)+5)+2s_L^2(2Q_f-3)+1\right.\nonumber\\
    &\left.-\frac{M_{W_R}^2}{4M_{\psi^0}^2} (2s_L^4(I_3(8I_3-5)+Q_f(8Q_f-15)+5)+s_L^2(10Q_f-2I_3-5)+1)\right.\nonumber\\
    &\left.+\frac{M_{W_R}^4}{32M_{\psi^0}^4}(s_L^4(8I_3(4I_3-1)+8Q_f(4Q_f-3)+5)+2s_L^2(4Q_f-4I_3-1)+1)\right],\nonumber\\
    \sigma_0v|_{W_LW_L}&=\frac{\pi\alpha_2^2}{12M_{\psi^0}^2}\frac{c_L^4}{(1-2s_L^2)^2}\left(1-\frac{1+s_L^2}{c_L^2}\frac{M_{W_R}^2}{4M_{\psi^0}^2}\right)^{2}\left(1-\frac{M_{Z_R}^2}{4M_{\psi^0}^2}\right)^{-2},\nonumber \\
    \sigma_0v|_{W_RW_R}&=\frac{\pi\alpha_2^2}{12 M_{\psi^0}^2}\frac{c_L^4}{(1-2 s_L^2)^2}\left(1-\frac{M_{W_R}^2}{M_{\psi^0}^2}\right)^{3/2}\left(1-\frac{M_{W_R}^2}{2M_{\psi^0}^2}\right)^{-2}\nonumber\\
    &\times\left(1+5\frac{M_{W_R}^2}{M_{\psi^0}^2}+\frac{3}{4}\frac{M_{W_R}^4}{M_{\psi^0}^4}\right)\left(1-\frac{s_L^2}{c_L^2}\frac{M_{W_R}^2}{M_{\psi^0}^2}\right)^{2}\left(1-\frac{M_{Z_R}^2}{4M_{\psi^0}^2}\right)^{-2},\nonumber\\
    \sigma_0v|_{Z_Lh}&=\frac{\pi\alpha_2^2}{12M_{\psi^0}^2}\frac{(1-3s_L)^2}{(1-2s_L^2)^2}\left(1-\frac{M_{W_R}^2}{4M_{\psi^0}^2}\right)^{2}\left(1-\frac{M_{Z_R}^2}{4M_{\psi^0}^2}\right)^{-2}.\nonumber
\end{align}
\subsection{$S_0$, $S_1$ and $S_2$}

\subsubsection{$S_0\ol{S_0}$}
For an initial state with spin-0,
\begin{align}
    \sigma_0v|_{W_LW_L}&=\frac{2\pi\alpha_2^2}{M_{\psi^0}^2},~
    \sigma_0v|_{W_RW_R}=\frac{8\pi\alpha_2^2}{M_{\psi^0}^2}\left(1-\frac{M_{W_R}^2}{M_{\psi^0}^2}\right)^{3/2}\left(1-\frac{M_{W_R}^2}{2M_{\psi^0}^2}\right)^{-2},\\
    \sigma_0v|_{Z_LZ_L}&=\frac{4\pi\alpha_2^2}{c_L^4M_{\psi^0}^2},~
    \sigma_0v|_{Z_LZ_R}=\frac{8\pi\alpha_2^2}{M_{\psi^0}^2}\frac{s_L^4}{c_L^4(1-2s_L^2)}\left(1-\frac{M_{Z_R}^2}{4M_{\psi^0}^2}\right),\nonumber\\
    \sigma_0v|_{Z_RZ_R}&=\frac{4\pi\alpha_2^2}{M_{\psi^0}^2}\frac{s_L^8}{c_L^4(1-2 s_L^2)^2}\left(1-\frac{M_{Z_R}^2}{M_{\psi^0}^2}\right)^{3/2}\left(1-\frac{M_{Z_R}^2}{2M_{\psi^0}^2}\right)^{-2}.\nonumber
\end{align}

For an initial state with spin-1,
\begin{align}
    \sigma_0v|_{f\ol{f}}&=\frac{\pi\alpha_2^2}{6M_{\psi^0}^2}\frac{1}{(1-2s_L^2)^2}\left(1-\frac{M_{Z_R}^2}{4M_{\psi^0}^2}\right)^{-2}\\
    &\times\left[8s_L^4((I_3-1)^2+Q_f^2)-4s_L^2(Q_f-I_3+1)+1\right.\nonumber \\
    &-\frac{M_{W_R}^2}{2M_{\psi^0}^2}(s_L^4(8I_3^2-12I_3+8Q_f^2+4)+s_L^2(4I_3-4Q_f-3)+1)\nonumber\\
    &\left.+\frac{M_{W_R}^4}{16M_{\psi^0}^4}(s_L^2(2s_L^2(1-2I_3)^2+4I_3+8s_L^2Q_f^2-4Q_f-2)+1)\right],\nonumber\\
    \sigma_0v|_{W_LW_L}&=\frac{\pi\alpha_2^2}{12M_{\psi^0}^2}\frac{(1-3s_L^2)^2}{(1-2s_L^2)^2}\left(1-\frac{1-3s_L^2}{1-2s_L^2}\frac{M_{W_R}^2}{4M_{\psi^0}^2}\right)^{2}\left(1-\frac{M_{Z_R}^2}{4M_{\psi^0}^2}\right)^{-2},\nonumber\\
    \sigma_0v|_{W_RW_R}&=\frac{\pi\alpha_2^2}{12 M_{\psi^0}^2}\frac{s_L^4}{(1-2 s_L^2)^2}\left(1-\frac{M_{W_R}^2}{M_{\psi^0}^2}\right)^{3/2}\left(1+5\frac{M_{W_R}^2}{M_{\psi^0}^2}+\frac{3}{4}\frac{M_{W_R}^4}{M_{\psi^0}^4}\right)\left(1-\frac{M_{Z_R}^2}{4M_{\psi^0}^2}\right)^{-2},\nonumber\\
    \sigma_0v|_{Z_Lh}&=\frac{\pi\alpha_2^2}{12 M_{\psi^0}^2}\frac{c_L^4}{(1-2 s_L^2)^2}\left(1-\frac{1}{c_L^2}\frac{M_{W_R}^2}{4M_{\psi^0}^2}\right)^{2}\left(1-\frac{M_{Z_R}^2}{4M_{\psi^0}^2}\right)^{-2}. \nonumber
\end{align}

\subsubsection{$S_0\ol{S_1}$}
For an initial state with spin-0,
\begin{align}
    \sigma_0v|_{W_LZ_L}&=\frac{2\pi \alpha_2^2}{M_{\psi^0}^2}\frac{(s_L^2+1)^2}{c_L^2},~
    \sigma_0v|_{W_LZ_R}=\frac{8\pi\alpha_2^2}{M_{\psi^0}^2}\frac{s_L^4}{c_L^2(1-2 s_L^2)}\left(1-\frac{M_{Z_R}^2}{4M_{\psi^0}^2}\right),~
    \sigma_0v|_{W_L\gamma}=\frac{2\pi \alpha_2^2s_L^2}{M_{\psi^0}^2}.
\end{align}

For an initial state with spin-1,
\begin{align}
    \sigma_0v|_{f\ol{f'}}&=\frac{\pi \alpha_2^2}{3M_{\psi^0}^2},~\sigma_0v|_{W_Lh}=\frac{\pi \alpha_2^2}{12M_{\psi^0}^2},~\sigma_0v|_{W_LZ_L}=\frac{\pi \alpha_2^2}{12M_{\psi^0}^2}.
\end{align}

\subsubsection{$S_0\ol{S_2}$}
For an initial state with spin-0,
\begin{align}
    \sigma_0v|_{W_LW_L}=\frac{4\pi \alpha_2^2}{ M_{\psi^0}^2}.
\end{align}

\subsubsection{$S_1\ol{S_1}$}
For an initial state with spin-0,
\begin{align}
    \sigma_0v|_{W_LW_L}&=\frac{8\pi \alpha_2^2}{M_{\psi^0}^2},~
    \sigma_0v|_{W_RW_R}=\frac{8\pi\alpha_2^2}{M_{\psi^0}^2}\left(1-\frac{M_{W_R}^2}{M_{\psi^0}^2}\right)^{3/2}\left(1-\frac{M_{W_R}^2}{2M_{\psi^0}^2}\right)^{-2},~
    \sigma_0v|_{Z_LZ_L}=\frac{4\pi \alpha_2^2}{M_{\psi^0}^2}\frac{s_L^8}{c_L^4},\nonumber \\
    \sigma_0v|_{Z_LZ_R}&=\frac{8\pi\alpha_2^2}{M_{\psi^0}^2}\frac{s_L^8}{c_L^4(1-2s_L^2)}\left(1-\frac{M_{W_R}^2}{4M_{\psi^0}^2}\right),~
    \sigma_0v|_{Z_L\gamma}=\frac{8\pi \alpha_2^2}{M_{\psi^0}^2}\frac{s_L^6}{c_L^2},\\
    \sigma_0v|_{Z_RZ_R}&=\frac{4\pi\alpha_2^2}{M_{\psi^0}^2}\frac{s_L^8}{c_L^4(1-2s_L^2)^2}\left(1-\frac{M_{Z_R}^2}{M_{\psi^0}^2}\right)^{3/2}\left(1-\frac{M_{Z_R}^2}{2M_{\psi^0}^2}\right)^{-2},\nonumber\\
    \sigma_0v|_{Z_R\gamma}&=\frac{8\pi\alpha_2^2}{M_{\psi^0}^2}\frac{s_L^6}{c_L^2(1-2s_L^2)}\left(1-\frac{M_{W_R}^2}{4M_{\psi^0}^2}\right),~
    \sigma_0v|_{\gamma\gamma}=\frac{4\pi \alpha_2^2s_L^4}{M_{\psi^0}^2}.\nonumber
\end{align}

For an initial state with spin-1,
\begin{align}
    \sigma_0v|_{f\ol{f}}&=\frac{2\pi\alpha_2^2}{3M_{\psi^0}^2}\frac{s_L^4}{(1-2s_L^2)^2}\left(1-\frac{M_{Z_R}^2}{4M_{\psi^0}^2}\right)^{-2}\\
    &\times\left[2I_3(I_3-1)+2Q_f(Q_f-1)+1\right.\nonumber\\
    &-\frac{M_{W_R}^2}{4M_{\psi^0}^2}(I_3(4I_3-3)+Q_f(4Q_f-3)+1)\nonumber\\
    &\left.+\frac{M_{W_R}^4}{64M_{\psi^0}^4}(4I_3(2I_3-1)+4Q_f(2Q_f-1)+1)\right],\nonumber\\
    \sigma_0v|_{W_LW_L}&=\frac{\pi\alpha_2^2}{12M_{\psi^0}^2}\frac{s_L^4}{(1-2s_L^2)^2}\left(1-\frac{M_{W_R}^2}{4M_{\psi^0}^2}\right)^{2}\left(1-\frac{M_{Z_R}^2}{4M_{\psi^0}^2}\right)^{-2},\nonumber\\
    \sigma_0v|_{W_RW_R}&=\frac{\pi\alpha_2^2}{12 M_{\psi^0}^2}\frac{s_L^4}{(1-2 s_L^2)^2}\left(1-\frac{M_{W_R}^2}{M_{\psi^0}^2}\right)^{3/2}\left(1+5\frac{M_{W_R}^2}{M_{\psi^0}^2}+\frac{3}{4}\frac{M_{W_R}^4}{M_{\psi^0}^4}\right)\left(1-\frac{M_{Z_R}^2}{4M_{\psi^0}^2}\right)^{-2},\nonumber\\
    \sigma_0v|_{Z_Lh}&=\frac{\pi\alpha_2^2}{12M_{\psi^0}^2}\frac{s_L^4}{(1-2s_L^2)^2}\left(1-\frac{M_{W_R}^2}{4M_{\psi^0}^2}\right)^{2}\left(1-\frac{M_{Z_R}^2}{4M_{\psi^0}^2}\right)^{-2}.\nonumber
\end{align}

\subsubsection{$S_1\ol{S_2}$}
For an initial state with spin-0,
\begin{align}
    \sigma_0v|_{W_LZ_L}&=\frac{2\pi \alpha_2^2}{M_{\psi^0}^2}\frac{(1-3s_L^2)^2}{c_L^2},~
    \sigma_0v|_{W_LZ_R}=\frac{8\pi \alpha_2^2}{M_{\psi^0}^2}\frac{s_L^4}{c_L^2(1-2s_L^2)}\left(1-\frac{M_{Z_R}^2}{4M_{\psi^0}^2}\right),~
    \sigma_0v|_{W_L\gamma}=\frac{18\pi \alpha_2^2s_L^2}{M_{\psi^0}^2}. 
\end{align}
For an initial state with spin-1,
\begin{align}
    \sigma_0v|_{f\ol{f'}}&=\frac{\pi \alpha_2^2}{3M_{\psi^0}^2},~\sigma_0v|_{W_Lh}=\frac{\pi \alpha_2^2}{12M_{\psi^0}^2},~\sigma_0v|_{W_LZ_L}=\frac{\pi \alpha_2^2}{12M_{\psi^0}^2}.
\end{align}

\subsubsection{$S_2\ol{S_2}$}
For an initial state with spin-0,
\begin{align}
    \sigma_0v|_{W_LW_L}&=\frac{2\pi \alpha_2^2}{ M_{\psi^0}^2},~
    \sigma_0v|_{W_RW_R}=\frac{8\pi \alpha_2^2}{ M_{\psi^0}^2}\left(1-\frac{M_{W_R}^2}{M_{\psi^0}^2}\right)^{3/2}\left(1-\frac{M_{W_R}^2}{2M_{\psi^0}^2}\right)^{-2},\\
    \sigma_0v|_{Z_LZ_L}&=\frac{4\pi \alpha_2^2}{ M_{\psi^0}^2}\frac{(1-2s_L^2)^4}{c_L^4},~
    \sigma_0v|_{Z_LZ_R}=\frac{8\pi \alpha_2^2}{ M_{\psi^0}^2}\frac{s_L^4(1-2s_L^2)}{c_L^4}\left(1-\frac{M_{Z_R}^2}{4M_{\psi^0}^2}\right),\nonumber\\
    \sigma_0v|_{Z_L\gamma}&=\frac{32\pi \alpha_2^2}{ M_{\psi^0}^2}\frac{s_L^2(1-2s_L^2)^2}{c_L^2},~
    \sigma_0v|_{Z_RZ_R}=\frac{4\pi \alpha_2^2}{ M_{\psi^0}^2}\frac{s_L^8}{c_L^4(1-2s_L^2)^2}\left(1-\frac{M_{Z_R}^2}{M_{\psi^0}^2}\right)^{3/2}\left(1-\frac{M_{Z_R}^2}{2M_{\psi^0}^2}\right)^{-2},\nonumber\\
    \sigma_0v|_{Z_R\gamma}&=\frac{32\pi \alpha_2^2}{ M_{\psi^0}^2}\frac{s_L^6}{c_L^2(1-2s_L^2)}\left(1-\frac{M_{Z_R}^2}{4M_{\psi^0}^2}\right),~
    \sigma_0v|_{\gamma\gamma}=\frac{64\pi \alpha_2^2s_L^4}{ M_{\psi^0}^2}.\nonumber
\end{align}

For an initial state with spin-1,
\begin{align}
    \sigma_0v|_{f\ol{f}}&=\frac{\pi\alpha_2^2}{6M_{\psi^0}^2}\frac{1}{(1-2s_L^2)^2}\left(1-\frac{M_{Z_R}^2}{4M_{\psi^0}^2}\right)^{-2}\\
    &\times\left[8s_L^4(I_3^2+(Q_f-1)^2)+4s_L^2(Q_f-I_3-1)+1\right.\nonumber\\
    &-\frac{M_{W_R}^2}{2M_{\psi^0}^2}(s_L^4(8I_3^2+4Q_f(2Q_f-3)+4)+s_L^2(4Q_f-4I_3-3)+1)\nonumber\\
    &\left.+\frac{M_{W_R}^4}{16M_{\psi^0}^4}(s_L^4(8I_3^2+8Q_f(Q_f-1)+2)+s_L^2(4Q_f-4I_3-2)+1)\right],\nonumber\\
    \sigma_0v|_{W_LW_L}&=\frac{\pi\alpha_2^2}{12M_{\psi^0}^2}\frac{c_L^4}{(1-2s_L^2)^2}\left(1-\frac{1}{c_L^2}\frac{M_{W_R}^2}{8M_{\psi^0}^2}\right)^{2}\left(1-\frac{M_{Z_R}^2}{4M_{\psi^0}^2}\right)^{-2},\nonumber\\
    \sigma_0v|_{W_RW_R}&=\frac{\pi\alpha_2^2}{12M_{\psi^0}^2}\frac{s_L^4}{(1-2s_L^2)^2}\left(1-\frac{M_{W_R}^2}{M_{\psi^0}^2}\right)^{1/2}\left(1-\frac{M_{Z_R}^2}{4M_{\psi^0}^2}\right)^{-2}\nonumber\\
    &\times\left(1+4\frac{M_{W_R}^2}{M_{\psi^0}^2}-\frac{17}{4}\frac{M_{W_R}^4}{M_{\psi^0}^4}-\frac{3}{4}\frac{M_{W_R}^6}{M_{\psi^0}^6}\right),\nonumber\\
    \sigma_0v|_{Z_Lh}&=\frac{\pi\alpha_2^2}{12M_{\psi^0}^2}\frac{(1-3 s_L^2)^2}{(1-2s_L^2)^2}\left(1-\frac{1-2s_L^2}{1-3s_L^2}\frac{M_{W_R}^2}{4M_{\psi^0}^2}\right)^{2}\left(1-\frac{M_{Z_R}^2}{4M_{\psi^0}^2}\right)^{-2}.\nonumber
\end{align}

\subsection{$\psi^+$, $\psi^-$ and $\psi^0$}
\subsubsection{$\psi^+\ol{\psi^+}$}
For an initial state with spin-0,
\begin{align}
    \sigma_0v|_{W_LW_L}&=\frac{2\pi \alpha_2^2}{ M_{\psi^0}^2},~
    \sigma_0v|_{W_RW_R}=\frac{2\pi \alpha_2^2}{ M_{\psi^0}^2}\left(1-\frac{M_{W_R}^2}{M_{\psi^0}^2}\right)^{3/2}\left(1-\frac{M_{W_R}^2}{2M_{\psi^0}^2}\right)^{-2},\\
    \sigma_0v|_{Z_LZ_L}&=\frac{4\pi \alpha_2^2c_L^4}{ M_{\psi^0}^2},~
    \sigma_0v|_{Z_LZ_R}=\frac{8\pi \alpha_2^2}{ M_{\psi^0}^2}\frac{c_L^4}{(1-2s_L^2)}\left(1-\frac{M_{Z_R}^2}{4M_{\psi^0}^2}\right),~
    \sigma_0v|_{Z_L\gamma}=\frac{8\pi \alpha_2^2s_L^2c_L^2}{ M_{\psi^0}^2},\nonumber\\
    \sigma_0v|_{Z_RZ_R}&=\frac{4\pi \alpha_2^2}{ M_{\psi^0}^2}\frac{c_L^4}{(1-2s_L^2)^2}\left(1-\frac{M_{Z_R}^2}{M_{\psi^0}^2}\right)^{3/2}\left(1-\frac{M_{Z_R}^2}{2M_{\psi^0}^2}\right)^{-2},\nonumber\\
    \sigma_0v|_{Z_R\gamma}&=\frac{8\pi \alpha_2^2}{ M_{\psi^0}^2}\frac{c_L^2s_L^2}{(1-2s_L^2)}\left(1-\frac{M_{Z_R}^2}{4M_{\psi^0}^2}\right),~
    \sigma_0v|_{\gamma\gamma}=\frac{4\pi \alpha_2^2s_L^4}{ M_{\psi^0}^2}.\nonumber
\end{align}

For an initial state with spin-1,
\begin{align}
    \sigma_0v|_{f\ol{f}}&=\frac{\pi\alpha_2^2}{3 M_{\psi^0}^2}\frac{1}{(1-2s_L^2)^2}\left(1-\frac{M_{Z_R}^2}{4M_{\psi^0}^2}\right)^{-2}\\
    &\times\left[s_L^4(4I_3(I_3+1)+4(Q_f-1)Q_f+2)-2s_L^2 (2I_3+1)+1\right.\nonumber\\
    &-\frac{M_{Z_R}^2}{4M_{\psi^0}^2}(1-2s_L^2)(2s_L^2(Q_f-I_3-1)+1)\nonumber\\
    &\left.+\frac{M_{Z_R}^4}{32 M_{\psi^0}^4}(1-2s_L^2)^2\right]\nonumber\\
    \sigma_0v|_{W_LW_L}&=\frac{\pi\alpha_2^2}{12M_{\psi^0}^2}\frac{c_L^4}{(1-2s_L^2)^2}\left(1-\frac{M_{W_R}^2}{4M_{\psi^0}^2}\right)^{2}\left(1-\frac{M_{Z_R}^2}{4M_{\psi^0}^2}\right)^{-2},\nonumber\\
    \sigma_0v|_{W_RW_R}&=\frac{\pi\alpha_2^2}{12M_{\psi^0}^2}\frac{(1-3s_L^2)^2}{(1-2s_L^2)^2}\left(1-\frac{M_{W_R}^2}{M_{\psi^0}^2}\right)^{1/2}\left(1-\frac{M_{W_R}^2}{2M_{\psi^0}^2}\right)^{-2}\left(1-\frac{M_{Z_R}^2}{4M_{\psi^0}^2}\right)^{-2}\nonumber\\
    &\times\left(1+4\frac{M_{W_R}^2}{M_{\psi^0}^2}-\frac{17}{4}\frac{M_{W_R}^4}{M_{\psi^0}^4}-\frac{3}{4}\frac{M_{W_R}^6}{M_{\psi^0}^6}\right),\nonumber\\
    \sigma_0v|_{Z_Lh}&=\frac{\pi\alpha_2^2}{12M_{\psi^0}^2}\frac{(1-3 s_L^2)^2}{(1-2s_L^2)^2}\left(1-\frac{c_L^2}{1-3s_L^2}\frac{M_{W_R}^2}{4M_{\psi^0}^2}\right)^{2}\left(1-\frac{M_{Z_R}^2}{4M_{\psi^0}^2}\right)^{-2}.\nonumber
\end{align}

\subsubsection{$\psi^+\ol{\psi^-}$}
For an initial state with spin-0,
\begin{align}
    \sigma_0v|_{W_LW_L}&=\frac{4\pi \alpha_2^2}{ M_{\psi^0}^2}.
\end{align}

\subsubsection{$\psi^+\ol{\psi^0}$}
For an initial state with spin-0,
\begin{align}
    \sigma_0v|_{W_LZ_L}&=\frac{2\pi \alpha_2^2c_L^2}{ M_{\psi^0}^2},~
    \sigma_0v|_{W_LZ_R}=\frac{8\pi \alpha_2^2}{ M_{\psi^0}^2}\frac{c_L^2}{(1-2s_L^2)}\left(1-\frac{M_{Z_R}^2}{4M_{\psi^0}^2}\right),~
    \sigma_0v|_{W_L\gamma}=\frac{2\pi \alpha_2^2s_L^2}{ M_{\psi^0}^2}.
\end{align}

For an initial state with spin-1,
\begin{align}
    \sigma_0v|_{f\ol{f'}}&=\frac{\pi \alpha_2^2}{3M_{\psi^0}^2},~\sigma_0v|_{W_Lh}=\frac{\pi \alpha_2^2}{12M_{\psi^0}^2},~\sigma_0v|_{W_LZ_L}=\frac{\pi \alpha_2^2}{12M_{\psi^0}^2}.
\end{align}

\subsubsection{$\psi^-\ol{\psi^-}$}
For an initial state with spin-0,
\begin{align}
    \sigma_0v|_{W_LW_L}&=\frac{2\pi \alpha_2^2}{ M_{\psi^0}^2},~
    \sigma_0v|_{W_RW_R}=\frac{2\pi \alpha_2^2}{ M_{\psi^0}^2}\left(1-\frac{M_{W_R}^2}{M_{\psi^0}^2}\right)^{3/2}\left(1-\frac{M_{W_R}^2}{2M_{\psi^0}^2}\right)^{-2},~
    \sigma_0v|_{Z_LZ_L}=\frac{4\pi \alpha_2^2c_L^4}{ M_{\psi^0}^2},\nonumber\\
    \sigma_0v|_{Z_LZ_R}&=\frac{8\pi \alpha_2^2}{ M_{\psi^0}^2}\frac{c_L^4}{(1-2s_L^2)}\left(1-\frac{M_{Z_R}^2}{4M_{\psi^0}^2}\right),~
    \sigma_0v|_{Z_L\gamma}=\frac{8\pi \alpha_2^2s_L^2c_L^2}{ M_{\psi^0}^2},\\
    \sigma_0v|_{Z_RZ_R}&=\frac{4\pi \alpha_2^2}{ M_{\psi^0}^2}\frac{c_L^4}{(1-2s_L^2)^2}\left(1-\frac{M_{Z_R}^2}{M_{\psi^0}^2}\right)^{3/2}\left(1-\frac{M_{Z_R}^2}{2M_{\psi^0}^2}\right)^{-2},\nonumber\\
    \sigma_0v|_{Z_R\gamma}&=\frac{8\pi \alpha_2^2}{ M_{\psi^0}^2}\frac{c_L^2s_L^2}{(1-2s_L^2)}\left(1-\frac{M_{Z_R}^2}{4M_{\psi^0}^2}\right),~
    \sigma_0v|_{\gamma\gamma}=\frac{4\pi \alpha_2^2s_L^4}{ M_{\psi^0}^2}. \nonumber
\end{align}

For an initial state with spin-1,
\begin{align}
    \sigma_0v|_{f\ol{f}}&=\frac{\pi\alpha_2^2}{3 M_{\psi^0}^2}\frac{1}{(1-2s_L^2)^2}\left(1-\frac{M_{Z_R}^2}{4M_{\psi^0}^2}\right)^{-2}\\
    &\times\left[s_L^4(4I_3(I_3-1)+4(Q_f+1)Q_f+2)-2s_L^2 (2Q_f+1)+1\right.\nonumber\\
    &+\frac{M_{Z_R}^2}{4M_{\psi^0}^2}(1-2s_L^2)(2s_L^2(Q_f-I_3+1)-1)\nonumber\\
    &\left.+\frac{M_{Z_R}^4}{32 M_{\psi^0}^4}(1-2s_L^2)^2\right]\nonumber\\
    \sigma_0v|_{W_LW_L}&=\frac{\pi\alpha_2^2}{12M_{\psi^0}^2}\frac{c_L^4}{(1-2s_L^2)^2}\left(1-\frac{M_{W_R}^2}{4M_{\psi^0}^2}\right)^{2}\left(1-\frac{M_{Z_R}^2}{4M_{\psi^0}^2}\right)^{-2},\nonumber\\
    \sigma_0v|_{W_RW_R}&=\frac{\pi\alpha_2^2}{12M_{\psi^0}^2}\frac{(1-3s_L^2)^2}{(1-2s_L^2)^2}\left(1-\frac{M_{W_R}^2}{M_{\psi^0}^2}\right)^{1/2}\left(1-\frac{M_{W_R}^2}{2M_{\psi^0}^2}\right)^{-2}\left(1-\frac{M_{Z_R}^2}{4M_{\psi^0}^2}\right)^{-2}\nonumber\\
    &\times\left(1+4\frac{M_{W_R}^2}{M_{\psi^0}^2}-\frac{17}{4}\frac{M_{W_R}^4}{M_{\psi^0}^4}-\frac{3}{4}\frac{M_{W_R}^6}{M_{\psi^0}^6}\right),\nonumber\\
    \sigma_0v|_{Z_Lh}&=\frac{\pi\alpha_2^2}{12M_{\psi^0}^2}\frac{(1-3 s_L^2)^2}{(1-2s_L^2)^2}\left(1-\frac{c_L^2}{1-3s_L^2}\frac{M_{W_R}^2}{4M_{\psi^0}^2}\right)^{2}\left(1-\frac{M_{Z_R}^2}{4M_{\psi^0}^2}\right)^{-2}.\nonumber
\end{align}

\subsubsection{$\psi^-\ol{\psi^0}$}
For an initial state with spin-0,
\begin{align}
    \sigma_0v|_{W_LZ_L}&=\frac{2\pi \alpha_2^2c_L^2}{ M_{\psi^0}^2},~
    \sigma_0v|_{W_LZ_R}=\frac{8\pi \alpha_2^2}{ M_{\psi^0}^2}\frac{c_L^2}{(1-2s_L^2)}\left(1-\frac{M_{Z_R}^2}{4M_{\psi^0}^2}\right),~
    \sigma_0v|_{W_L\gamma}&=\frac{2\pi \alpha_2^2s_L^2}{ M_{\psi^0}^2}.
\end{align}

For an initial state with spin-1,
\begin{align}
    \sigma_0v|_{f\ol{f'}}&=\frac{\pi \alpha_2^2}{3M_{\psi^0}^2},~\sigma_0v|_{W_Lh}=\frac{\pi \alpha_2^2}{12M_{\psi^0}^2},~\sigma_0v|_{W_LZ_L}=\frac{\pi \alpha_2^2}{12M_{\psi^0}^2}.
\end{align}

\subsubsection{$\psi^0\ol{\psi^0}$}
For an initial state with spin-0,
\begin{align}
    \sigma_0v|_{W_LW_L}&=\frac{8\pi \alpha_2^2}{ M_{\psi^0}^2},~
    \sigma_0v|_{W_RW_R}=\frac{2\pi \alpha_2^2}{ M_{\psi^0}^2}\left(1-\frac{M_{W_R}^2}{M_{\psi^0}^2}\right)^{3/2}\left(1-\frac{M_{W_R}^2}{2M_{\psi^0}^2}\right)^{-2},\\
    \sigma_0v|_{Z_LZ_L}&=\frac{64\pi \alpha_2^2s_L^8}{c_L^4M_{\psi^0}^2},~
    \sigma_0v|_{Z_RZ_R}=\frac{4\pi \alpha_2^2}{ M_{\psi^0}^2}\frac{c_L^4}{(1-2s_L^2)^2}\left(1-\frac{M_{Z_R}^2}{M_{\psi^0}^2}\right)^{3/2}\left(1-\frac{M_{Z_R}^2}{2M_{\psi^0}^2}\right)^{-2}. \nonumber
\end{align}

For an initial state with spin-1,
\begin{align}
    \sigma_0v|_{f\ol{f}}&=\frac{\pi\alpha_2^2}{6M_{\psi^0}^2}\frac{1}{(1-2s_L^2)^2}\left(1-\frac{M_{Z_R}^2}{4M_{\psi^0}^2}\right)^{-2}\left(8s_L^4(I_3^2+Q_f^2)-4s_L^2(I_3+Q_f)+1\right),\\
    \sigma_0v|_{W_LW_L}&=\frac{\pi \alpha_2^2}{ 12M_{\psi^0}^2}\frac{s_L^4}{(1-2s_L^2)^2}\left(1-\frac{M_{Z_R}^2}{4M_{\psi^0}^2}\right)^{-2},\nonumber\\
    \sigma_0v|_{W_RW_R}&=\frac{\pi\alpha_2^2}{12 M_{\psi^0}^2}\frac{(1-3s_L^2)^2}{(1-2 s_L^2)^2}\left(1-\frac{M_{W_R}^2}{M_{\psi^0}^2}\right)^{3/2}\left(1-\frac{M_{Z_R}^2}{4M_{\psi^0}^2}\right)^{-2}\left(1-\frac{M_{W_R}^2}{2M_{\psi^0}^2}\right)^{-2}\nonumber\\
    &\times\left(1+5\frac{M_{W_R}^2}{M_{\psi^0}^2}+\frac{3}{4}\frac{M_{W_R}^4}{M_{\psi^0}^4}\right),\nonumber\\
    \sigma_0v|_{Z_Lh}&=\frac{\pi \alpha_2^2}{ 12M_{\psi^0}^2}\frac{s_L^4}{(1-2s_L^2)^2}\left(1-\frac{M_{Z_R}^2}{4M_{\psi^0}^2}\right)^{-2}. \nonumber
\end{align}

\subsection{$R_{1,2,3}$ and $S_{0,1,2}$ mixing}
\subsubsection{$R_1\ol{S_0}$}
For an initial state with spin-0,
\begin{align}
    \sigma_0v|_{W_RZ_L}&=\frac{2\pi \alpha_2^2}{ M_{\psi^0}^2}\frac{(s_L^2+1)^2}{c_L^2}\left(1-\frac{M_{W_R}^2}{4M_{\psi^0}^2}\right),~\sigma_0v|_{W_R\gamma}=\frac{2\pi \alpha_2^2s_L^2}{ M_{\psi^0}^2}\left(1-\frac{M_{W_R}^2}{4M_{\psi^0}^2}\right),\\
    \sigma_0v|_{W_RZ_R}&=\frac{2\pi\alpha_2^2}{M_{\psi^0}^2}\frac{(1-4s_L^2)^2}{c_L^2(1-2 s_L^2)}\left(1+\frac{2-3s_L^2}{1-2s_L^2}\frac{M_{W_R}^2}{2M_{\psi^0}^2}+\frac{s_L^4}{(1-2s_L^2)^2}\frac{M_{W_R}^4}{16M_{\psi^0}^4}\right)^{3/2}\nonumber\\
    &\times\left(1-\frac{2-3s_L^2}{1-2s_L^2}\frac{M_{W_R}^2}{4M_{\psi^0}^2}\right)^{-2}.\nonumber
\end{align}

For an initial state with spin-1,
\begin{align}
    \sigma_0v|_{f\ol{f'}}&=\frac{\pi \alpha_2^2}{3M_{\psi^0}^2}\left(1-\frac{M_{W_R}^2}{4M_{\psi^0}^2}\right)^{-2},\\
    \sigma_0v|_{W_RZ_R}&=\frac{\pi\alpha_2^2}{12M_{\psi^0}^2}\frac{1}{c_L^{10}}\left(1-\frac{M_{W_R}^2}{4M_{\psi^0}^2}\right)^{-2}\left(1-\frac{2-3s_L^2}{c_L^2}\frac{M_{Z_R}^2}{4M_{\psi^0}^2}\right)^{-2}\left(1-\frac{M_{W_R}^2}{2M_{\psi^0}^2}+t_L^4\frac{M_{Z_R}^4}{16M_{\psi^0}^4}\right)^{1/2}\nonumber\\
    &\times\left(c_L^8+2c_L^6(2-3s_L^2)\frac{M_{Z_R}^2}{M_{\psi^0}^2}-c_L^2(77s_L^4-102s_L^2+34)\frac{M_{Z_R}^4}{8M_{\psi^0}^4}\right.\nonumber\\
    &\left.-(15s_L^8-52s_L^6+64s_L^4-33s_L^2+6)\frac{M_{Z_R}^6}{8M_{\psi^0}^6}+(25s_L^8-36s_L^6+12s_L^4)\frac{M_{Z_R}^8}{256M_{\psi^0}^8}\right).\nonumber
\end{align}

\subsubsection{$R_1\ol{S_1}$}
For an initial state with spin-0,
\begin{align}
    \sigma_0v|_{W_RZ_L}&=\frac{8\pi \alpha_2^2}{M_{\psi^0}^2}\left(1-\frac{M_{W_R}^2}{4M_{\psi^0}^2}\right).
\end{align}

\subsubsection{$R_2\ol{S_0}$}
For an initial state with spin-0,
\begin{align}
    \sigma_0v|_{W_LW_R}&=\frac{8\pi \alpha_2^2}{M_{\psi^0}^2}\left(1-\frac{M_{W_R}^2}{4M_{\psi^0}^2}\right).
\end{align}

\subsubsection{$R_2\ol{S_1}$}
For an initial state with spin-0,
\begin{align}
    \sigma_0v|_{W_RZ_L}&=\frac{18\pi \alpha_2^2}{M_{\psi^0}^2}\frac{s_L^4}{c_L^2}\left(1-\frac{M_{W_R}^2}{4M_{\psi^0}^2}\right),~\sigma_0v|_{W_R\gamma}=\frac{18\pi \alpha_2^2s_L^2}{M_{\psi^0}^2}\left(1-\frac{M_{W_R}^2}{4M_{\psi^0}^2}\right),\\
    \sigma_0v|_{W_RZ_R}&=\frac{2\pi\alpha_2^2}{M_{\psi^0}^2}\frac{(1-4s_L^2)^2}{c_L^2(1-2 s_L^2)}\left(1+\frac{2-3s_L^2}{1-2s_L^2}\frac{M_{W_R}^2}{2M_{\psi^0}^2}+\frac{s_L^4}{(1-2s_L^2)^2}\frac{M_{W_R}^4}{16M_{\psi^0}^4}\right)^{3/2}\nonumber\\
    &\times\left(1-\frac{2-3s_L^2}{1-2s_L^2}\frac{M_{W_R}^2}{4M_{\psi^0}^2}\right)^{-2}.\nonumber
\end{align}

For an initial state with spin-1,
\begin{align}
    \sigma_0v|_{f\ol{f'}}&=\frac{\pi \alpha_2^2}{3M_{\psi^0}^2}\left(1-\frac{M_{W_R}^2}{4M_{\psi^0}^2}\right)^{-2},\\
    \sigma_0v|_{W_RZ_R}&=\frac{\pi\alpha_2^2}{12M_{\psi^0}^2}\frac{1}{c_L^{10}}\left(1-\frac{M_{W_R}^2}{4M_{\psi^0}^2}\right)^{-2}\left(1-\frac{2-3s_L^2}{c_L^2}\frac{M_{Z_R}^2}{4M_{\psi^0}^2}\right)^{-2}\left(1-\frac{M_{W_R}^2}{2M_{\psi^0}^2}+t_L^4\frac{M_{Z_R}^4}{16M_{\psi^0}^4}\right)^{1/2}\nonumber\\
    &\times\left(c_L^8+2c_L^6(2-3s_L^2)\frac{M_{Z_R}^2}{M_{\psi^0}^2}-c_L^2(77s_L^4-102s_L^2+34)\frac{M_{Z_R}^4}{8M_{\psi^0}^4}\right.\nonumber\\
    &\left.-(15s_L^8-52s_L^6+64s_L^4-33s_L^2+6)\frac{M_{Z_R}^6}{8M_{\psi^0}^6}+(25s_L^8-36s_L^6+12s_L^4)\frac{M_{Z_R}^8}{256M_{\psi^0}^8}\right).\nonumber
\end{align}

\subsubsection{$R_2\ol{S_2}$}
For an initial state with spin-0,
\begin{align}
    \sigma_0v|_{W_LW_R}&=\frac{8\pi \alpha_2^2}{M_{\psi^0}^2}\left(1-\frac{M_{W_R}^2}{4M_{\psi^0}^2}\right).
\end{align}
\subsubsection{$R_3\ol{S_1}$}
For an initial state with spin-0,
\begin{align}
    \sigma_0v|_{W_LW_R}&=\frac{8\pi \alpha_2^2}{M_{\psi^0}^2}\left(1-\frac{M_{W_R}^2}{4M_{\psi^0}^2}\right).
\end{align}

\subsubsection{$R_3\ol{S_2}$}
For an initial state with spin-0,
\begin{align}
    \sigma_0v|_{W_RZ_L}&=\frac{2\pi \alpha_2^2}{M_{\psi^0}^2}\frac{(5s_L^2-2)^2}{c_L^2}\left(1-\frac{M_{W_R}^2}{4M_{\psi^0}^2}\right),~\sigma_0v|_{W_R\gamma}=\frac{50\pi \alpha_2^2}{M_{\psi^0}^2s_L^2}\left(1-\frac{M_{W_R}^2}{4M_{\psi^0}^2}\right),\\
    \sigma_0v|_{W_RZ_R}&=\frac{2\pi\alpha_2^2}{M_{\psi^0}^2}\frac{(1-4s_L^2)^2}{c_L^2(1-2 s_L^2)}\left(1+\frac{2-3s_L^2}{1-2s_L^2}\frac{M_{W_R}^2}{2M_{\psi^0}^2}+\frac{s_L^4}{(1-2s_L^2)^2}\frac{M_{W_R}^4}{16M_{\psi^0}^4}\right)^{3/2}\nonumber\\
    &\times\left(1-\frac{2-3s_L^2}{1-2s_L^2}\frac{M_{W_R}^2}{4M_{\psi^0}^2}\right)^{-2}.\nonumber
\end{align}

For an initial state with spin-1,
\begin{align}
    \sigma_0v|_{f\ol{f'}}&=\frac{\pi \alpha_2^2}{3M_{\psi^0}^2}\left(1-\frac{M_{W_R}^2}{4M_{\psi^0}^2}\right)^{-2},\\
    \sigma_0v|_{W_RZ_R}&=\frac{\pi\alpha_2^2}{12M_{\psi^0}^2}\frac{1}{c_L^{10}}\left(1-\frac{M_{W_R}^2}{4M_{\psi^0}^2}\right)^{-2}\left(1-\frac{2-3s_L^2}{c_L^2}\frac{M_{Z_R}^2}{4M_{\psi^0}^2}\right)^{-2}\left(1-\frac{M_{W_R}^2}{2M_{\psi^0}^2}+t_L^4\frac{M_{Z_R}^4}{16M_{\psi^0}^4}\right)^{1/2}\nonumber\\
    &\times\left(c_L^8+2c_L^6(2-3s_L^2)\frac{M_{Z_R}^2}{M_{\psi^0}^2}-c_L^2(77s_L^4-102s_L^2+34)\frac{M_{Z_R}^4}{8M_{\psi^0}^4}\right.\nonumber\\
    &\left.-(15s_L^8-52s_L^6+64s_L^4-33s_L^2+6)\frac{M_{Z_R}^6}{8M_{\psi^0}^6}+(25s_L^8-36s_L^6+12s_L^4)\frac{M_{Z_R}^8}{256M_{\psi^0}^8}\right).\nonumber
\end{align}

\subsection{$R_{1,2,3}$ and $\psi^{\pm,0}$ mixing}
\subsubsection{$R_1\ol{\psi^-}$}
For an initial state with spin-0,
\begin{align}
    \sigma_0v|_{W_RW_R}&=\frac{4\pi \alpha_2^2}{M_{\psi^0}^2}\left(1-\frac{M_{W_R}^2}{M_{\psi^0}^2}\right)^{3/2}\left(1-\frac{M_{W_R}^2}{2M_{\psi^0}^2}\right)^{-2}.
\end{align}

\subsubsection{$R_2\ol{\psi^0}$}
For an initial state with spin-0,
\begin{align}
    \sigma_0v|_{W_RW_R}&=\frac{4\pi \alpha_2^2}{M_{\psi^0}^2}\left(1-\frac{M_{W_R}^2}{M_{\psi^0}^2}\right)^{3/2}\left(1-\frac{M_{W_R}^2}{2M_{\psi^0}^2}\right)^{-2}.
\end{align}

\subsubsection{$R_3\ol{\psi^+}$}
For an initial state with spin-0,
\begin{align}
    \sigma_0v|_{W_RW_R}&=\frac{4\pi \alpha_2^2}{M_{\psi^0}^2}\left(1-\frac{M_{W_R}^2}{M_{\psi^0}^2}\right)^{3/2}\left(1-\frac{M_{W_R}^2}{2M_{\psi^0}^2}\right)^{-2}.
\end{align}

\subsection{$S_{0,1,2}$ and $\psi^{\pm,0}$ mixing}

\subsubsection{$S_0\ol{\psi^-}$}
For an initial state with spin-0,
\begin{align}
    \sigma_0v|_{W_RZ_L}&=\frac{2\pi \alpha_2^2}{M_{\psi^0}^2}\frac{(2-s_L^2)^2}{c_L^2}\left(1-\frac{M_{W_R}^2}{4M_{\psi^0}^2}\right),~ \sigma_0v|_{W_R\gamma}=\frac{2\pi \alpha_2^2s_L^2}{M_{\psi^0}^2}\left(1-\frac{M_{W_R}^2}{4M_{\psi^0}^2}\right),\\
    \sigma_0v|_{W_RZ_R}&=\frac{2\pi\alpha_2^2}{M_{\psi^0}^2}\frac{1}{c_L^2(1-2 s_L^2)}\left(1+\frac{2-3s_L^2}{1-2s_L^2}\frac{M_{W_R}^2}{2M_{\psi^0}^2}+\frac{s_L^4}{(1-2s_L^2)^2}\frac{M_{W_R}^4}{16M_{\psi^0}^4}\right)^{3/2}\nonumber\\
    &\times\left(1-\frac{2-3s_L^2}{1-2s_L^2}\frac{M_{W_R}^2}{4M_{\psi^0}^2}\right)^{-2}.\nonumber
\end{align}

For an initial state with spin-1,
\begin{align}
    \sigma_0v|_{f\ol{f'}}&=\frac{\pi \alpha_2^2}{3M_{\psi^0}^2}\left(1-\frac{M_{W_R}^2}{4M_{\psi^0}^2}\right)^{-2},\\
    \sigma_0v|_{W_RZ_R}&=\frac{\pi\alpha_2^2}{12M_{\psi^0}^2}\frac{1}{c_L^{10}}\left(1-\frac{M_{W_R}^2}{4M_{\psi^0}^2}\right)^{-2}\left(1-\frac{2-3s_L^2}{c_L^2}\frac{M_{Z_R}^2}{4M_{\psi^0}^2}\right)^{-2}\left(1-\frac{M_{W_R}^2}{2M_{\psi^0}^2}+t_L^4\frac{M_{Z_R}^4}{16M_{\psi^0}^4}\right)^{1/2}\nonumber\\
    &\times\left(c_L^8+2c_L^6(2-3s_L^2)\frac{M_{Z_R}^2}{M_{\psi^0}^2}-c_L^2(77s_L^4-102s_L^2+34)\frac{M_{Z_R}^4}{8M_{\psi^0}^4}\right.\nonumber\\
    &\left.-(15s_L^8-52s_L^6+64s_L^4-33s_L^2+6)\frac{M_{Z_R}^6}{8M_{\psi^0}^6}+(25s_L^8-36s_L^6+12s_L^4)\frac{M_{Z_R}^8}{256M_{\psi^0}^8}\right).\nonumber
\end{align}

\subsubsection{$S_0\ol{\psi^0}$}
For an initial state with spin-0,
\begin{align}
    \sigma_0v|_{W_LW_R}&=\frac{8\pi \alpha_2^2}{ M_{\psi^0}^2}\left(1-\frac{M_{W_R}^2}{4 M_{\psi^0}^2}\right).
\end{align}

\subsubsection{$S_1\ol{\psi^+}$}
For an initial state with spin-0,
\begin{align}
    \sigma_0v|_{W_LW_R}=\frac{8\pi \alpha_2^2}{ M_{\psi^0}^2}\left(1-\frac{M_{W_R}^2}{4 M_{\psi^0}^2}\right).
\end{align}

\subsubsection{$S_1\ol{\psi^-}$}
For an initial state with spin-0,
\begin{align}
    \sigma_0v|_{W_LW_R}=\frac{8\pi \alpha_2^2}{ M_{\psi^0}^2}\left(1-\frac{M_{W_R}^2}{4 M_{\psi^0}^2}\right).
\end{align}

\subsubsection{$S_1\ol{\psi^0}$}
For an initial state with spin-0,
\begin{align}
    \sigma_0v|_{W_RZ_L}&=\frac{2\pi \alpha_2^2}{ M_{\psi^0}^2}\frac{s_L^4}{c_L^2}\left(1-\frac{M_{W_R}^2}{4 M_{\psi^0}^2}\right),~\sigma_0v|_{W_R\gamma}=\frac{2\pi \alpha_2^2s_L^2}{ M_{\psi^0}^2}\left(1-\frac{M_{W_R}^2}{4 M_{\psi^0}^2}\right),\\
    \sigma_0v|_{W_RZ_R}&=\frac{2\pi\alpha_2^2}{M_{\psi^0}^2}\frac{1}{c_L^2(1-2 s_L^2)}\left(1+\frac{2-3s_L^2}{1-2s_L^2}\frac{M_{W_R}^2}{2M_{\psi^0}^2}+\frac{s_L^4}{(1-2s_L^2)^2}\frac{M_{W_R}^4}{16M_{\psi^0}^4}\right)^{3/2}\nonumber\\
    &\times\left(1-\frac{2-3s_L^2}{1-2s_L^2}\frac{M_{W_R}^2}{4M_{\psi^0}^2}\right)^{-2}.\nonumber
\end{align}

For an initial state with spin-1,
\begin{align}
    \sigma_0v|_{f\ol{f'}}&=\frac{\pi \alpha_2^2}{ 3M_{\psi^0}^2}\left(1-\frac{M_{W_R}^2}{4 M_{\psi^0}^2}\right)^{-2},\\
    \sigma_0v|_{W_RZ_R}&=\frac{\pi\alpha_2^2}{12M_{\psi^0}^2}\frac{1}{c_L^{10}}\left(1-\frac{M_{W_R}^2}{4M_{\psi^0}^2}\right)^{-2}\left(1-\frac{2-3s_L^2}{c_L^2}\frac{M_{Z_R}^2}{4M_{\psi^0}^2}\right)^{-2}\left(1-\frac{M_{W_R}^2}{2M_{\psi^0}^2}+t_L^4\frac{M_{Z_R}^4}{16M_{\psi^0}^4}\right)^{1/2}\nonumber\\
    &\times\left(c_L^8+2c_L^6(2-3s_L^2)\frac{M_{Z_R}^2}{M_{\psi^0}^2}-c_L^2(77s_L^4-102s_L^2+34)\frac{M_{Z_R}^4}{8M_{\psi^0}^4}\right. \nonumber\\
    &\left.-(15s_L^8-52s_L^6+64s_L^4-33s_L^2+6)\frac{M_{Z_R}^6}{8M_{\psi^0}^6}+(25s_L^8-36s_L^6+12s_L^4)\frac{M_{Z_R}^8}{256M_{\psi^0}^8}\right). \nonumber
\end{align}

\subsubsection{$S_2\ol{\psi^+}$}
For an initial state with spin-0,
\begin{align}
    \sigma_0v|_{W_RZ_L}&=\frac{2\pi \alpha_2^2}{ M_{\psi^0}^2}\frac{(2-3s_L^2)^2}{c_L^2}\left(1-\frac{M_{W_R}^2}{4 M_{\psi^0}^2}\right),~\sigma_0v|_{W_R\gamma}=\frac{18\pi \alpha_2^2s_L^2}{ M_{\psi^0}^2}\left(1-\frac{M_{W_R}^2}{4 M_{\psi^0}^2}\right),\\
    \sigma_0v|_{W_RZ_R}&=\frac{2\pi\alpha_2^2}{M_{\psi^0}^2}\frac{1}{c_L^2(1-2 s_L^2)}\left(1+\frac{2-3s_L^2}{1-2s_L^2}\frac{M_{W_R}^2}{2M_{\psi^0}^2}+\frac{s_L^4}{(1-2s_L^2)^2}\frac{M_{W_R}^4}{16M_{\psi^0}^4}\right)^{3/2} \nonumber\\
    &\times\left(1-\frac{2-3s_L^2}{1-2s_L^2}\frac{M_{W_R}^2}{4M_{\psi^0}^2}\right)^{-2}.\nonumber
\end{align}

For an initial state with spin-1,
\begin{align}
    \sigma_0v|_{f\ol{f'}}&=\frac{\pi \alpha_2^2}{3 M_{\psi^0}^2}\left(1-\frac{M_{W_R}^2}{4 M_{\psi^0}^2}\right)^{-2},\\
    \sigma_0v|_{W_RZ_R}&=\frac{\pi\alpha_2^2}{12M_{\psi^0}^2}\frac{1}{c_L^{10}}\left(1-\frac{M_{W_R}^2}{4M_{\psi^0}^2}\right)^{-2}\left(1-\frac{2-3s_L^2}{c_L^2}\frac{M_{Z_R}^2}{4M_{\psi^0}^2}\right)^{-2}\left(1-\frac{M_{W_R}^2}{2M_{\psi^0}^2}+t_L^4\frac{M_{Z_R}^4}{16M_{\psi^0}^4}\right)^{1/2} \nonumber\\
    &\times\left(c_L^8+2c_L^6(2-3s_L^2)\frac{M_{Z_R}^2}{M_{\psi^0}^2}-c_L^2(77s_L^4-102s_L^2+34)\frac{M_{Z_R}^4}{8M_{\psi^0}^4}\right.\nonumber\\
    &\left.-(15s_L^8-52s_L^6+64s_L^4-33s_L^2+6)\frac{M_{Z_R}^6}{8M_{\psi^0}^6}+(25s_L^8-36s_L^6+12s_L^4)\frac{M_{Z_R}^8}{256M_{\psi^0}^8}\right).\nonumber
\end{align}

\subsubsection{$S_2\ol{\psi^0}$}
For an initial state with spin-0,
\begin{align}
    \sigma_0v|_{W_LW_R}&=\frac{8\pi \alpha_2^2}{ M_{\psi^0}^2}\left(1-\frac{M_{W_R}^2}{4 M_{\psi^0}^2}\right).
\end{align}

\section{Long-range potentials for $(\mathbf{3},\mathbf{3},0)$}\label{App:long_range_potentials_330}

In this appendix, we give the long-range potentials relevant for the Sommerfeld effect for $(\mathbf{3},\mathbf{3},0)$ DM. We omit diagonal contributions from particle mass splittings for brevity, but take these into account in numerical computations. 

\subsection{$\chi^0$ and $\chi^\pm$}
\subsubsection{  $\chi^0 \chi^-$}
\begin{align}
    V = - \alpha_2 \frac{e^{-M_{W_L} r}}{r}.
\end{align}

\subsubsection{$\chi^+ \chi^-$ and $\chi^0$ $\chi^0$}
\begin{align}
    V = \begin{pmatrix}
            - \frac{\alpha}{r} - \alpha_2 c_L^2 \frac{e^{-M_{Z_L} r}}{r} & - \sqrt{2} \alpha_2 \frac{e^{-M_{W_L} r}}{r} \\
            - \sqrt{2} \alpha_2 \frac{e^{-M_{W_L} r}}{r}                            & 0
        \end{pmatrix},
\end{align}
in the basis $(\chi^+ \chi^-,\chi^0\chi^0)$.

\subsubsection{$\chi^- \chi^-$}
\begin{align}
    V = \frac{\alpha}{r} + \alpha_2 c_L^2 \frac{e^{-M_{Z_L} r}}{r}.
\end{align}

\subsection{$D$, $E$ and $N$}

\subsubsection{$\ol{N} D$}
\begin{align}
    V = \alpha_2 \frac{c_L^2-s_L^2}{c_L^2} \frac{e^{-M_{Z_L} r}}{r}.
\end{align}

\subsubsection{$\ol{D} D$, $\ol{E}E$ and $\ol{N} N$}
\begin{align}
    V = \begin{pmatrix}
             - \frac{4 \alpha}{r} - \alpha_2 \frac{(c_L^2 - s_L^2)^2}{c_L^2} \frac{e^{-M_{Z_L} r}}{r} & -  \alpha_2 \frac{e^{-M_{W_L} r}}{r}                                                    & 0                                              \\
            - \alpha_2 \frac{e^{-M_{W_L} r}}{r}                                                                     &- \frac{\alpha}{r} - \alpha_2 \frac{s_L^4}{c_L^2} \frac{e^{-M_{Z_L} r}}{r} & - \alpha_2 \frac{e^{-M_{W_L} r}}{r}                \\
            0                                                                                                   & - \alpha_2 \frac{e^{-M_{W_L} r}}{r}                                                     & - \alpha_2 \frac{1}{c_L^2}\frac{e^{-M_{Z_L} r}}{r}
        \end{pmatrix},
\end{align}
in the basis $(\ol{D} D,\ol{E}E,\ol{N} N)$.

\subsubsection{$D\ol{E}$ and $E \ol{N}$}
\begin{align}
    V = \begin{pmatrix}
           - \frac{2 \alpha}{r} + \alpha_2 \frac{s_L^2(c_L^2 - s_L^2)}{c_L^2} \frac{e^{-M_{Z_L} r}}{r} & + \alpha_2 \frac{e^{-M_{W_L} r}}{r}                    \\
            + \alpha_2 \frac{e^{-M_{W_L} r}}{r}                                                                   & -\alpha_2 \frac{s_L^2}{c_L^2} \frac{e^{-M_{Z_L} r}}{r}
        \end{pmatrix},
\end{align}
in the basis $(D\ol{E},E \ol{N})$.

\subsection{$\chi^0, \chi^\pm$ and $D,E$, $N$ mixing}

\subsubsection{$\chi^- E$ and $\chi^0 D$}
\begin{align}
    V(r) = \begin{pmatrix}
              \frac{\alpha}{r} - \frac{\alpha_2 s_L^2}{r} e^{-M_{Z_L} r} & -\alpha_2 \frac{e^{-M_{W_L} r}}{r} \\
               -\alpha_2 \frac{e^{-M_{W_L} r}}{r}                                               & 0
           \end{pmatrix},
\end{align}
in the basis $(\chi^- E, \chi^0D )$.

\subsubsection{$\chi^+ D$, $\chi^0 E$ and $N \chi^-$}
\begin{align}
    V(r) = \begin{pmatrix}
               - 2 \frac{\alpha}{r} - \frac{\alpha_2(c_L^2 - s_L^2)}{r} e^{-M_{Z_L} r} & - \alpha_2 \frac{e^{-M_{W_L} r}}{r} & 0                               \\
               - \alpha_2 \frac{e^{-M_{W_L} r}}{r}                                               & 0           &  \frac{\alpha_2}{r} e^{-M_{W_L} r} \\
               0                                                                             &  \frac{\alpha_2}{r} e^{-M_{W_L} r} & - \frac{\alpha_2}{r} e^{-M_{Z_L} r}
           \end{pmatrix},
\end{align}
in the basis $(\chi^+ D$, $\chi^0 E$, $\chi^- N)$.

\subsubsection{$\chi^- \ol{E}$ and $\chi^0 \ol{N}$}
\begin{align}
    V(r) = \begin{pmatrix}
               - \frac{\alpha}{r} + \frac{\alpha_2 s_L^2}{r} e^{-M_{Z_L} r} & - \frac{\alpha_2}{r} e^{-M_{W_L} r} \\
               - \frac{\alpha_2}{r} e^{-M_{W_L} r}                                               & 0
           \end{pmatrix},
\end{align}
in the basis $(\chi^- \ol{E}$, $\chi^0 \ol{N})$.

\section{Long-range potentials for $(\mathbf{3},\mathbf{3},1)$}\label{App:long_range_potentials_331}
In this appendix, we give the long-range potentials relevant for Sommerfeld effect for $(\mathbf{3},\mathbf{3},1)$ DM. We omit diagonal contributions from particle mass splittings for brevity, but take these into account in numerical computations.

\subsection{$R_1$, $R_2$ and $R_3$}
\subsubsection{$R_1\ol{R_1}$, $R_2\ol{R_2}$ and $R_3\ol{R_3}$}
\begin{align}
    V(r) = \begin{pmatrix}
               -\frac{\alpha}{r}-\frac{\alpha_2 (s_L^2+1)^2}{c_L^2}\frac{e^{-M_{Z_L} r}}{r} & - \frac{\alpha_2}{r} e^{-M_{W_L} r}&0 \\
               - \frac{\alpha_2}{r} e^{-M_{W_L} r}                                               &  -4\frac{\alpha}{r}-4\frac{\alpha_2 s_L^4}{c_L^2}\frac{e^{-M_{Z_L} r}}{r} &- \frac{\alpha_2}{r} e^{-M_{W_L} r}\\
               0&- \frac{\alpha_2}{r} e^{-M_{W_L} r}&-9\frac{\alpha}{r}-\frac{\alpha_2 (1-3s_L^2)^2}{c_L^2}\frac{e^{-M_{Z_L} r}}{r}
           \end{pmatrix},
\end{align}
in the basis $(R_1\ol{R_1},R_2\ol{R_2},R_3\ol{R_3})$.

\subsubsection{$R_1\ol{R_2}$ and $R_2\ol{R_3}$}
\begin{align}
    V(r) = \begin{pmatrix}
               -2\frac{\alpha}{r}-2\frac{\alpha_2 s_L^2(s_L^2+1)}{c_L^2}\frac{e^{-M_{Z_L} r}}{r} & \frac{\alpha_2}{r} e^{-M_{W_L} r}&\\
               \frac{\alpha_2}{r} e^{-M_{W_L} r}                                                &-6\frac{\alpha}{r}+2\frac{\alpha_2 s_L^2(1-3s_L^2)}{c_L^2}\frac{e^{-M_{Z_L} r}}{r}
           \end{pmatrix},
\end{align}
in the basis $(R_1\ol{R_2},R_2\ol{R_3})$.

\subsubsection{$R_1\ol{R_3}$}
\begin{align}
    V(r) = -3\frac{\alpha}{r}+\frac{\alpha_2 s_L^2(s_L^2+1)(1-3s_L^2)}{c_L^2}\frac{e^{-M_{Z_L} r}}{r}.
\end{align}

\subsection{$S_0$, $S_1$ and $S_2$}
\subsubsection{$S_0\ol{S_0}$, $S_1\ol{S_1}$ and $S_2\ol{S_2}$}
\begin{align}
    V(r) = \begin{pmatrix}
               -\frac{\alpha_2}{c_L^2}\frac{e^{-M_{Z_L} r}}{r} &-\alpha_2\frac{e^{-M_{W_L} r}}{r} & 0\\
              -\alpha_2\frac{e^{-M_{W_L} r}}{r}&-\frac{\alpha}{r}-\frac{\alpha_2s_L^4}{c_L^2}\frac{e^{-M_{Z_L} r}}{r}&-\alpha_2\frac{e^{-M_{W_L} r}}{r}\\
              0&-\alpha_2\frac{e^{-M_{W_L} r}}{r}&-4\frac{\alpha}{r}-\frac{\alpha_2(1-2s_L^2)^2}{c_L^2}\frac{e^{-M_{Z_L} r}}{r}
           \end{pmatrix},
\end{align}
in the basis $(S_0\ol{S_0},S_1\ol{S_1},S_2\ol{S_2})$.

\subsubsection{$S_0\ol{S_1}$ and $S_1\ol{S_2}$}
\begin{align}
    V(r) = \begin{pmatrix}
               -\frac{\alpha_2s_L^2}{c_L^2}\frac{e^{-M_{Z_L} r}}{r} &\alpha_2\frac{e^{-M_{W_L} r}}{r} \\
              \alpha_2\frac{e^{-M_{W_L} r}}{r}&-2\frac{\alpha}{r}+\frac{\alpha_2s_L^2(1-2s_L^2)}{c_L^2}\frac{e^{-M_{Z_L} r}}{r}
           \end{pmatrix},
\end{align}
in the basis $(S_0\ol{S_1},S_1\ol{S_2})$.

\subsubsection{$S_0\ol{S_2}$}
\begin{align}
    V(r) = \frac{\alpha_2(1-2s_L^2)}{c_L^2}\frac{e^{-M_{Z_L} r}}{r}.
\end{align}

\subsection{$\psi^+$, $\psi^-$ and $\psi^0$}
\subsubsection{$\psi^+\ol{\psi^+}$, $\psi^-\ol{\psi^-}$ and $\psi^0\ol{\psi^0}$}
\begin{align}
    V(r) = \begin{pmatrix}
               -\frac{\alpha}{r}-\alpha_2c_L^2\frac{e^{-M_{Z_L}r}}{r} &0&-\alpha_2\frac{e^{-M_{W_L} r}}{r} \\
               0&-\frac{\alpha}{r}-\alpha_2c_L^2\frac{e^{-M_{Z_L}r}}{r}&-\alpha_2\frac{e^{-M_{W_L} r}}{r}\\
              -\alpha_2\frac{e^{-M_{W_L} r}}{r}  &-\alpha_2\frac{e^{-M_{W_L} r}}{r}&0
           \end{pmatrix},
\end{align}
in the basis $(\psi^+\ol{\psi^+},\psi^-\ol{\psi^-},\psi^0\ol{\psi^0})$.

\subsubsection{$\psi^+\ol{\psi^-}$}
\begin{align}
    V(r) = \frac{\alpha}{r}+\alpha_2c_L^2\frac{e^{-M_{Z_L}r}}{r}.
\end{align}

\subsection{$R_{1,2,3}$ and  $S_{0,1,2}$ mixing}
\subsubsection{$R_1\ol{S_0}$, $R_2\ol{S_1}$ and $R_3\ol{S_2}$}
\begin{align}
    V(r) = \begin{pmatrix}
               -\frac{\alpha_2 (s_L^2+1)}{c_L^2}\frac{e^{-M_{Z_L} r}}{r} & - \frac{\alpha_2}{r} e^{-M_{W_L} r}&0 \\
               - \frac{\alpha_2}{r} e^{-M_{W_L} r}                                               &  -2\frac{\alpha}{r}-2\frac{\alpha_2 s_L^4}{c_L^2}\frac{e^{-M_{Z_L} r}}{r} &- \frac{\alpha_2}{r} e^{-M_{W_L} r}\\
               0&- \frac{\alpha_2}{r} e^{-M_{W_L} r}&-6\frac{\alpha}{r}-\frac{\alpha_2 (1-2s_L^2)(1-3s_L^2)}{c_L^2}\frac{e^{-M_{Z_L} r}}{r}
           \end{pmatrix},
\end{align}
in the basis $(R_1\ol{S_0},R_2\ol{S_1},R_3\ol{S_2})$.

\subsubsection{$R_1\ol{S_1}$ and $R_2\ol{S_2}$}
\begin{align}
    V(r) = \begin{pmatrix}
               -\frac{\alpha}{r}-\frac{\alpha_2 s_L^2(s_L^2+1)}{c_L^2}\frac{e^{-M_{Z_L} r}}{r} & \frac{\alpha_2}{r} e^{-M_{W_L} r}&\\
               \frac{\alpha_2}{r} e^{-M_{W_L} r}                                                &-4\frac{\alpha}{r}+2\frac{\alpha_2 s_L^2(1-2s_L^2)}{c_L^2}\frac{e^{-M_{Z_L} r}}{r}
           \end{pmatrix},
\end{align}
in the basis $(R_1\ol{S_1},R_2\ol{S_2})$.

\subsubsection{$R_2\ol{S_0}$ and $R_3\ol{S_1}$}
\begin{align}
    V(r) = \begin{pmatrix}
               -2\frac{\alpha_2 s_L^2}{c_L^2}\frac{e^{-M_{Z_L} r}}{r} & \frac{\alpha_2}{r} e^{-M_{W_L} r}&\\
               \frac{\alpha_2}{r} e^{-M_{W_L} r}                                                &-3\frac{\alpha}{r}+\frac{\alpha_2 s_L^2(1-3s_L^2)}{c_L^2}\frac{e^{-M_{Z_L} r}}{r}
           \end{pmatrix},
\end{align}
in the basis $(R_2\ol{S_0},R_3\ol{S_1})$.

\subsection{$S_{0,1,2}$ and $\psi^{\pm,0}$ mixing}
\subsubsection{$S_0\ol{\psi^-}$, $S_1\ol{\psi^0}$ and $S_2\ol{\psi^+}$}
\begin{align}
    V(r) = \begin{pmatrix}
               -\alpha_2\frac{e^{-M_{Z_L} r}}{r} &-\alpha_2\frac{e^{-M_{W_L} r}}{r}& 0\\
              -\alpha_2\frac{e^{-M_{W_L} r}}{r} &0&-\alpha_2\frac{e^{-M_{W_L} r}}{r}\\
              0&-\alpha_2\frac{e^{-M_{W_L} r}}{r}&-2\frac{\alpha}{r}-\alpha_2(1-2s_L^2)\frac{e^{-M_{Z_L} r}}{r}
           \end{pmatrix},
\end{align}
in the basis $(S_0\ol{\psi^-},S_1\ol{\psi^0},S_2\ol{\psi^+})$.

\subsubsection{$S_0\ol{\psi^0}$ and $S_1\ol{\psi^+}$}
\begin{align}
    V(r) = \begin{pmatrix}
               0&\alpha_2\frac{e^{-M_{W_L} r}}{r} \\
              \alpha_2\frac{e^{-M_{W_L} r}}{r} & -\frac{\alpha}{r}+\alpha_2s_L^2\frac{e^{-M_{Z_L}r}}{r}
           \end{pmatrix},
\end{align}
in the basis $(S_0\ol{\psi^0},S_1\ol{\psi^+})$.

\subsubsection{$S_1\ol{\psi^-}$ and $S_2\ol{\psi^0}$}
\begin{align}
    V(r) = \begin{pmatrix}
               \frac{\alpha}{r}-\alpha_2s_L^2\frac{e^{-M_{Z_L}r}}{r} &\alpha_2\frac{e^{-M_{W_L} r}}{r} \\
              \alpha_2\frac{e^{-M_{W_L} r}}{r} &0
           \end{pmatrix},
\end{align}
in the basis $(S_1\ol{\psi^-},S_2\ol{\psi^0})$.

\small{\bibliography{DM-Parity}}

\end{document}